\documentclass[aps,pra,twocolumn,superscriptaddress,10pt,floatfix,footinbib]{revtex4-1}

\usepackage[utf8]{inputenc}
\usepackage[TS1,T1]{fontenc}
\usepackage{microtype}
\usepackage[english]{babel}
\usepackage{amsmath}
\usepackage{amssymb}
\usepackage{graphicx}
\usepackage{scalerel}
\usepackage{dsfont}
\usepackage{fixmath}
\usepackage{braket}
\usepackage{xcolor}
\usepackage{siunitx}
\usepackage{hyperref}
\hypersetup{hypertexnames=false,breaklinks=true,colorlinks,linkcolor=blue,citecolor=blue,urlcolor=blue,final}
\usepackage{cleveref}
\newcommand{\csubref}[2]{\namecref{#1}~\hyperref[#1]{\ref*{#1}(#2)}}
\newcommand{\Csubref}[2]{\nameCref{#1}~\hyperref[#1]{\ref*{#1}(#2)}}
\newcommand{\subref}[2]{\hyperref[#1]{\ref*{#1}#2}}
\crefname{section}{Sec.}{Secs.}
\crefname{appendix}{Appendix}{Appendices}
\crefname{equation}{Eq.}{Eqs.}
\crefname{table}{Table}{Tables}
\crefname{figure}{Fig.}{Figs.}
\Crefname{figure}{Figure}{Figures}

\newcommand{\orcidicon}[1]{\href{https://orcid.org/#1}{\includegraphics[width=0.8em]{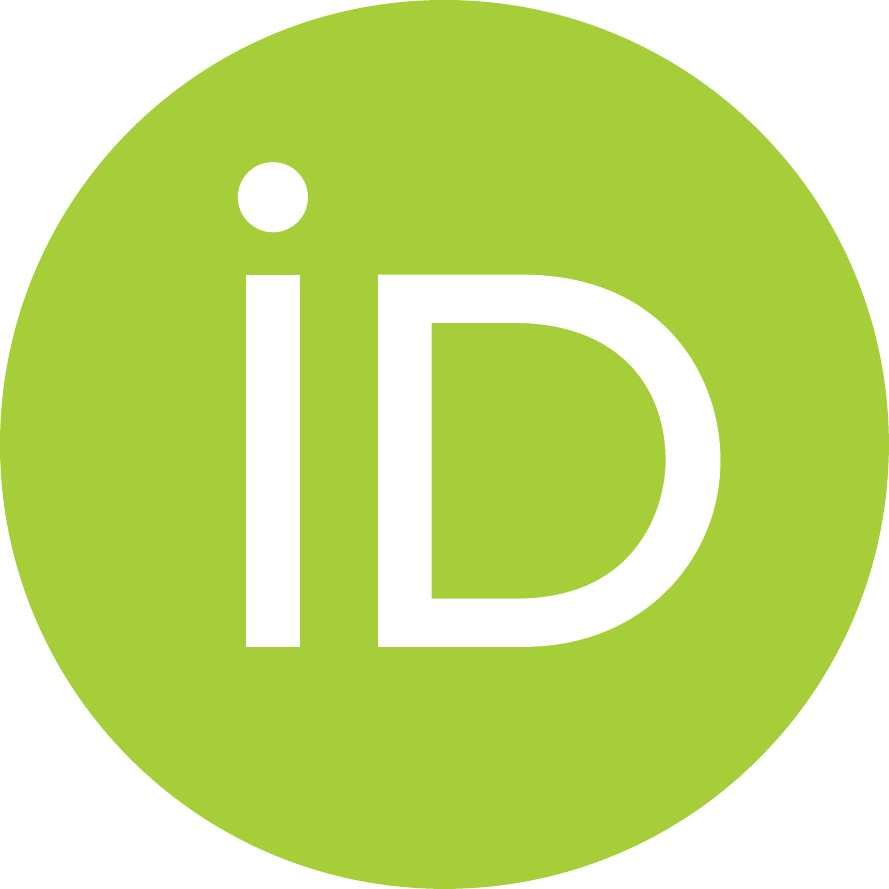}}}
\let\Im\relax
\DeclareMathOperator{\Im}{Im}
\let\Re\relax
\DeclareMathOperator{\Re}{Re}
\newcommand{\freq}{\omega}
\newcommand{\atindex}{\xi}
\newcommand{\atindexb}{{\atindex'}}
\newcommand{\atLinewidth}{\Gamma}
\newcommand{\atfreq}{\omega}
\newcommand{\atintfreq}{\omega_0}
\newcommand{\pot}{V}
\newcommand{\adsorption}{\text{ad}}
\newcommand{\optical}{\text{opt}}
\newcommand{\rpos}{r}
\newcommand{\phipos}{\varphi}
\newcommand{\zpos}{z}
\newcommand{\rad}{R}
\newcommand{\len}{L}
\newcommand{\temp}{T}
\newcommand{\op}[1]{\hat{#1}}
\newcommand{\Hamilop}{\op{\Hamilfunc}}
\newcommand{\Hamilfunc}{H}
\newcommand{\motional}{\text{ext}}
\newcommand{\electronic}{\text{int}}
\newcommand{\vibrational}{\text{phn}}
\newcommand{\atphonint}{{\motional\text{-}\vibrational}}
\newcommand{\effective}{\text{eff}}
\newcommand{\phonindex}{\mu}
\newcommand{\phonindexb}{{\phonindex'}}
\newcommand{\ccsign}{*}
\newcommand{\cconj}[1]{{#1}^\ccsign}
\newcommand{\hconj}[1]{{#1}^\dagger}
\newcommand{\phonaop}{\op{b}}
\newcommand{\phonfreq}{\omega}
\newcommand{\atradwf}{\psi}
\newcommand{\atmass}{M}
\newcommand{\pare}[1]{\left( {#1} \right)}
\newcommand{\spare}[1]{\left[ {#1} \right]}
\newcommand{\cpare}[1]{\left\{ {#1} \right\}}
\newcommand{\positionsymbol}{r}
\newcommand{\pos}{\vec{\positionsymbol}}
\renewcommand{\vec}[1]{\mathbold{#1}}

\newcommand{\atphong}{g}
\newcommand{\hc}{\text{H.c.}}
\newcommand{\atomoverlap}{\mathcal{A}}
\newcommand{\dens}{\rho}
\newcommand{\dd}{\mathop{}\!\mathrm{d}}
\newcommand{\atmomsymbol}{p}
\newcommand{\atmom}{\vec{\atmomsymbol}}
\newcommand{\atmomop}{\op\atmom}
\newcommand{\atpossymbol}{r}
\newcommand{\atpos}{\vec{\atpossymbol}}
\newcommand{\atposop}{\op\atpos}
\newcommand{\wmodecomp}{w}
\newcommand{\wmode}{\vec{\wmodecomp}}
\newcommand{\wmodercomp}{\mathcal{W}}
\newcommand{\wmoder}{\boldsymbol{\wmodercomp}}
\newcommand{\im}{i}
\newcommand{\phonl}{j}
\newcommand{\phonlb}{{\phonl'}}
\newcommand{\phonk}{p}
\newcommand{\phonkb}{{\phonk'}}

\newcommand{\cFL}{v}

\newcommand{\YoungE}{E}

\newcommand{\minimum}{\text{min}}
\newcommand{\CPpotstrength}{C}
\newcommand{\Ppotstrength}{D}
\newcommand{\wavelen}{\lambda}
\newcommand{\power}{P}
\newcommand{\ufieldcomp}{u}
\newcommand{\ufield}{\vec{\ufieldcomp}}

\newcommand{\permitt}{\epsilon}
\newcommand{\relpermitt}{\permitt}

\newcommand{\atn}{\nu}
\newcommand{\atnb}{{\atn'}}
\newcommand{\atl}{l}
\newcommand{\atk}{q}
\newcommand{\atwf}{\Psi}
\newcommand{\R}{\mathds{R}}
\newcommand{\Z}{\mathds{Z}}
\newcommand{\N}{\mathds{N}}
\newcommand{\kronecker}{\delta}
\newcommand{\post}{\pos,\tm}
\newcommand{\tm}{t}
\newcommand{\ufieldmodedens}{U}
\newcommand{\ufieldop}{\op{\ufield}}
\newcommand{\ufieldcompop}{\op{\ufieldcomp}}
\newcommand{\atphipos}{\phipos}
\newcommand{\atphiposop}{\op\atphipos}
\newcommand{\atzpos}{\zpos}
\newcommand{\atzposop}{\op\atzpos}
\newcommand{\atrpos}{\rpos}
\newcommand{\atrposop}{\op\atrpos}
\newcommand{\boltzmann}{k_B}
\newcommand{\nbar}{\bar{n}}
\newcommand{\trappos}{\atpos_0}

\newcommand{\trapfreq}{\omega}
\newcommand{\rtrapfreq}{\trapfreq_{\rpos}}

\newcommand{\itrapfreq}{\trapfreq_{ i}}
\newcommand{\trapr}{\atrpos_0}
\newcommand{\trapphi}{\atphipos_0}
\newcommand{\trapz}{\atzpos_0}
\newcommand{\ataop}{\hat{a}}

\newcommand{\DOS}{\rho}
\newcommand{\atlb}{{\atl'}}
\newcommand{\dirac}{\delta}
\newcommand{\atkb}{{\atk'}}

\newcommand{\photk}{k}

\newcommand{\photfreqoutin}{\omega}
\newcommand{\dipolecomp}{d}
\newcommand{\dipolecompop}{\op{\dipolecomp}}
\newcommand{\dipolevec}{\vec{\dipolecomp}}
\newcommand{\dipolevecop}{\op{\dipolevec}}
\DeclareMathOperator{\tr}{tr}
\newcommand{\FCtensorcomp}{\mathcal{F}}

\newcommand{\laserdetuning}{\Delta}
\newcommand{\atintdecayrate}{\atLinewidth_0}
\newcommand{\Smatrixcomp}{S}
\newcommand{\photindex}{\eta}
\newcommand{\photindexin}{{\photindex_p}}
\newcommand{\photindexout}{{\photindex_s}}
\newcommand{\vacpermitt}{\permitt_0}
\newcommand{\Efieldmodedens}{E}
\newcommand{\emodercomp}{\mathcal{E}}
\newcommand{\emoder}{\boldsymbol{\emodercomp}}
\newcommand{\emodecomp}{e}
\newcommand{\emode}{\vec{\emodecomp}}
\newcommand{\photonic}{\text{pht}}
\newcommand{\atphotint}{{\electronic\text{-}\photonic}}
\newcommand{\photfreq}{\omega}
\newcommand{\photaop}{\op{a}}
\newcommand{\Efieldcomp}{E}
\newcommand{\Efield}{\vec{\Efieldcomp}}
\newcommand{\Efieldop}{\op{\Efield}}
\newcommand{\photfreqin}{\photfreq_p}
\newcommand{\photfreqout}{\photfreq_s}
\newcommand{\photl}{m}
\newcommand{\HEmode}{\text{HE}}

\newcommand{\besselK}[1]{K_{#1}}

\newcommand{\photaR}{\alpha}
\newcommand{\besselJ}[1]{J_{#1}}
\newcommand{\phota}{a}
\newcommand{\cvac}{c}
\newcommand{\cbody}{v}
\newcommand{\photbb}{b}
\newcommand{\ketbra}[2]{\ket{#1}\!\bra{#2}}
\newcommand{\photfreqcutoff}{\photfreq_c}

\newcommand{\photlb}{\photl'}

\newcommand{\reddetuned}{r}
\newcommand{\LOshift}{\Delta\photfreq}
\newcommand{\bscoupling}{G}
\newcommand{\sqcoupling}{K}
\newcommand{\atomoverlapFirstOrder}{\mathcal{A}^{(1)}}
\newcommand{\atomoverlapSecOrder}{\mathcal{A}^{(2)}}
\newcommand{\decayrate}{\Gamma}
\newcommand{\linewidthT}{\Gamma}
\newcommand{\linewidthH}{\Gamma^{(1)}}
\newcommand{\linewidthIH}{\Gamma^{(2)}}
\newcommand{\phononDecayRate}{\kappa}
\newcommand{\phonQfactor}{Q}
\newcommand{\decayrateDown}{\Gamma^-}
\newcommand{\decayrateUp}{\Gamma^+}
\newcommand{\dephasingRate}{\Gamma^z}
\newcommand{\depopulationRate}{\Gamma^d}
\newcommand{\transitionFreq}{\atfreq_{\atnb\atn}}
\newcommand{\fundamentalphonfreq}{\phonfreq_1}
\newcommand{\phonm}{m}
\newcommand{\zerovec}{\boldsymbol{0}}
\newcommand{\atfreqdiffcorr}{\freq_0}
\newcommand{\Lamb}{L}
\newcommand{\lambshift}{\Delta_\Lamb}
\newcommand{\atfreqFinal}{\freq_\effective}

\newcommand{\Pauliz}{{\op{\sigma}^z}}

\newcommand{\Paulip}{{\op{\sigma}^+}}

\newcommand{\Paulim}{{\op{\sigma}^-}}

\newcommand{\Liouvillian}{\mathcal{L}}

\newcommand{\dissipator}{\mathcal{D}}
\newcommand{\densop}{\op{\rho}}
\newcommand{\com}[2]{\left[ {#1},{#2} \right]}

\newcommand{\densopsyscomp}{\mu}
\newcommand{\densopsys}{\op{\densopsyscomp}}
\newcommand{\densopsysss}{\op{\densopsyscomp}_\text{ss}}
\newcommand{\densopbath}{\op{\alpha}}
\newcommand{\densopbathss}{\densopbath_\text{ss}}
\newcommand{\correlator}{K}
\newcommand{\drive}{d}
\newcommand{\RabiFrequency}{\Omega}
\newcommand{\detuning}{\Delta}

\begin{document}
\selectlanguage{english}

\title{Probing Surface-Bound Atoms with Quantum Nanophotonics}

\author{Daniel Hümmer\,\orcidicon{0000-0002-0228-2887}}
\affiliation{Institute for Quantum Optics and Quantum Information of the Austrian Academy of Sciences, 6020 Innsbruck, Austria}
\affiliation{Institute for Theoretical Physics, University of Innsbruck, 6020 Innsbruck, Austria}
\author{Oriol Romero-Isart\,\orcidicon{0000-0003-4006-3391}}
\affiliation{Institute for Quantum Optics and Quantum Information of the Austrian Academy of Sciences, 6020 Innsbruck, Austria}
\affiliation{Institute for Theoretical Physics, University of Innsbruck, 6020 Innsbruck, Austria}
\author{Arno Rauschenbeutel\,\orcidicon{0000-0003-2467-4029}}
\affiliation{Department of Physics, Humboldt-Universität zu Berlin, 10099 Berlin, Germany}
\author{Philipp Schneeweiss\,\orcidicon{0000-0002-1485-7502}}
\affiliation{Department of Physics, Humboldt-Universität zu Berlin, 10099 Berlin, Germany}
\affiliation{Atominstitut, TU Wien, 1020 Vienna, Austria}
\date{\today}

\begin{abstract}

Quantum control of atoms at ultrashort distances from surfaces would open a new paradigm in quantum optics and offer a novel tool for the investigation of near-surface physics. Here, we investigate the motional states of atoms that are bound weakly to the surface of a hot optical nanofiber. We theoretically demonstrate that with optimized mechanical properties of the nanofiber these states are quantized despite phonon-induced decoherence. We further show that it is possible to influence their properties with additional nanofiber-guided light fields and suggest heterodyne fluorescence spectroscopy to probe the spectrum of the quantized atomic motion. Extending the optical control of atoms to smaller atom-surface separations could create opportunities for quantum communication and instigate the convergence of surface physics, quantum optics, and the physics of cold atoms.

\end{abstract}

\maketitle

Obtaining optical control over individual atoms close to surfaces would enable significant advances in fundamental research. For instance, trapping atoms closer to a waveguide increases their coupling to the guided light fields. The increased emission into the waveguide aids the exploration of novel effects in quantum optics~\cite{chang_colloquium:_2018} and benefits powerful light-matter interfaces useful for quantum communication~\cite{corzo_waveguide-coupled_2019}. Moreover, the measurement precision of effects in surface and near-surface physics such as dispersion forces could profit from isotopically clean atomic probes with well-defined initial conditions and long interrogation times~\cite{dalvit_casimir_2011,gierling_cold-atom_2011,schneeweiss_dispersion_2012,yang_scanning_2017,fichet_exploring_2007,peyrot_measurement_2019}. A detailed understanding of atom-surface interactions is paramount, for example, in the search for post-Newtonian forces~\cite{onofrio_casimir_2006} or surface-induced friction~\cite{intravaia_friction_2015}. Precise control over the motional and electronic degrees of freedom of atoms near surfaces would, therefore, provide advantages for quantum optics and surface physics and could ultimately enable the transfer of techniques between these two disparate fields. At present, cold atoms can be optically trapped at distances of a few hundred nanometers from surfaces~\cite{hammes_cold-atom_2002,stehle_plasmonically_2011,thompson_coupling_2013,goban_superradiance_2015,vetsch_optical_2010,goban_demonstration_2012,beguin_generation_2014,kato_strong_2015,lee_inhomogeneous_2015,corzo_large_2016}. At shorter distances, attractive dispersion forces dominate over conventional traps and can lead to adsorption~\cite{desjonqueres_concepts_2012}. Conversely, the omnipresence of dispersion forces has stimulated ideas to exploit them for trapping atoms in the first place~\cite{hung_trapped_2013,chang_trapping_2014,gonzalez-tudela_subwavelength_2015}. In previous works on the optical control of adsorbed atoms~\cite{lima_long-range_2000,silans_laser-induced_2006,afanasiev_laser-induced_2007,afanasev_atomic_2008,soares_2d_2009,nayak_optical_2007}, it remained unclear whether the motional states are quantized despite decoherence~\cite{gortel_desorption_1980,kreuzer_physisorption_1986,kien_phonon-mediated_2007}, and how to optimally probe and manipulate this system.

\begin{figure}[!hb]
  \centering
  \includegraphics[width=245.453245pt]{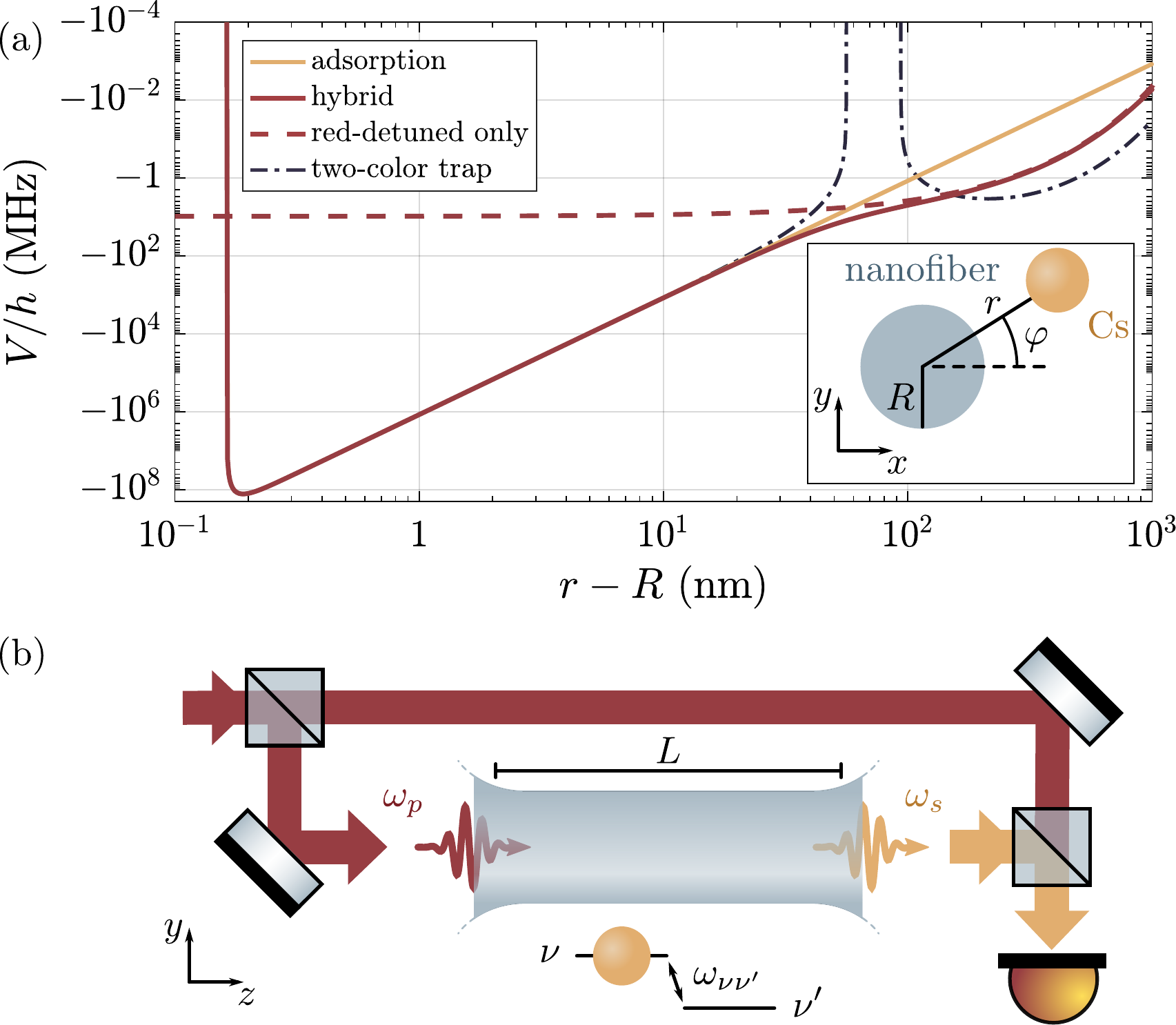}
  \caption{Panel~(a) shows the adiabatic potential as a function of the atom-surface separation. The yellow line represents the adsorption potential $\pot_\adsorption$, the red line a hybrid light- and surface-induced potential, and the red dashed line the contribution of the optical potential $\pot_\optical$. The dash-dotted line corresponds to a typical two-color optical trap for comparison. The inset illustrates some key experimental parameters. Panel~(b) outlines the proposed setup for heterodyne fluorescence spectroscopy of the motional states.}
  \label{fig: setup}
\end{figure}

Here, we propose an experiment to optically detect the quantized motion of atoms bound directly to the surface of a waveguide. We consider two cases: adsorbed atoms and surface-bound atoms in a hybrid potential created by adding an attractive optical force. We focus on weakly bound motional states with binding energies corresponding to a few megahertz since these states can efficiently be probed with light. We account for the finite linewidth of transitions between motional states, which is caused by thermal vibrations (phonons) of the waveguide. We identify a parameter regime in which the atomic motion normal to the surface is quantized despite the interaction with phonons. Interestingly, the linewidths are limited by phonon-induced dephasing rather than state depopulation. We further show that the spectrum of the quantized atomic motion can be resolved using heterodyne fluorescence spectroscopy.

We consider cesium atoms bound to a silica nanofiber~\cite{nieddu_optical_2016,solano_chapter_2017,nayak_nanofiber_2018} for the sake of concreteness. The existence of adsorbed states of cesium on silica is undisputed~\cite{stephens_study_1994,bouchiat_electrical_1999,freitas_spectroscopy_2002}. However, the quantization of the adatoms' motion normal to the surface can only be observed if transitions between different motional states have linewidths smaller than the splitting between the transition frequencies in the absence of vibrations. The interaction with phonons is the dominant mechanism causing depopulation both for adsorbed~\cite{gortel_desorption_1980,kreuzer_physisorption_1986} and optically trapped atoms~\cite{hummer_heating_2019}, and leads to dephasing as well. We assume that the nanofiber forms a phonon cavity of length $\len$. Such a cavity provides control over the nanofiber phonon modes and could, for instance, be realized by optimizing the nanofiber tapers~\cite{pennetta_tapered_2016}. To calculate the total linewidth of transitions between the motional states of an individual atom, we describe the coupled dynamics of the atomic motion and the nanofiber phonons using the Hamiltonian
\begin{equation}\label{eqn: hamiltonian}
  \Hamilop = \Hamilop_\motional + \Hamilop_\vibrational + \Hamilop_\atphonint.
\end{equation}
The atom Hamiltonian $\Hamilop_\motional = \atmomop^2/(2\atmass) + \pot(\atrposop)$ describes the motion of the atom of mass $\atmass$ in the adiabatic potential $\pot(\rpos)$. The operator $\atrposop$ represents the distance of the atom from the axis of nanofiber and $\atmomop$ the momentum of the atom. The term $\Hamilop_\vibrational$ describes the dynamics of the nanofiber phonons, and the term $\Hamilop_\atphonint$ accounts for the atom-phonon coupling. It is sufficient to treat each atom individually since the far-detuned probe laser subsequently used for the spectroscopy does not induce long-ranged atom-atom interactions mediated by the exchange of resonant waveguide photons \cite{solano_super-radiance_2017,le_kien_nanofiber-mediated_2017,olmos_interaction_2020}.

\begin{figure}[ht]
  \centering
  \includegraphics[width=235pt]{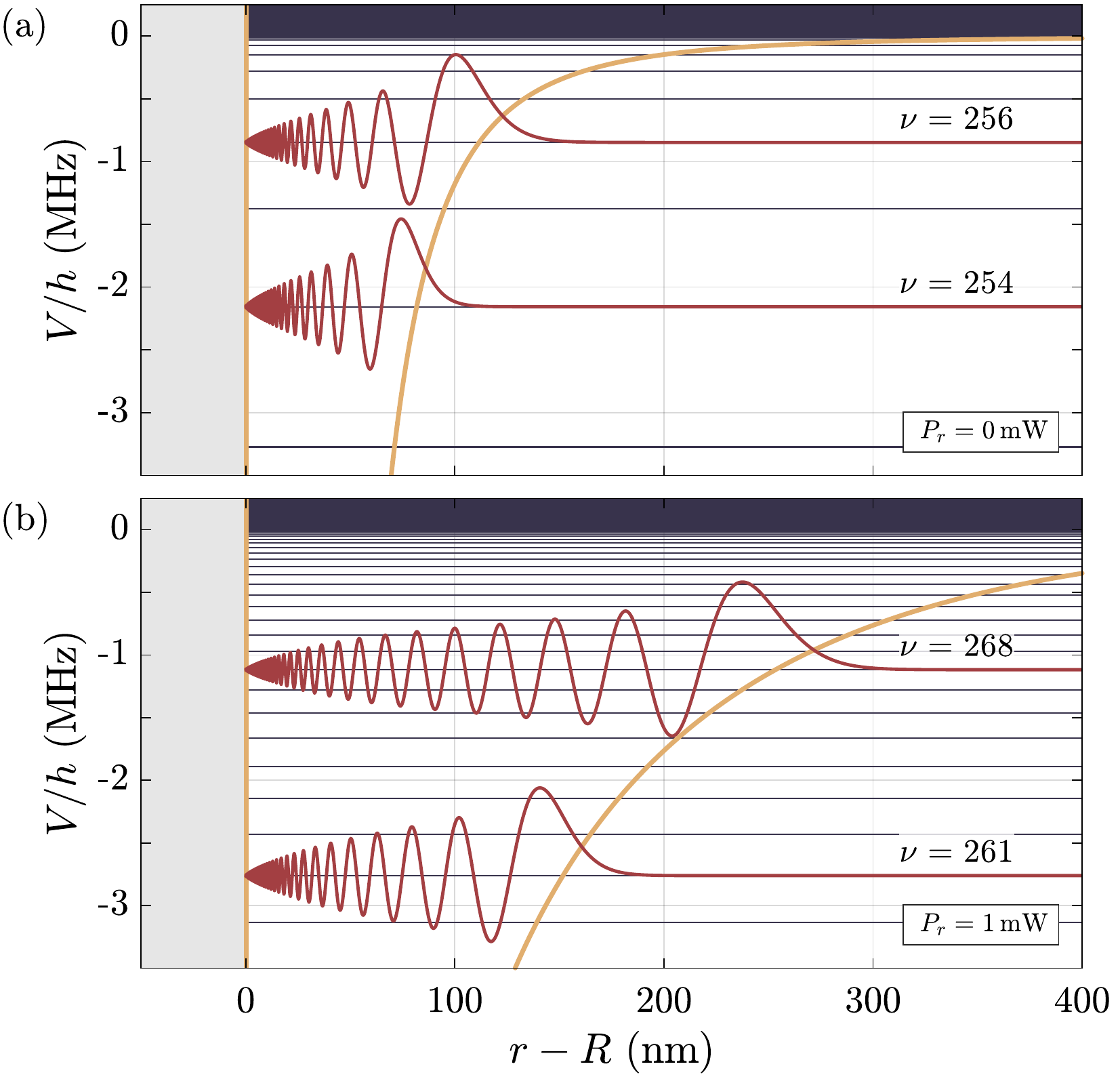}
  \caption{Radial motional states of a cesium atom bound to a silica nanofiber. Panel~(a) shows adsorbed states, panel~(b) hybrid surface-bound states.  We plot the corresponding potential $\pot$ (yellow) generated at power $\power_\reddetuned$ of the fiber-guided light beam, the spectrum $\atfreq_\atn/2\pi$ of motional states (dark blue), and two examples of the atom wave function (red) in arbitrary units. The gray area at $\rpos-\rad<0$ marks the position of the nanofiber.}
  \label{fig: motional states}
\end{figure}

\begin{figure}
  \raggedright
  \includegraphics[width=244pt]{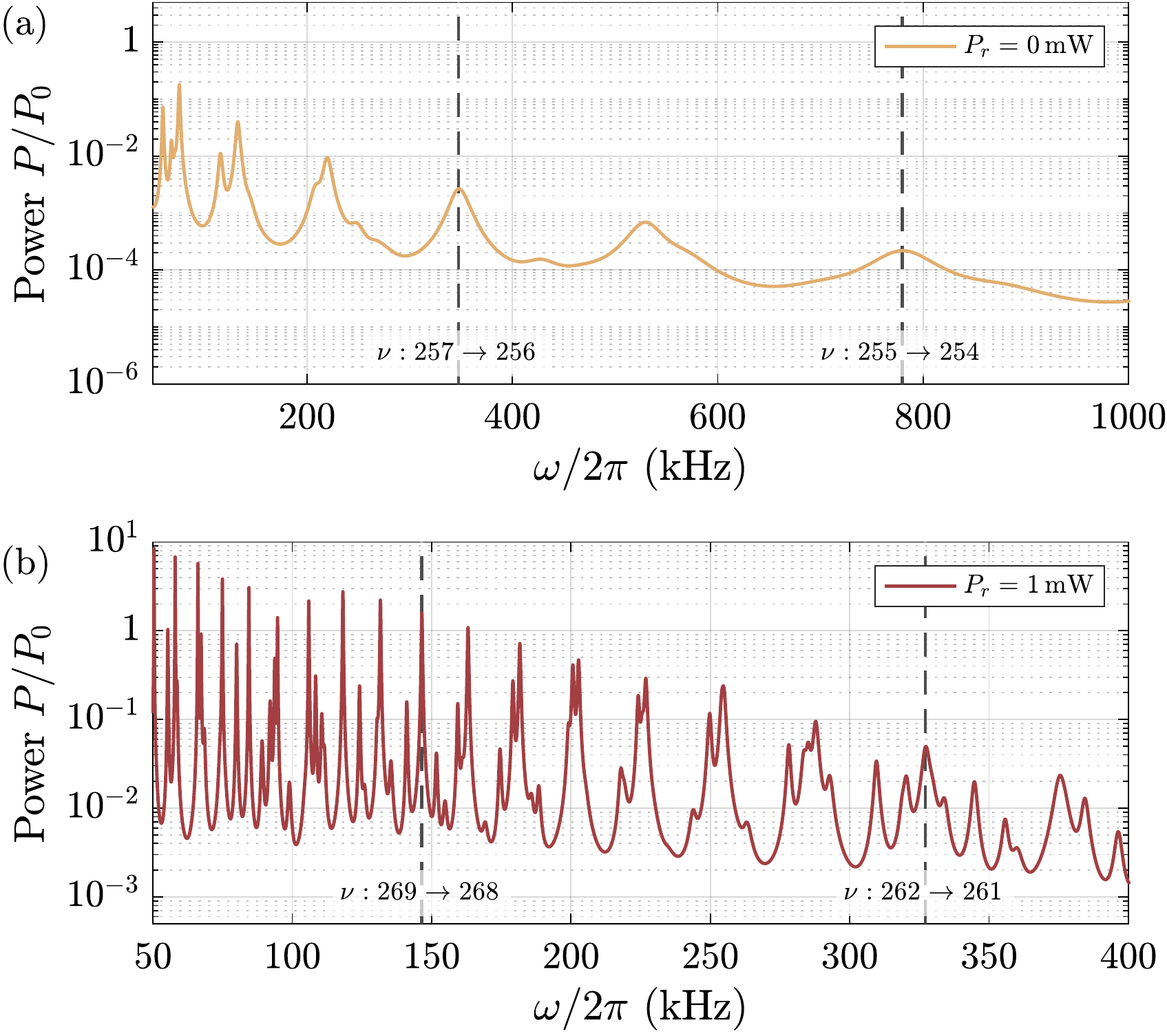}
  \caption{Spectrum of light inelastically scattered by nanofiber-bound atoms. We plot the power of the scattered light as a function of the frequency difference $\photfreqoutin = \photfreqout - \photfreqin$ of the probe photon and the scattered photon. The scale $P_0$ is  explained in the text. Panels (a) and (b) show sidebands due to transitions between the states in \csubref{fig: motional states}{a} and \csubref{fig: motional states}{b}, respectively.}
  \label{fig: heterodyne spectrum}
\end{figure}

The potential $\pot(\rpos)$ arises from both optical dipole forces~\cite{dowling_evanescent_1996,le_kien_atom_2004} and surface effects~\cite{kien_phonon-mediated_2007,buhmann_dispersion_2012}. We approximate the total potential as $\pot(\rpos) = \pot_\optical(\rpos) + \pot_\adsorption(\rpos)$. Nonadditive corrections are only relevant for sufficiently strong light fields~\cite{fuchs_nonadditivity_2018}. The potential $\pot_\optical(\rpos)$ can be calculated~\cite{le_kien_atom_2004,le_kien_dynamical_2013,le_kien_state-dependent_2013}. In contrast to nanofiber-based two-color traps~\cite{vetsch_optical_2010,goban_demonstration_2012}, we consider a cylindrically symmetric potential without a repulsive optical force to prevent the atom from accessing the nanofiber surface. The adsorption potential $\pot_\adsorption(\rpos)$ is determined by the choice of atom species and nanofiber material. It is predominantly due to the Casimir-Polder interaction and the exchange interaction~\cite{zaremba_theory_1977,zangwill_physics_1988,desjonqueres_concepts_2012}. The attractive Casimir-Polder force (dispersion force) dominates over optical forces at atom-surface separations below a few tens of nanometers~\cite{le_kien_atom_2004,buhmann_dispersion_2012}. The exchange interaction becomes relevant when electrons orbiting the atom begin to overlap with electrons in the nanofiber surface~\cite{zaremba_theory_1977,hoinkes_physical_1980,desjonqueres_concepts_2012}. It causes a strong repulsion of the atom immediately at the nanofiber surface. We model the adsorption potential as
\begin{equation}\label{eqn: adsorption potential}
  \pot_\adsorption(\rpos) = - \CPpotstrength (\rpos-\rad)^{-3} + \Ppotstrength (\rpos-\rad)^{-12}.
\end{equation}
Here, $\rpos$ is the radial distance of the atom from the nanofiber axis and $\rad$ is the radius of the nanofiber; see the inset in \csubref{fig: setup}{a}. The first term in \cref{eqn: adsorption potential} is the dispersion force between an atom and a half-space. This simplified model neglects effects such as retardation and the nanofiber's cylindrical geometry, which do not qualitatively alter the results presented in the following~
\footnote{
Precise calculations of the dispersion force need to account for the full complexity and imperfections of the atom-surface system~\cite{klimchitskaya_casimir_2009}. While it is possible to calculate the exact form of the dispersion force between an atom and a dielectric cylinder from first principles~\cite{schmeits_physical_1977,boustimi_van_2002,nabutovskii_interaction_1979}, we are here mainly interested in scenarios where the dispersion force is only dominant at atom-surface separations smaller than the radius of the nanofiber. In this limit, the exact solution can be approximated by the nonretarded dispersion force near a half-space~\cite{boustimi_van_2002,le_kien_atom_2004}.}.
The constant $\CPpotstrength>0$ can be calculated~\cite{mclachlan_van_1964,schmeits_physical_1977,wylie_quantum_1984} and determined experimentally. For a cesium atom and a silica surface $\CPpotstrength/h= \SI{1.18}{\tera\hertz\,\nano\meter^3}$~\cite{stern_simulations_2011}, where $h$ is Planck's constant. The second term in \cref{eqn: adsorption potential} is a standard heuristic model for the exchange energy~\cite{hoinkes_physical_1980}. The constant $\Ppotstrength>0$ can be inferred from the minimum $\pot_\minimum$ of the adsorption potential $\pot_\adsorption(\rpos)$. We use $\pot_\minimum/h = \SI{-128}{\tera\hertz}$~\cite{stephens_study_1994,bouchiat_electrical_1999}, which yields $\Ppotstrength/h = \SI{96.5}{\kilo\hertz\,\nano\meter^{12}}$. Importantly, the bound state energies and spectral peaks presented in \cref{fig: motional states,fig: heterodyne spectrum} quantitatively depend on the parameters $\CPpotstrength$, $\pot_\minimum$, and the exponent $p=-12$ used in \cref{eqn: adsorption potential} and hence provide information about the atom-surface interaction. At the same time, our findings are qualitatively independent of these details and still hold when using alternative models like an exponential barrier \cite{hoinkes_physical_1980} for the short-range repulsive interaction~%
\footnote{Our findings do not change appreciably when using an exponential barrier $\Ppotstrength \exp\spare{ - (\rpos-\rad) p }$ instead of the polynomial in \cref{eqn: adsorption potential}, for instance with $\CPpotstrength/h = \SI{1.56}{\tera\hertz\,\nano\meter^3}$, $\pot_\minimum/h = \SI{-128}{\tera\hertz}$, repulsive amplitude $\Ppotstrength/h = \SI{1.6e6}{\tera\hertz}$, and decay length $p = \SI{53}{\nano\meter^{-1}}$ as suggested in Ref.~\cite{kien_phonon-mediated_2007}.}.

In \csubref{fig: setup}{a}, we plot the potential $\pot(\rpos)$. The hybrid light- and surface-induced potential is realized by launching into the nanofiber a circularly polarized, guided, running-wave light field with a free-space wavelength of $\SI{1064}{\nano\meter}$ (red detuned relative to the cesium $D_2$ line) and a power $\power_\reddetuned = \SI{1}{\milli\watt}$. We also show the potential of a typical nanofiber-based two-color optical dipole trap for comparison; see the Supplemental Material for details~\footnote{See Supplemental Material appended at the end of this article for an extended discussion of the photonic and phononic nanofiber modes, the calculation of the motional linewidths, and the heterodyne fluorescence spectroscopy scheme, which includes Refs.~\cite{achenbach_wave_1973,gurtin_linear_1984,cohen-tannoudji_photons_2004,armenakas_free_1969,glauber_quantum_1991,messiah_quantum_2014,cirac_laser_1992,breuer_theory_2002,reitz_coherence_2013,albrecht_fictitious_2016,cirac_spectrum_1993,sague_cold-atom_2007,patterson_spectral_2018,lindberg_resonance_1986,cohen-tannoudji_atom-photon_1998}.}.
We assume a relative permittivity of $\relpermitt = 2.1$~\cite{bass_handbook_2001} and a nanofiber radius of $\rad = \SI{305}{\nano\meter}$~\footnote{This is the largest radius compatible with the single-mode regime for the light fields of the two-color trap}.

The radial motional states have frequencies $\atfreq_\atn$ and wave functions $\atradwf_\atn(\rpos)\equiv \sqrt{\rpos} \braket{\rpos|\atn}$ that are obtained by solving the time-independent Schrödinger equation
\begin{equation}\label{eqn: Schroedinger equation}
    \spare{ - \frac{\hbar^2}{2\atmass} \partial_\rpos^2 + \pot(\rpos)} \atradwf_\atn(\rpos) = \hbar \atfreq_\atn \atradwf_\atn(\rpos).
\end{equation}
Here, the index $\atn$ counts the motional quanta in radial direction. The motion in azimuthal and axial direction can be neglected~\cite{Note3}, so $\Hamilop_\motional = \hbar \sum_\atn \atfreq_\atn \ketbra{\atn}{\atn}$. We solve \cref{eqn: Schroedinger equation} numerically~
  \footnote{We use the commercial COMSOL Multiphysics\textsuperscript{\textregistered} software package~\cite{comsol_inc_comsol_2016}. Unbound states are obtained by approximating free space with an interval sufficiently large so as not to influence any of the results presented in this Letter.}.
In \cref{fig: motional states}, we plot the spectrum $\atfreq_\atn$ and some example wave functions $\atradwf_\atn(\rpos)$ using $\atmass = \SI{2.21e-25}{\kilo\gram}$~\cite{meija_atomic_2016}. \Csubref{fig: motional states}{a} shows weakly bound states with binding energies up to a few megahertz. \Csubref{fig: motional states}{b} shows surface-bound states in the hybrid light- and surface-induced potential. While the expected center-of-mass position of an atom in these states is on the order of $\SI{100}{\nano\meter}$, there is no potential barrier to keep the atom from accessing the surface.

The phonon Hamiltonian is $\Hamilop_\vibrational = \hbar \sum_\phonindex \phonfreq_\phonindex \hconj{\phonaop}_\phonindex \phonaop_\phonindex$, where $\phonindex$ is an index labeling the phonon modes and $\phonaop_\phonindex$ are the corresponding bosonic ladder operators. The phonon modes of a nanofiber can be calculated analytically~\cite{meeker_guided_1964,Note3}. The depopulation of the motional states in nanofiber-based two-color traps is dominated by their interaction with flexural phonon modes~\cite{hummer_heating_2019}. The coupling primarily arises because the moving nanofiber surface displaces the adiabatic potential~\cite{hummer_heating_2019}. The atom experiences the shifted potential $\pot[\atrposop - \ufieldcompop^\rpos(\rad,\atphiposop,\atzposop)]$~\cite{kreuzer_physisorption_1986,kien_phonon-mediated_2007}, where $\ufieldcompop^\rpos$ is the radial displacement of the nanofiber surface and $\atposop = (\atrposop,\atphiposop,\atzposop)$ is the position operator of the atom in cylindrical coordinates. To describe depopulation and dephasing, we expand the potential to second order in the phonon field. The zero-order term appears in $\Hamilop_\motional$, while higher orders form the interaction Hamiltonian $\Hamilop_\atphonint \simeq \Hamilop_\atphonint^{(1)} + \Hamilop_\atphonint^{(2)}$. At first order,
\begin{equation}\label{eqn: interaction hamiltonian order 1}
  \Hamilop^{(1)}_\atphonint = \hbar \sum_{\phonindex \atnb \atn} \pare{ \atphong_{\phonindex \atnb \atn} \phonaop_\phonindex \ketbra{\atnb}{\atn} + \hc }.
\end{equation}
At second order, we only retain terms describing resonant elastic two-photon scattering, which yield the principal second-order contribution to the broadening of motional transitions~\cite{Note3}:
\begin{equation}\label{eqn: interaction hamiltonian order 2}
  \Hamilop^{(2)}_\atphonint = \hbar \sum_{\phonindex \atn} \bscoupling_{\phonindex \atn} \hconj\phonaop_\phonindex \phonaop_\phonindex \ketbra{\atn}{\atn}.
\end{equation}
The coupling rates are
\begin{align}
  \label{eqn: atom-phonon coupling constants}
  \atphong_{ \phonindex \atnb \atn} &= \frac{ \im }{\sqrt{2\pi}} \frac{\atomoverlapFirstOrder_{\atnb\atn}}{\sqrt{\hbar\dens \phonfreq_\phonindex \len}\rad} &
  \bscoupling_{\phonindex \atn} &= \frac{1}{2\pi} \frac{\atomoverlapSecOrder_{\atn\atn}}{ \dens\phonfreq_\phonindex\len\rad^2}
\end{align}
where $\dens$ is the density of the nanofiber ($\dens = \SI{2.20}{\gram/\centi\meter^3}$ for fused silica~\cite{bass_handbook_2001}), and we define the phonon-induced overlap between different states
\begin{equation}
  \atomoverlap_{\atnb\atn}^{(i)} \equiv \int_0^\infty \cconj\atradwf_\atnb(\rpos)\atradwf_\atn(\rpos) \partial_\rpos^i \pot(\rpos) \dd\rpos.
\end{equation}
The wave functions $\atradwf_\atn(\rpos)$ are normalized according to the orthonormality condition $\int_0^\infty \cconj\atradwf_\atn(\rpos)\atradwf_\atnb(\rpos) \dd \rpos = \kronecker_{\atn\atnb}$, where $\kronecker$ is the Kronecker symbol. The coupling rates are small compared to the transition frequencies $\atfreq_{\atnb\atn} \equiv \atfreq_\atnb - \atfreq_\atn$; that is $|\atfreq_{\atnb\atn}| \gg |\atphong_{\phonindex\atnb\atn}|, |\bscoupling_{\phonindex\atn}|$. Assuming further that the phonon modes have large decay rates $\phononDecayRate_\phonindex \gg |\atphong_{\phonindex\atnb\atn}|, |\bscoupling_{\phonindex\atn}|$ compared to the coupling rates, the phonon modes can be adiabatically eliminated to obtain an effective description of the atom motion in the presence of the thermal phonon bath~\cite{Note3}.

One can then show that if a transition $\atnb \leftrightarrow \atn$ between different motional states is externally driven, its resonance has a finite phonon-induced linewidth (full width at half maximum) of
\begin{equation}\label{eqn: total linewidth}
  \linewidthT_{\atnb\atn} = \linewidthH_{\atnb\atn} + \linewidthIH_{\atnb\atn};
\end{equation}
see \cite{Note3}. Here, $\linewidthH_{\atnb\atn} = \depopulationRate_{\atnb} + \depopulationRate_{\atn}$ is the broadening due to depopulation of the two motional states caused by phonon absorption and emission through $\Hamilop^{(1)}_\atphonint$. The depopulation rate $\depopulationRate_{\atn} \simeq \decayrateDown_{\atn} + \decayrateUp_{\atn}$ of each state is dominated by transitions to the nearest neighboring states. It is beneficial to work with a short phonon cavity to minimize $\linewidthT_{\atnb\atn}$. For our case study, we choose a cavity sufficiently small such that the frequency $\fundamentalphonfreq = \pi^2\rad\sqrt{\YoungE/\dens} /(2\len^2)$ of the fundamental cavity mode $\phonindex_1$ is larger than the transition frequencies $|\atfreq_{(\atn\pm1)\atn}|$ of interest. Here, $\YoungE$ is the Young's modulus of the nanofiber ($\YoungE = \SI{72.6}{\giga\pascal}$ for fused silica~\cite{bass_handbook_2001}). In this limit, $\decayrate^{\pm}_{\atn}$ is determined by the nonresonant coupling to the fundamental mode. As a result,
\begin{equation}\label{eqn: homogeneous linewidth}
  \decayrate^{\pm}_{\atn} \simeq 4 \nbar \frac{|\atphong_{\phonindex_1(\atn\pm1)\atn}|^2}{\fundamentalphonfreq} \frac{1}{\phonQfactor},
\end{equation}
where $\nbar$ is the thermal population and $\phonQfactor = \fundamentalphonfreq/\phononDecayRate_1$ the quality factor. In deriving \cref{eqn: homogeneous linewidth}, we assume $\nbar \simeq \boltzmann \temp /\hbar \fundamentalphonfreq \gg 1$ where $\temp$ is the temperature of the nanofiber and $\boltzmann$ is the Boltzmann constant. The second contribution in \cref{eqn: total linewidth},
\begin{equation}
  \linewidthIH_{\atnb\atn}
  \simeq 16 \nbar^2 \frac{|\bscoupling_{\phonindex_1\atnb\atn}|^2}{\fundamentalphonfreq} \phonQfactor,
\end{equation}
is primarily caused by dephasing between the motional states due to the resonant coupling to the fundamental mode through $\Hamilop^{(2)}_\atphonint$. Here, $\bscoupling_{\phonindex_1\atnb\atn} \equiv (\bscoupling_{\phonindex_1\atnb} - \bscoupling_{\phonindex_1\atn} )/2$. We assume a cavity of length $\len = \SI{5}{\micro\meter}$ and quality factor $\phonQfactor = 100$. In this case, the linewidth is limited by dephasing; that is, $\linewidthIH_{\atnb\atn} \gg \linewidthH_{\atnb\atn}$. Remarkably, $\linewidthT_{\atnb\atn}$ can be small enough such that transitions between the motional states shown in \cref{fig: motional states} can be resolved as we now argue.

We propose to measure the spectrum of the quantized nanofiber-bound states using heterodyne fluorescence spectroscopy, see~\csubref{fig: setup}{b}, which allows the observation of the quantized motion of atoms in optical potentials~\cite{jessen_observation_1992}. To this end, a cloud of laser-cooled atoms is prepared around the nanofiber. The nanofiber-bound states are in a thermal equilibrium~\cite{gortel_desorption_1980,kreuzer_physisorption_1986}. Laser light with a frequency $\photfreqin$ far detuned from resonance with the atom is split into a probe beam and a local oscillator; see \csubref{fig: setup}{b}. The probe beam is coupled into the nanofiber with circular polarization. A guided probe photon can be scattered inelastically by a bound atom through the evanescent electric field, changing its frequency to $\photfreqout$ and causing the atom to change its motional state from $\atn$ to $\atnb$. This process creates sidebands in the spectrum of the probe beam. After the transmission through the nanofiber, the probe beam is recombined with the local oscillator. The beat signal is detected with a photodetector. The frequency of the local oscillator is shifted by an offset $\LOshift$ to separate the Stokes and anti-Stokes sidebands, and its polarization is matched to that of the probe beam. This setup is only sensitive to the radial motion of bound atoms~\cite{Note3}. The power $\power$ of the scattered light as a function of the difference $\photfreqoutin \equiv \photfreqout - \photfreqin$ can be inferred from the spectrum of the photocurrent.

The spectroscopy can be modeled by the Hamiltonian
\begin{equation}\label{eqn: spectroscopy hamiltonian}
 \Hamilop' = \Hamilop + \Hamilop_\electronic + \Hamilop_\photonic + \Hamilop_\atphotint,
\end{equation}
where $\Hamilop_\photonic = \hbar \sum_\photindex \photfreq_\photindex \hconj{\photaop}_\photindex \photaop_\photindex$ describes the nanofiber-guided photon modes $\photindex$ and $\Hamilop_\atphotint = - \dipolevecop \cdot \Efieldop(\atposop)$ is the dipole coupling~\cite{Note3}. Here, $\Efieldop$ is the electric field and $\dipolevecop$ is the dipole moment of a single atom. One can show that the power of scattered light as a function of the frequency difference $\photfreqoutin$ is approximately~\cite{Note3}
\begin{equation}\label{eqn: fluorescence spectrum}
  \power(\photfreqoutin) \propto \sum_{\atn,\atnb\neq\atn} \frac{ \atLinewidth_{\atnb\atn}/2 }{ \pare{ \atfreq_{\atnb\atn} - \photfreqoutin }^2 + \pare{ \atLinewidth_{\atnb\atn}/2 }^2 } n(\atn) \left| \FCtensorcomp_{\atnb\atn} \right|^2.
\end{equation}
Since the potential $\pot(\rpos)$ is not harmonic, this spectrum contains a separate sideband for each transition $\atn \leftrightarrow \atnb$. The amplitude of each sideband is proportional to the Franck-Condon factor
\begin{equation}\label{eqn: Franck Condon factors}
  \FCtensorcomp_{\atnb\atn} \equiv \frac{\Efieldmodedens_\photindexout \Efieldmodedens_\photindexin}{(2\pi)^2}  \int_0^\infty \cconj\atradwf_{\atnb}(\rpos) \, \cconj{\emoder}_\photindexout(\rpos) \cdot \emoder_\photindexin(\rpos) \, \atradwf_{\atn}(\rpos) \dd \rpos,
\end{equation}
where we define $\Efieldmodedens_\photindex \equiv \sqrt{ \hbar \vacpermitt \photfreq_\photindex/2}$. Here, $\vacpermitt$ is the vacuum permittivity, the index $\photindexin$ ($\photindexout$) comprises the quantum numbers of the nanofiber-guided probe (scattered) photon, and $\emoder_\photindex(\rpos)$ is the radial partial wave of the corresponding electric mode field of the fundamental $\HEmode_{11}$ mode of a nanofiber~\cite{marcuse_light_1982,snyder_optical_2012,le_kien_field_2004}.

In \csubref{fig: heterodyne spectrum}{a}, we plot the anti-Stokes sidebands corresponding to downward transitions between the adsorbed states shown in \csubref{fig: motional states}{a}, assuming a nanofiber temperature of $\temp = \SI{300}{\kelvin}$. The spectrum in \csubref{fig: heterodyne spectrum}{b} corresponds to the hybrid surface-bound states shown in \csubref{fig: motional states}{b}, assuming $\temp = \SI{420}{\kelvin}$ based on the power $\power_\reddetuned$~\cite{wuttke_thermalization_2013}. In both cases, transitions between neighboring levels are resolved. Examples of such transitions are indicated by the dashed lines. Transitions between levels that are further separated in $\atn$ appear as smaller, interstitial peaks. In plotting \cref{fig: heterodyne spectrum}, we choose a wavelength of $\SI{1000}{\nano\meter}$ for the probe laser and approximate the occupation of all relevant states as equal since the frequency interval they cover is much smaller than $\boltzmann \temp$. The signal decreases for larger $\photfreqoutin$ since the involved states have a smaller spatial extent, resulting in lower Franck-Condon factors. For this reason, we focus on states with binding energies of a few megahertz. The additional red-detuned light field increases the scattering probability in \csubref{fig: motional states}{b} by widening the wave functions: The resonances highlighted in \csubref{fig: motional states}{a} and \csubref{fig: heterodyne spectrum}{b} involve states with similar binding energies, but the signal is increased in the latter case, boosting resonances above $\power/\power_0 = 1$. Here, $\power_0$ is the power of the sideband corresponding to transitions between the first excited state $\atn=1$ and the ground state $\atn=0$ in the regular nanofiber-based two-color trap shown in \cref{fig: setup}~\cite{Note3}, a signal that has already been observed experimentally~\cite{meng_near-ground-state_2018}.

In summary, we analyze the spectrum and phonon-induced linewidths of the motional states of a cesium atom bound directly to the surface of an optical nanofiber. We find that the phonon-induced linewidth of transitions between states with binding energies of a few megahertz can be smaller than the spacing of the transitions, allowing one to resolve quantized motional states. We further propose to probe these states using heterodyne fluorescence spectroscopy. An additional attractive light field enhances the expected signal compared to purely adsorbed atoms. When working at room temperature, it is necessary to optimize the nanofiber's mechanical properties to resolve the quantization of the motional states, which could explain why it has not previously been observed. The proposed technique can be adapted for other waveguide geometries, including chip-based implementations, and is expected to work for other combinations of atom species and waveguide materials.

\begin{acknowledgments}
  We thank Jürgen Volz and Carlos Gonzalez-Ballestero for helpful discussions. Support by the Austrian Academy of Sciences (ÖAW, ESQ Discovery Grant QuantSurf), the Studienstiftung des Deutschen Volkes, and the Alexander von Humboldt Foundation in the framework of the Alexander von Humboldt Professorship endowed by the Federal Ministry of Education and Research is gratefully acknowledged.
\end{acknowledgments}

\clearpage

\onecolumngrid
\setcounter{page}{1}
\setcounter{figure}{0}
\setcounter{equation}{0}
\renewcommand{\thefigure}{S\arabic{figure}}
\renewcommand{\theequation}{S\arabic{equation}}
\renewcommand\thesection{S\arabic{section}}

\thispagestyle{empty}

\begin{center}
  \large
  \textbf{Supplemental Material for `Probing Surface-Bound Atoms with Quantum Nanophotonics'}
\end{center}

\begin{center}
  Daniel Hümmer\,\orcidicon{0000-0002-0228-2887},\textsuperscript{1,\,2} %
  Oriol Romero-Isart\,\orcidicon{0000-0003-4006-3391},\textsuperscript{1,\,2} %
  Arno Rauschenbeutel\,\orcidicon{0000-0003-2467-4029},\textsuperscript{3} %
  and Philipp Schneeweiss\,\orcidicon{0000-0002-1485-7502}\textsuperscript{3,\,4}
\end{center}

\begin{center}
  \small
  \textsuperscript{1}\textit{Institute for Quantum Optics and Quantum Information of the Austrian Academy of Sciences, 6020 Innsbruck, Austria}\\
  \textsuperscript{2}\textit{Institute for Theoretical Physics, University of Innsbruck, 6020 Innsbruck, Austria}\\
  \textsuperscript{3}\textit{Department of Physics, Humboldt-Universität zu Berlin, 10099 Berlin, Germany}\\
  \textsuperscript{4}\textit{Atominstitut, TU Wien, 1020 Vienna, Austria}
\end{center}

\vspace{1em}

In this supplement, we provide details on the calculation of the phonon-induced linewidths and the fluorescence spectra. In \cref{sec: fiber eigenmodes}, we summarize the relevant phononic and photonic modes of the nanofiber. In \cref{sec: linewidths surface-bound and adsorbed}, we discuss the motional states of adsorbed and surface-bound atoms shown in Fig.~2 of the paper. We describe how they couple to flexural cavity phonons and how to calculate the resulting finite linewidths of transitions between motional states. In \cref{sec: linewidths optical trap}, we discuss motional states of atoms in nanofiber-based two-color traps. We describe how they couple to traveling flexural phonons and how to calculate the resulting depopulation rates of motional states, both numerically and analytically in the limit of a harmonic trap potential. We use these results to verify our numerical calculations and as a benchmark for the power of the spectroscopy signal from surface-bound atoms. In \cref{sec: spectroscopy}, we derive the spectra of light scattered by nanofiber-bound atoms when probed with a nanofiber-guided light field. These spectra are shown in Fig.~3 of the paper.

\section{Nanofiber Modes}
\label{sec: fiber eigenmodes}

It is useful to quantize both the displacement field $\ufieldop(\pos)$ and the electric field $\Efieldop(\pos)$ in terms of eigenmodes of the nanofiber, modeled as a cylinder of radius $\rad$.

\subsection{Flexural Phonons}
\label{sec: nanofiber phonons}

The thermal vibrations of a nanofiber can be described using linear elasticity theory. The dynamical quantity of linear elasticity theory is the displacement field $\ufield(\post)$ that indicates how far and in which direction each point $\pos$ of a body is displaced from its equilibrium position~\cite{S_achenbach_wave_1973,gurtin_linear_1984}. Canonical quantization of linear elasticity theory in terms of a set of vibrational eigenmodes can be performed in the usual way~\cite{S_cohen-tannoudji_photons_2004}. The resulting displacement field operator in the Schrödinger picture is
\begin{equation}
  \ufieldop(\pos)=\sum_\phonindex \ufieldmodedens_\phonindex \spare{\wmode_\phonindex(\pos) \phonaop_\phonindex +\hc}.
\end{equation}
Here, $\wmode_\phonindex(\pos)$ are the mode fields associated with the phonon modes, $\phonindex$ is a multiindex suitable for labeling the modes, $\phonaop_\phonindex$ are the corresponding bosonic ladder operators, and $\hc$ indicates the Hermitian conjugate. The mode density is $\ufieldmodedens_\phonindex \equiv \sqrt{\hbar/2\dens\phonfreq_\phonindex}$, where $\dens$ denotes the mass density of the nanofiber and $\phonfreq_\phonindex$ are the phonon frequencies. The phonon Hamiltonian takes the form $\Hamilop_\vibrational = \hbar \sum_\phonindex \phonfreq_\phonindex \hconj{\phonaop}_\phonindex \phonaop_\phonindex$. The eigenmodes of a nanofiber (modeled as a homogeneous, and isotropic cylinder) are well known~\cite{S_achenbach_wave_1973,meeker_guided_1964,armenakas_free_1969}. In cylindrical coordinates $(\rpos,\phipos,\zpos)$, the mode fields factorize into partial waves
\begin{align}\label{eqn-S:displacement radial partial wave decomposition}
  \wmode_\phonindex(\pos) &=  \frac{\wmoder_\phonindex(\rpos)}{2\pi}  e^{\im (\phonl\phipos + \phonk \zpos)}
  &\text{or}&&
  \wmode_\phonindex(\pos) &= \frac{\wmoder_\phonindex(\rpos)}{\sqrt{\pi\len}}e^{\im \phonl \phipos} \sin(\phonk \zpos),
\end{align}
where $\phonk$ is the propagation constant along the nanofiber axis and $\phonl\in\Z$. The left expression corresponds to the mode fields of an infinitely long nanofiber. It models traveling phonons on a long nanofiber that are not reflected at its tapered ends. In this case, $\phonk\in\R$. The right expression models the standing waves of a finite nanofiber (a phonon cavity) located at $\zpos \in [0,\len]$ with fixed ends that reflect phonons. Such a cavity supports phonons with $\phonk = \pi\phonm/\len$, where $\phonm=1,2,\dots$. Transitions between motional states in a nanofiber-based two-color trap are dominated by flexural phonon modes with $\phonl=\pm1$~\cite{S_hummer_heating_2019}. The continuum of traveling flexural phonons can be labeled by $\phonindex = (\phonk,\phonl)$, and the discrete set of cavity modes by $\phonindex = (\phonm,\phonl)$. Flexural phonons with $\si{\kilo\hertz}$ to $\si{\mega\hertz}$ frequencies that are relevant here have wavelengths much larger than the radius of the nanofiber. In this limit, the radial partial waves $\wmoder_\phonindex(\pos)$ have vector components
\begin{align}\label{eqn-S:displacement F modes low-frequency limit}
      \wmodercomp^\rpos_{\phonindex}(\rpos) &= \frac{1}{\rad}, &
      \wmodercomp^\phipos_{\phonindex}(\rpos) &= \frac{\im \phonl}{\rad}, &
      \wmodercomp^\zpos_{\phonindex}(\rpos) &= -\frac{\im\phonk}{\rad}\rpos,
\end{align}
which are normalized according to $\int_0^\rad \rpos|\wmoder_\phonindex(\rpos)|^2 \dd\rpos=1$ to leading order in $\phonk\rad$. These flexural modes form a single band in the $(\phonfreq_\phonindex,\phonk)$ plane with a dispersion relation $\phonfreq_\phonindex = \cFL\rad\phonk^2/2$ that is quadratic in the low frequency limit~\cite{S_hummer_heating_2019}. In the case of a flexural mode cavity, the phonon spectrum is hence $\phonfreq_\phonindex = \phonm^2 \pi^2  \rad\sqrt{\YoungE/\dens} / (2 \len^2) $. The effective speed of sound is $\cFL = \sqrt{\YoungE/\dens}$, where $\YoungE$ is the Young modulus of the nanofiber material. For fused silica, $\YoungE = \SI{72.6}{\giga\pascal}$ and $\dens = \SI{2.20}{\gram/\centi\meter^3}$ such that $\cFL=\SI{5.74e3}{\meter/\second}$~\cite{S_bass_handbook_2001}.

\subsection{Nanofiber-guided Photons}
\label{sec: nanofiber photons}

In the paper, we propose to perform fluorescence spectroscopy of surface-bound states using a nanofiber-guided probe laser. We need to describe nanofiber-guided photons to model this spectroscopy scheme. The electromagnetic field in the presence of the nanofiber can be quantized based on the photonic eigenmodes of the system~\cite{S_glauber_quantum_1991,cohen-tannoudji_photons_2004}. The photonic eigenmodes of a nanofiber (modeled as a cylindrical step-index waveguide with relative electric permittivity $\relpermitt$) are well known~\cite{S_marcuse_light_1982,snyder_optical_2012}. The resulting Hamiltonian is $\Hamilop_\photonic = \hbar \sum_\photindex \photfreq_\photindex \hconj{\photaop}_\photindex \photaop_\photindex$, where $\photindex$ is a multi-index suitable for labeling the eigenmodes, $\photfreq_\photindex$ is the frequency of each eigenmode, and $\photaop_\photindex$ is the corresponding bosonic ladder operator. The electric field operator in the Schrödinger picture is
\begin{equation}
  \Efieldop(\pos) = \sum_\photindex \Efieldmodedens_\photindex \spare{\photaop_\photindex \,\emode_\photindex(\pos) + \hc },
\end{equation}
where we define the mode density $\Efieldmodedens_\photindex \equiv \sqrt{ \hbar \vacpermitt \photfreq_\photindex/2}$ and $\vacpermitt$ is the vacuum permittivity. The electric mode fields are of the form
\begin{equation}
  \emode_\photindex(\pos) = \frac{\emoder_\photindex(\rpos)}{2\pi} e^{\im(\photl\phipos + \photk \zpos)},
\end{equation}
with propagation constant $\photk \in \R$ and azimuthal order $\photl\in\Z$. These modes are quasi-circular polarized~\cite{S_le_kien_field_2004}. We are interested in photons in the single-mode regime of the nanofiber, that is, with frequencies below the cutoff frequency $\photfreqcutoff \simeq 2.405 \, \cvac / ( \rad \sqrt{\relpermitt-1})$~\cite{S_marcuse_light_1982}. Here, $\cvac$ is the vacuum speed of light. For fused silica, $\relpermitt = 2.1$~\cite{S_bass_handbook_2001} such that the silica nanofiber with a radius of $\rad = \SI{305}{\nano\meter}$ considered in our case study has a cutoff frequency corresponding to a free-space wavelength of $\wavelen_c = \SI{835.7}{\nano\meter}$. In the single-mode regime, only modes on the $\HEmode_{11}$ band with azimuthal order $\photl = \pm1$ are nanofiber-guided. For the setup considered in the paper, the fluorescence spectrum is independent of the sign of $\photl$ and we may choose $\photl=1$ without loss of generality. In this case, the radial partial waves of the electric mode field have vector components
\begin{align}
  \rpos &< \rad \text{ :} & \rpos &> \rad \text{ :} \nonumber \\
\emodercomp^{\rpos}_{\photindex}(\rpos) &= \frac{\im A_\photindex}{\phota^2} \spare{ \photk \phota \besselJ{1}'(\phota\rpos) - \frac{\photfreq_\photindex}{\cvac} \beta \frac{\besselJ{1}(\phota \rpos)}{\rpos} }, &
\emodercomp^{\rpos}_{\photindex}(\rpos) &= - \alpha \frac{\im A_\photindex}{\photbb^{2}}\spare{ \photk \photbb \besselK{1}'(\photbb\rpos) -  \beta \frac{\photfreq_\photindex}{\cvac} \frac{\besselK{1}(\photbb \rpos)}{\rpos} }, \nonumber \\
\emodercomp^{\phipos}_{\photindex}(\rpos) &= \frac{A_\photindex}{\photaR^2} \spare{ \beta \frac{\photfreq_\photindex}{\cvac}  \phota \besselJ{1}'(\phota\rpos) - \photk \frac{\besselJ{1}(\phota \rpos)}{\rpos} }, &
\emodercomp^{\phipos}_{\photindex}(\rpos) &= - \alpha \frac{A_\photindex}{\photbb^{2} }\spare{ \beta \frac{\photfreq_\photindex}{\cvac} \photbb \besselK{1}'(\photbb\rpos) - \photk \frac{\besselK{1}(\photbb \rpos)}{\rpos} }, \\
\emodercomp^{\zpos}_{\photindex}(\rpos) &= A_\photindex\besselJ{1}(\phota \rpos), &
\emodercomp^{\zpos}_{\photindex}(\rpos) &= \alpha A_\photindex \besselK{1}(\photbb \rpos), \nonumber
\end{align}
where $\phota \equiv \sqrt{\photfreq_\photindex^2/\cbody^2 - \photk^2}$, $\photbb \equiv \sqrt{\photk^2 - \photfreq_\photindex^2/\cvac^2 }$ and $\cbody =\cvac/\sqrt{\relpermitt}$ is the speed of light inside the nanofiber. The functions $\besselJ{\photl}$ and $\besselK{\photl}$ are Bessel functions and modified Bessel functions, respectively. The prime indicates the first derivative. We define
\begin{align}
  \alpha &\equiv \frac{\besselJ{1}(\phota\rad)}{\besselK{1}(\photbb\rad}, &
  \beta &\equiv \frac{(\relpermitt-1)}{\rad \cvac} \frac{ \photk \photfreq_\photindex}{\phota \photbb} \frac{\besselJ{1}(\phota \rad) \besselK{1}(\photbb \rad)}{ \phota \besselJ{1}(\phota \rad) \besselK{1}'(\photbb \rad) + \photbb \besselJ{1}'(\phota \rad) \besselK{1}(\photbb \rad) }.
\end{align}
The amplitude $A_\photindex$ is determined by the normalization condition $\vacpermitt^2\int_0^\infty \rpos\relpermitt(\rpos) \, \cconj{\emoder}_\photindex(\rpos) \cdot \emoder_\photindex(\rpos)\, \dd\rpos = 1$. Here, $\relpermitt(\rpos)$ is the relative permittivity as a function of the radial position. The dispersion relation $\photfreq_\photindex(\photk)$ is implicitly given by the frequency equation
\begin{equation}\label{eqn-S:guided mode frequency equation}
  \spare{ \phota \besselJ{1}(\phota \rad) \besselK{1}'(\photbb  \rad) + \photbb \besselK{1}(\photbb \rad) \besselJ{1}'(\phota \rad) }
   \spare{ \phota \besselJ{1}(\phota \rad) \besselK{1}'(\photbb \rad) + \relpermitt \photbb \besselK{1}(\photbb \rad) \besselJ{1}'(\phota \rad) }
  = \spare{ \frac{(\relpermitt-1)}{\rad \cvac} \frac{\photk \photfreq_\photindex}{ \phota \photbb}\besselJ{1}(\phota \rad) \besselK{1}(\photbb \rad) }^2.
\end{equation}
The frequency equation has only one zero $\photfreq_\photindex(\photk)$ in the single-mode regime.

\section{Linewidths for Adsorbed and Surface-Bound Atoms}
\label{sec: linewidths surface-bound and adsorbed}

We provide details on the calculation of the motional states of adsorbed and surface-bound atoms shown in Fig.~2 of the paper. We also summarize how to calculate the linewidths of transition between the motional states due to the interaction with flexural cavity phonons. These linewidths are used to plot the spectra in Fig.~3 of the paper.

\subsection{Motional States}

The potentials considered in the paper are cylindrically symmetric, that is, $\pot(\pos) = \pot(\rpos)$. The motional states $\ket{\atindex} \equiv \ket{\atn,\atl,\atk}$ of an atom in these potentials, therefore, have wavefunctions of the form
\begin{equation}\label{eqn-S:adsorbed eigenstates}
    \atwf_\atindex(\pos) = \braket{\pos|\atn,\atl,\atk} = \frac{\atradwf_{\atl\atn}(\rpos)}{2\pi\sqrt{\rpos}} e^{\im(\atl\phipos + \atk\zpos)}.
\end{equation}
The Hamiltonian describing the motion of the atom is $\Hamilop_\motional = \hbar \sum_\atindex \atfreq_\atindex \ketbra{\atindex}{\atindex}$. The corresponding frequencies are $ \atfreq_\atindex = \atfreq_{\atl\atn} + \hbar\atk^2/2\atmass$ for an atom of mass $\atmass$. Here, the quantum numbers $\atn\in\N$, $\atl\in\Z$, and $\atk\in\R$ label the excitations in radial, azimuthal, and axial direction, respectively. The radial partial waves $\atradwf_{\atl\atn}(\rpos)$ are obtained by solving the one-dimensional Schrödinger equation with the effective potential $\pot_\atl(\rpos)$~\cite{S_messiah_quantum_2014}:
\begin{align}\label{eqn-S:effective Schroedinger equation}
  &\spare{-\frac{\hbar^2}{2\atmass}\partial_\rpos^2 + \pot_\atl(\rpos)} \atradwf_{\atl\atn}(\rpos) = \hbar\atfreq_{\atl\atn}\atradwf_{\atl\atn}(\rpos), &
  \pot_\atl(\rpos) &\equiv \pot(\rpos) + \frac{\hbar^2}{2\atmass} \pare{\atl^2 - \frac{1}{4}}.
\end{align}
The second term in the above potential is an angular momentum barrier. It can be neglected for azimuthal orders $\atl$ up to of a few hundred for adsorbed cesium atoms in weakly bound states considered in this paper. In that case, there is no coupling between the atomic motion in radial and azimuthal direction and $\atradwf_{\atl\atn}(\rpos) = \atradwf_{\atn}(\rpos)$. \Cref{eqn-S:effective Schroedinger equation} then reduces to the Schrödinger equation
\begin{equation}\label{eqn-S:Schroedinger equation}
    \spare{ - \frac{\hbar^2}{2\atmass} \partial_\rpos^2 + \pot(\rpos)} \atradwf_\atn(\rpos) = \hbar \atfreq_\atn \atradwf_\atn(\rpos)
\end{equation}
that we solve to calculate the states shown in the paper.

The perfect cylindrical symmetry of the nanofiber is an idealization. In practice, the surface of a nanofiber is not perfectly smooth and may feature local imperfections. Moreover, the nanofiber cross section is not perfectly circular and varies both in size and exact shape over the length of the nanofiber. In consequence, the bound motional states of the surface-induced potential do not exhibit perfect cylindrical symmetry, either. However, the interaction between phonons and photons on the one side and atoms on the other is not significantly altered by such imperfections. In particular, they do not significantly affect the atoms' motion in radial direction, in particular for weakly bound states considered in our manuscript where the probability amplitude close to the nanofiber surface is low. Since the spectroscopy scheme we propose is only sensitive to the radial motion of the atoms and does not rely on a particular symmetry of the atom states, deviations from a perfect cylindrical symmetry in the atoms' motional state will not influence the predicted spectra in Fig.~3 of the paper.

\subsection{Atom-Phonon Interaction}

The coupling between atom motion and phonons arises because the phonons displace the potential, $\pot[\atrposop - \ufieldcompop^\rpos(\rad,\atphiposop,\atzposop)]$. The interaction Hamiltonian is obtained by expanding the shifted potential to second order around $\ufield=\zerovec$ and can be cast into the form $\Hamilop_\atphonint = \Hamilop^{(1)}_\atphonint + \Hamilop^{(2)}_\atphonint$ where
\begin{equation}
  \begin{split}
    \Hamilop^{(1)}_\atphonint &= \hbar \sum_{\phonindex \atindexb \atindex} \pare{ \atphong_{\phonindex \atindexb \atindex} \phonaop_\phonindex \ketbra{\atindexb}{\atindex} + \hc }, \\
    \Hamilop^{(2)}_\atphonint &= \hbar \sum_{\phonindexb \phonindex \atindexb \atindex} \pare{ \frac{\sqcoupling_{\phonindexb \phonindex \atindexb \atindex}}{2} \phonaop_\phonindexb \phonaop_\phonindex \ketbra{\atindexb}{\atindex} + \hc }
    + \hbar \sum_{\phonindexb \phonindex \atindexb \atindex} \bscoupling_{\phonindexb \phonindex \atindexb \atindex} \hconj\phonaop_\phonindexb \phonaop_\phonindex \ketbra{\atindexb}{\atindex}.
  \end{split}
\end{equation}
The coupling rates between atoms and cavity phonons are, at first order,
\begin{align}
  \label{eqn-S:def 1-phonon coupling constant cavity phonons}
  \atphong_{ \phonindex \atindexb \atindex} &= \atphong_{ \phonindex \atnb \atn} \kronecker_{(\atl+\phonl),\atlb} ~ \frac{1}{2}\cpare{ \dirac\spare{\atkb - (\atk+\phonk)} -  \dirac\spare{ \atkb - (\atk-\phonk) } }, &
  \atphong_{ \phonindex \atnb \atn} &= \frac{ \im }{\sqrt{2\pi}} \frac{\atomoverlapFirstOrder_{\atnb\atn}}{\sqrt{\hbar\dens \phonfreq_\phonindex \len}\rad},
\end{align}
and, at second order,
\begin{align}
\label{eqn-S:def 2-phonon coupling constant cavity phonons}
  \sqcoupling_{\phonindexb \phonindex \atindexb \atindex} &= \bscoupling_{\phonindexb \phonindex \atnb \atn} \kronecker_{\atlb,(\atl+\phonl+\phonlb)} [\dirac], &
  \bscoupling_{\phonindexb \phonindex \atindexb \atindex} &= \bscoupling_{\phonindexb \phonindex \atnb \atn} \kronecker_{(\atlb+\phonlb),(\atl+\phonl)} [\dirac], &
  \bscoupling_{\phonindexb \phonindex \atnb \atn} &= \frac{1}{2\pi} \frac{\atomoverlapSecOrder_{\atnb\atn}}{\dens\sqrt{\phonfreq_\phonindexb \phonfreq_\phonindex}\len\rad^2},
\end{align}
\begin{equation}
  [\delta] \equiv \frac{1}{4} \big\{ \dirac\spare{(\atkb+\phonkb) - (\atk+\phonk)} + \dirac\spare{(\atkb-\phonkb) - (\atk-\phonk)} - \dirac\spare{(\atkb-\phonkb) - (\atk+\phonk)} - \dirac\spare{(\atkb+\phonkb) - (\atk-\phonk)} \big\}.
\end{equation}
The wavefunction overlaps $\atomoverlapFirstOrder_{\atnb\atn}$ and $\atomoverlapSecOrder_{\atnb\atn}$ are defined in the paper.

We focus on the radial motion of the atoms. Since phonons carry only little momentum, we neglect changes in the momentum of the atomic motion in the axial and azimuthal direction. To infer how the presence of thermal phonons affects the radial atomic motion, let us at first select two states $\ket{\atn_1}$ and $\ket{\atn_2}=\ket{\atn_1+1}$ that are neighbors in frequency. For the time being, we neglect all other atom states. The dynamics of this simplified model can be described by
\begin{align}\label{eqn-S:minimal model}
  \Hamilop_\motional &= \hbar \frac{\atfreqdiffcorr}{2}  \Pauliz, &
  \Hamilop_\atphonint &= \hbar \sum_\phonindex \spare{ \pare{ \atphong_\phonindex  \Paulip - \cconj\atphong_\phonindex  \Paulim } \phonaop_\phonindex+ \hc } + \hbar \sum_\phonindex  \bscoupling_\phonindex ( \hconj\phonaop_\phonindex\phonaop_\phonindex - \nbar_\phonindex) \Pauliz.
\end{align}
We use Pauli matrices $\Paulip = \ketbra{\atn_2}{\atn_1}$, $\Paulim = \ketbra{\atn_1}{\atn_2}$, and $\Pauliz = \ketbra{\atn_2}{\atn_2} - \ketbra{\atn_1}{\atn_1}$. The coupling rates are $\atphong_\phonindex \equiv \atphong_{\phonindex\atn_2\atn_1}$ and $\bscoupling_\phonindex \equiv (\bscoupling_{\phonindex\phonindex\atn_2\atn_2} - \bscoupling_{\phonindex\phonindex\atn_1\atn_1} )/2 \in \R$. In deriving \cref{eqn-S:minimal model}, we have redefined $\Hamilop_\motional$ to include a correction $\Delta \atfreqdiffcorr \equiv \sum_\phonindex \bscoupling_\phonindex \nbar_\phonindex$ to the transition frequency $\atfreqdiffcorr \equiv \atfreq_{\atn_2}-\atfreq_{\atn_1} + \Delta \atfreqdiffcorr$. The correction arises from $\Hamilop_\atphonint^{(2)}$ due to the finite thermal population of the phonon modes. It can be neglected for the parameters used in the case study in the paper. We also neglect nonresonant terms (i.e., terms that are not energy conserving) in $\Hamilop_\atphonint^{(2)}$, since all phonon scattering, absorption, and emission processes are dominated by resonant terms. At this point, there are still terms proportional to $\Paulip$ and $\Paulim$ remaining, which lead to transitions between the two atom states through two-phonon absorption, emission, or inelastic scattering at first order in $\Hamilop_\atphonint^{(2)}$. These processes contribute to the broadening of the resonance when the transition $\atn_1 \leftrightarrow \atn_2$ is externally driven. However, the coupling constants are much smaller than for the elastic two-phonon scattering processes generated by the terms $\hconj\phonaop_\phonindex \phonaop_\phonindex \Pauliz$, which cause dephasing. As a result, the linewidth induced by $\Hamilop_\atphonint^{(2)}$ is dominated by dephasing due to the resonant $\Pauliz$ terms retained in \cref{eqn-S:minimal model}.

\subsection{Effective Evolution of the Atomic Motion}

In practice, the phonon modes have a thermal population and nonzero decay rates $\phononDecayRate_\phonindex$ due to internal losses and their interaction with the environment (e.g., through the absorption of guided laser light and the clamping of the nanofiber). We model the dynamics of the joint atom-phonon state operator $\densop$ using the Liouvillian $\Liouvillian = \Liouvillian_\motional + \Liouvillian_\vibrational + \Liouvillian_\atphonint$, where
\begin{align}
  \Liouvillian_\motional\densop &= -\frac{\im}{\hbar}[\Hamilop_\motional, \densop ], &
  \Liouvillian_\vibrational\densop &= -\frac{\im}{\hbar}[\Hamilop_\vibrational, \densop ] + \sum_\phonindex \phononDecayRate_\phonindex (\nbar_\phonindex+1) \dissipator_{\phonaop_\phonindex}\densop + \phononDecayRate_\phonindex \nbar_\phonindex \dissipator_{\hconj\phonaop_\phonindex}\densop, &
  \Liouvillian_\atphonint\densop &= -\frac{\im}{\hbar}[\Hamilop_\atphonint, \densop ],
\end{align}
and the dissipator is $\dissipator_{\phonaop_\phonindex}\densop = \phonaop_\phonindex \densop \hconj\phonaop_\phonindex - \{\hconj\phonaop_\phonindex \phonaop_\phonindex,\densop\}/2$. The steady-state of the phonon bath according to $\Liouvillian_\vibrational$ is the thermal state $\densopbathss = e^{-\Hamilop_\vibrational/(\boltzmann \temp)} / \tr[e^{-\Hamilop_\vibrational/(\boltzmann \temp)}]$ with thermal populations $\nbar_\phonindex$ determined by the Bose-Einstein distribution. Here, $\temp$ is the temperature of the nanofiber. Since the transition frequency $\atfreqdiffcorr \gg |\atphong_\phonindex|, |\bscoupling_\phonindex|$ is large compared to the coupling rates, it is possible to obtain an effective description of the atom motion alone. If we further assume $\phononDecayRate_\phonindex \gg |\atphong_\phonindex|, |\bscoupling_\phonindex|$, we can use adiabatic elimination to trace out the phonon modes~\cite{S_cirac_laser_1992,breuer_theory_2002}. The dynamics of the state operator $\densopsys$ of the atomic motion is then described by the Liouville–von Neumann equation $\partial_\tm \densopsys(\tm) = \Liouvillian_\effective\densopsys(\tm)$ with the effective Liouvillian
\begin{align}\label{eqn-S:effective Liouvillian}
  \Liouvillian_\effective\densopsys &= - \frac{\im}{\hbar} \com{\Hamilop_\effective }{ \densopsys } + \decayrateDown \dissipator_{\Paulim}\densopsys + \decayrateUp \dissipator_{\Paulip}\densopsys + \dephasingRate \dissipator_{\Pauliz}\densopsys, &
  \Hamilop_\effective &= \hbar\frac{\atfreqFinal}{2} \Pauliz .
\end{align}
Here, $\decayrateUp$ and $\decayrateDown$ are the phonon-induced depopulation rates of the states $\atn_1$ and $\atn_2$, respectively, and $\dephasingRate$ is the rate of phonon-induced dephasing between the two states:
\begin{align}
\label{eqn-S:TLS depopulation}
  \decayrateUp &= 2 \sum_\phonindex  |\atphong_\phonindex|^2 \Re \spare{ \nbar_\phonindex \correlator_\phonindex^- + (\nbar_\phonindex+1) \correlator_\phonindex^+ }, &
  \decayrateDown &= 2 \sum_\phonindex  |\atphong_\phonindex|^2 \Re \spare{ (\nbar_\phonindex+1) \correlator_\phonindex^- + \nbar_\phonindex \correlator_\phonindex^+ }, \\
  \label{eqn-S:TLS dephasing and correlator}
  \dephasingRate &= 2 \sum_\phonindex  \nbar_\phonindex (\nbar_\phonindex+1) \frac{\bscoupling_\phonindex^2}{\phononDecayRate_\phonindex}, &
  \correlator_\phonindex^\pm &\equiv \frac{\phononDecayRate_\phonindex/2}{(\phononDecayRate_\phonindex/2)^2 + ( \atfreqdiffcorr \pm \phonfreq_\phonindex )^2} + \im \frac{\atfreqdiffcorr \pm \phonfreq_\phonindex }{(\phononDecayRate_\phonindex/2)^2 + ( \atfreqdiffcorr \pm \phonfreq_\phonindex )^2}.
\end{align}
The transition frequency $\atfreqFinal \equiv \atfreqdiffcorr + \lambshift$ is subject to the Lamb shift $\lambshift \equiv \sum_\phonindex (2\nbar_\phonindex+1) |\atphong_\phonindex|^2 \Im \spare{ \correlator_\phonindex^- + \correlator_\phonindex^+ }$, which can be neglected in our case study.

\subsection{Linewidth of Transitions}

To determine the phonon-induced linewidth of the transition $\atn_1 \leftrightarrow \atn_2$, we can, for instance, add a driving term $\Hamilop_\drive(\tm) = \hbar \RabiFrequency \spare{ \Paulim e^{\im \freq_\drive \tm} + \hc }/2$ to \cref{eqn-S:effective Liouvillian}. In the limit of a driving that is weak compared the influence of the bath, $\RabiFrequency \ll (\decayrate^\pm, \dephasingRate)$, the steady-state population of the state $\ket{\atn_2}$ is
\begin{equation}
  \braket{\atn_2|\densopsysss|\atn_2} \simeq \frac{\RabiFrequency^2}{2 (\decayrateDown + \decayrateUp)} \frac{\decayrate/2}{\detuning^2 + (\decayrate/2)^2} + \frac{\decayrateUp}{\decayrateDown + \decayrateUp},
\end{equation}
where $\detuning \equiv \freq_\drive - \atfreqFinal$ is the detuning of the drive. The resonance in the population as a function of the detuning has a Lorentzian shape with linewidth (full width at half maximum) of
\begin{equation}\label{eqn-S:linewidth TLS}
  \decayrate = \decayrateDown + \decayrateUp + 4 \dephasingRate.
\end{equation}
The linewidth has two distinct contributions: $\linewidthH \equiv \decayrateDown + \decayrateUp$ due to the depopulation of the two involved states, and $\linewidthIH \equiv  4 \dephasingRate$ due to the dephasing of the two states. By construction of the model \cref{eqn-S:minimal model}, we neglect depopulation induced by $\Hamilop^{(2)}_\atphonint$ since it leads to a broadening that is smaller than $\linewidthIH$.

It is straightforward to generalize to transitions between any of the radial motional states $\ket{\atn}$. In analogy to \cref{eqn-S:linewidth TLS}, we model the linewidth of the transition $\atn \leftrightarrow \atnb$ between any two states as
\begin{align}
  \linewidthT_{\atnb\atn} &\equiv \linewidthH_{\atnb\atn} + \linewidthIH_{\atnb\atn}.
\end{align}
Here,
\begin{align}\label{eqn-S:dephasing broadening general}
  \linewidthIH_{\atnb\atn} &\equiv 8 \sum_\phonindex \nbar_\phonindex^2 \frac{ \bscoupling^2_{\phonindex\atnb\atn} }{\phononDecayRate_\phonindex}, &
  \bscoupling_{\phonindex\atnb\atn} &\equiv \frac{1}{4\pi} \frac{\atomoverlapSecOrder_{\atnb\atnb}-\atomoverlapSecOrder_{\atn\atn}}{ \dens \phonfreq_\phonindex \len\rad^2}
\end{align}
in analogy to \cref{eqn-S:TLS dephasing and correlator}. Note that $\bscoupling_{\phonindex\atnb\atn} \in \R$. The rate $\linewidthIH_{\atnb\atn}$ is dominated by the fundamental cavity mode $\phonindex_1$, since the coupling rates drop as $\phonfreq_\phonindex^{-2}$ with the phonon frequency. Hence,
\begin{equation}\label{eqn-S:dephasing broadening simplified}
    \linewidthIH_{\atnb\atn}
    \simeq 16 \nbar^2 \frac{\bscoupling^2_{\phonindex_1\atnb\atn} }{\fundamentalphonfreq} \phonQfactor
    = \frac{32}{\pi^{12}} \frac{\boltzmann^2 \temp^2 \len^8 \phonQfactor}{\hbar^2 \rad^9} \sqrt{\frac{\dens}{\YoungE^5}} \spare{\atomoverlapSecOrder_{\atnb\atnb}-\atomoverlapSecOrder_{\atn\atn}}^2,
\end{equation}
where $\nbar$ is the thermal population, $\fundamentalphonfreq$ the frequency, and $\phonQfactor = \fundamentalphonfreq/\phononDecayRate_1$ the quality factor of the fundamental cavity mode.

The broadening $\linewidthH_{\atnb\atn}$ is the sum of the depopulation rates of both states. In general, transitions to any other state contribute to the depopulation rates. In the limit of large thermal populations $\nbar_\phonindex \gg 1$, we obtain
\begin{align}\label{eqn-S:depopulation broadening general}
  \linewidthH_{\atnb\atn} &\equiv \depopulationRate_{\atnb} + \depopulationRate_{\atn}, &
  \depopulationRate_{\atn} & \equiv 2 \sum_{\atn''\neq\atn} \sum_\phonindex \nbar_\phonindex |\atphong_{\phonindex\atn''\atn}|^2 \Re \spare{ \correlator^-_{\phonindex\atn''\atn} + \correlator^+_{\phonindex\atn''\atn} }, &
  \Re\correlator_{\phonindex\atnb\atn}^\pm &\equiv \frac{\phononDecayRate_\phonindex/2}{(\phononDecayRate_\phonindex/2)^2 + ( |\transitionFreq| \pm \phonfreq_\phonindex )^2}
\end{align}
in analogy to \cref{eqn-S:TLS depopulation,eqn-S:TLS dephasing and correlator}. Here, $\transitionFreq \equiv \atfreq_\atnb - \atfreq_\atn$ is the transition frequency and $\atphong_{\phonindex\atnb\atn}$ is defined in \cref{eqn-S:def 1-phonon coupling constant cavity phonons}. The state overlaps $\atomoverlapFirstOrder_{\atnb\atn}$ quickly decay with increasing distance
$|\atnb-\atn|$. As a result, it is often sufficient to include transitions to the states $\atn''=\atn\pm1$ closest in frequency when calculating $\depopulationRate_{\atn}$. If the cavity is sufficiently small such that the fundamental cavity mode has a frequency $\fundamentalphonfreq$ larger than the relevant transition frequencies, $\linewidthH_{\atnb\atn}$ is dominated by the fundamental mode and we can approximate
\begin{align}\label{eqn-S:depopulation broadening small cavity}
  \linewidthH_{\atnb\atn} &\simeq \decayrateDown_{\atn} + \decayrateUp_{\atn} + \decayrateDown_{\atnb} + \decayrateUp_{\atnb}, &
  \decayrate^{\pm}_{\atn} &
  \equiv 4 \nbar \frac{|\atphong_{\phonindex_1(\atn\pm1)\atn}|^2}{\fundamentalphonfreq} \frac{1}{\phonQfactor}
  = \frac{16}{\pi^7} \frac{\boltzmann \temp \len^5}{\hbar^2 \rad^5 \phonQfactor } \sqrt{\frac{\dens}{\YoungE^3}} | \atomoverlapFirstOrder_{(\atn\pm1)\atn} |^2,
\end{align}
which corresponds to Eq.~(9) in the paper. We use \cref{eqn-S:depopulation broadening general,eqn-S:dephasing broadening simplified} to calculate the linewidths that appear in Fig.~3 of the paper, with relevant contributions only stemming from $\linewidthIH_{\atnb\atn}$.

\begin{figure}
  \centering
  \includegraphics[width=264.43122pt]{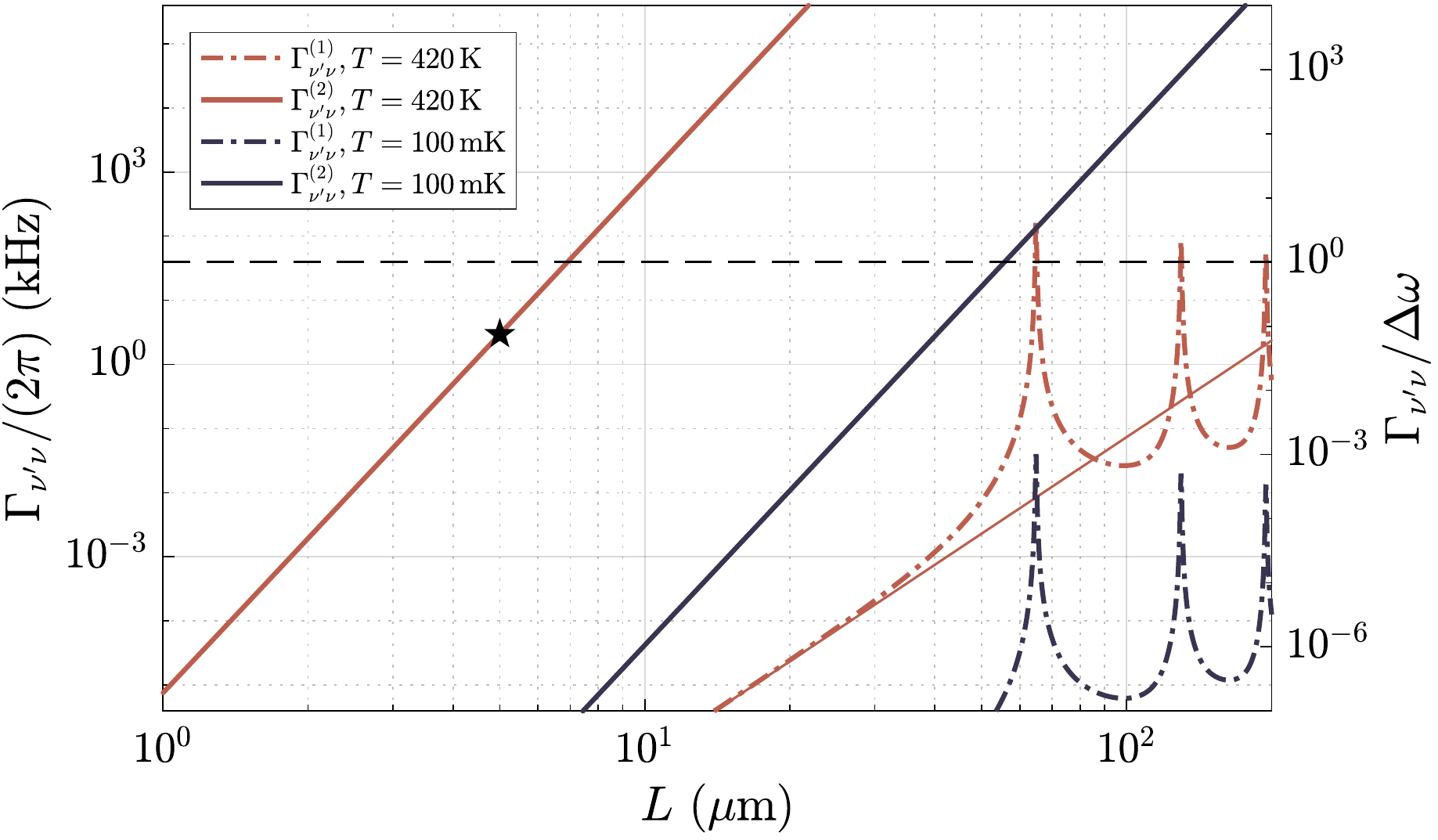}
  \caption{Contributions $\linewidthH_{\atnb\atn}$ and $\linewidthIH_{\atnb\atn}$ to the transition linewidth as a function of the cavity length $\len$ and for two different nanofiber temperatures. As an example, we select the transition between the states $\atn=261 \leftrightarrow \atnb=262$ shown in Fig.~2b of the paper. The transition frequency is $\transitionFreq = 2\pi \times \SI{327}{\kilo\hertz}$. The separation to the neighboring transition $\atn=262 \leftrightarrow \atnb=263$ is $\Delta \omega = 2\pi \times \SI{39}{\kilo\hertz}$. We assume a quality factor of $\phonfreq_\phonindex/\phononDecayRate_\phonindex=100$ for all phonon modes. The solid lines represent $\linewidthIH_{\atnb\atn}$, calculated from \cref{eqn-S:dephasing broadening simplified}. The dashed-dotted lines represent $\linewidthH_{\atnb\atn}$, calculated from \cref{eqn-S:depopulation broadening nearest neighbor}. We also plot the asymptote \cref{eqn-S:depopulation broadening nearest neighbor small cavity} for the limit $\fundamentalphonfreq \gg \transitionFreq$. The resonances visible in $\linewidthH_{\atnb\atn}$ occur whenever a cavity mode is resonant with the transition. The star indicates the parameters we use to plot the spectra in Fig.~3 of the paper. Below the horizontal dashed line $\linewidthT_{\atnb\atn}/\Delta \omega < 1$, which indicates that transitions between motional states can be resolved.}
  \label{fig-S: depopulation vs dephasing broadening}
\end{figure}

In the heterodyne fluorescence spectroscopy scheme we propose in the paper, transitions between all motional states are driven simultaneously. Transitions between states $\atn$ and $\atnb=\atn+1$ that are nearest neighbors in frequency are most likely and lead to resonances of the largest power, see Fig.~3 in the paper. Therefore, it is useful to focus on nearest-neighbor transitions to determine for which parameters the motional quantization can be resolved. For nearest-neighbor transitions, \cref{eqn-S:depopulation broadening general} simplifies to
\begin{equation}
  \label{eqn-S:depopulation broadening nearest neighbor}
  \linewidthH_{(\atn+1)\atn} \simeq 16 \sum_{\phonm=1}^\infty \nbar_\phonindex |\atphong_{\phonindex(\atn+1)\atn}|^2 \Re \spare{ \correlator_{\phonindex(\atn+1)\atn}^- + \correlator_{\phonindex(\atn+1)\atn}^+ }.
\end{equation}
In deriving \cref{eqn-S:depopulation broadening nearest neighbor}, we approximate the upward and downward depopulation rates of each state as equal. In this case, \cref{eqn-S:depopulation broadening small cavity} further simplifies to
\begin{equation}
  \label{eqn-S:depopulation broadening nearest neighbor small cavity}
  \linewidthH_{(\atn+1)\atn}
  \simeq 16 \nbar \frac{|\atphong_{\phonindex_1\atnb\atn}|^2}{\fundamentalphonfreq} \frac{1}{\phonQfactor}
  = \frac{64}{\pi^7} \frac{\boltzmann \temp \len^5}{\hbar^2 \rad^5 \phonQfactor } \sqrt{\frac{\dens}{\YoungE^3}} | \atomoverlapFirstOrder_{\atnb\atn} |^2.
\end{equation}
In \cref{fig-S: depopulation vs dephasing broadening}, we plot the contributions $\linewidthH_{\atnb\atn}$ and $\linewidthIH_{\atnb\atn}$ to the linewidth as a function of the cavity length $\len$ using \cref{eqn-S:depopulation broadening nearest neighbor,eqn-S:depopulation broadening nearest neighbor small cavity}. We select the transition between the states $\atn=261$ and $\atnb=262$ shown in Fig.~2b of the paper. Below the horizontal dashed line, the linewidth $\linewidthT_{\atnb\atn}$ is smaller than the separation $\Delta \atfreq$ to the next nearest-neighbor transition. In the regime $\linewidthT_{\atnb\atn}/\Delta \atfreq \ll 1$, transitions between motional states can be resolved. This regime can be realized either by choosing a sufficiently small cavity, or by working at sufficiently low nanofiber temperatures. For the parameters chosen in \cref{fig-S: depopulation vs dephasing broadening}, the contribution $\linewidthH_{\atnb\atn}$ can be neglected compared to $\linewidthIH_{\atnb\atn}$. Note that, for simplicity, we assume a constant quality factor $\phonQfactor = \phonfreq_\phonindex/\phononDecayRate_\phonindex  = 100$ for all modes (in particular the fundamental mode decisive for the linewidth). This assumption cannot hold for arbitrarily large cavities: It is to be expected that the quality factor is reduced for modes with longer wavelengths, which in turn lowers $\linewidthIH_{\atnb\atn}$ compared to a simple extrapolation of \cref{fig-S: depopulation vs dephasing broadening}.

The ideal length $\len$ optimizes between the absolute strength and the signal-to-noise ratio of the spectroscopy signal. Our analysis predicts that shorter nanofibers lead to a better signal-to-noise ratio; see \cref{fig-S: depopulation vs dephasing broadening}. However, the number of atoms close to the nanofiber is proportional to the nanofiber length. Shorter nanofibers will therefore reduce the absolute signal strength and require, for instance, longer measurement times. The length of $\SI{5}{\micro\meter}$ chosen in our case study represents the longest nanofiber compatible with resolving weakly bound atoms, assuming that the nanofiber is heated to a temperature of $\SI{420}{\kelvin}$ by the transmitted laser beam of power $\power_\reddetuned=\SI{1}{\milli\watt}$~\cite{S_wuttke_thermalization_2013}. Achieving lower nanofiber temperatures is difficult since the thermal coupling of the nanofiber to its environment is very low \cite{S_wuttke_thermalization_2013}, but would allow to work with longer nanofibers.

\section{Linewidths for Optically Trapped Atoms}
\label{sec: linewidths optical trap}

We derive the phonon-induced depopulation rate of radial motional states of atoms that are trapped in a two-color trap and interact with the traveling flexural phonons of a long nanofiber. This model is able to explain the heating rates observed in existing nanofiber-based atom trap setups~\cite{S_hummer_heating_2019}. We calculate the depopulation rates using the numerical methods also applied to the adsorbed and surface-bound states. We use these results to verify our numerical calculations by comparing them with analytical results obtained in the limit of a harmonic trap.

\begin{figure}
  \centering
  \includegraphics[width=225.3267pt]{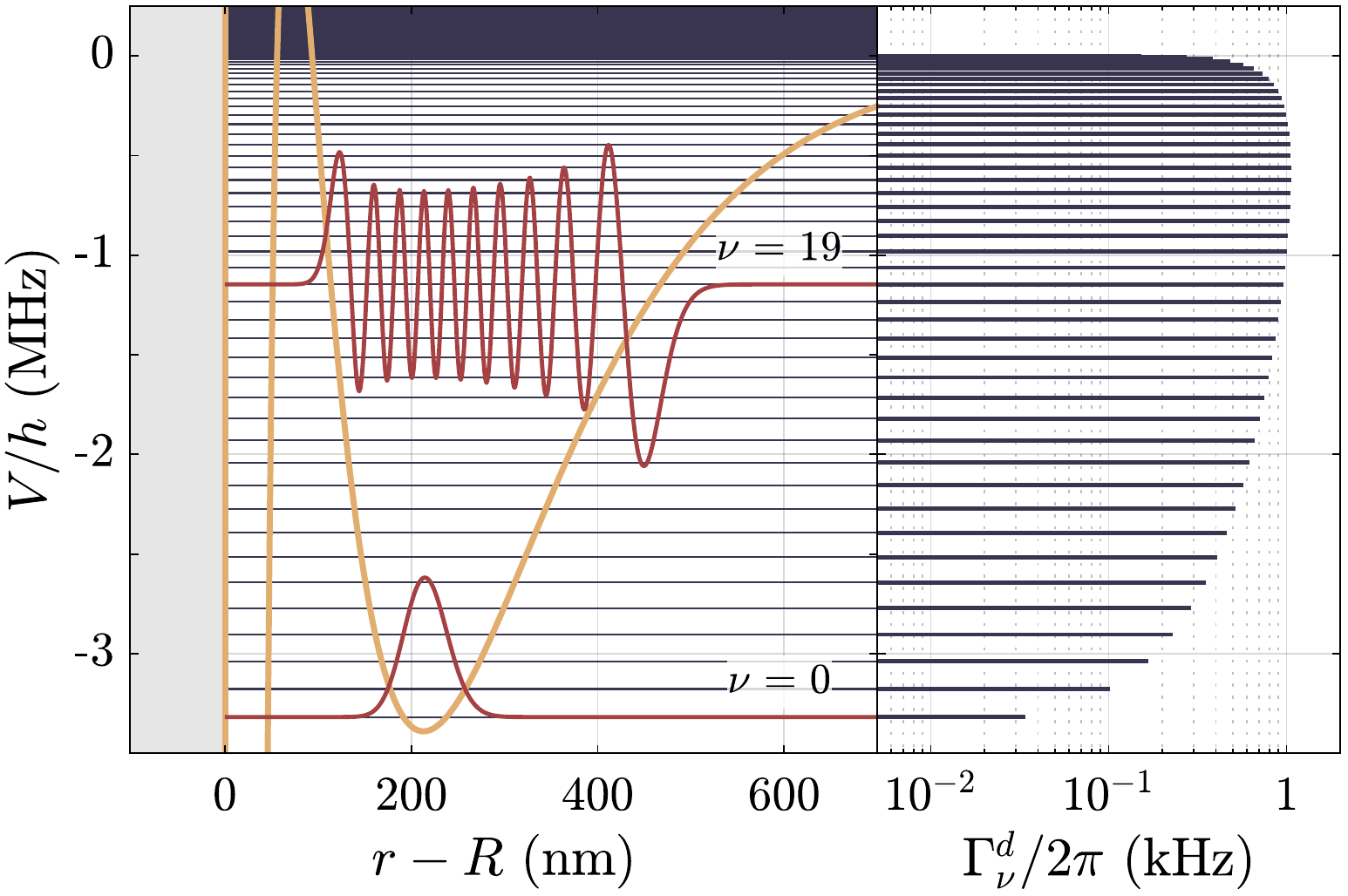}
  \caption{Radial states and their phonon-induced linewidths of a cesium atom in a nanofiber-based two-color optical trap. The states are obtained by solving \cref{eqn-S:Schroedinger equation}. We neglect the coupling between the motion in radial, azimuthal, and axial direction. On the left-hand side, we plot the corresponding potential $\pot$ (yellow), the spectrum $\atfreq_\atn/2\pi$ of motional states (dark blue), and two examples of the atom wavefunction (red) in arbitrary units. The gray area at $\rpos-\rad<0$ marks the position of the nanofiber. On the right-hand side, we plot the phonon-induced linewidths $\atLinewidth_\atn$ of the motional states, assuming a temperature of $\temp=\SI{600}{\kelvin}$.}
  \label{fig-S: optical trap linewidths}
\end{figure}

Fig.~1 of the paper shows a typical two-color trap potential. It is realized by launching two counterpropagating beams with a free-space wavelength of $\SI{1064}{\nano\meter}$ (red detuned with respect to the cesium $D_2$ line) and a combined power of $2\times\SI{2}{\milli\watt}$ into the nanofiber, as well as a running-wave light field with a wavelength of $\SI{840}{\nano\meter}$ (blue detuned) and a power of $\SI{4.5}{\milli\watt}$. All beams are linearly polarized, with a $\pi/2$ angle between the polarization planes of the blue- and red-detuned light fields. All other parameters are as in the case study presented in the paper. The trap minima are located in the polarization plane of the red-detuned light field. Close to the ground state of the trap, the radial motion of the atom decouples from its motion in the axial and azimuthal direction.

The radial motional states $\ket{\atn}$ can be obtained by solving \cref{eqn-S:Schroedinger equation}. We plot two examples of the corresponding wavefunctions in \cref{fig-S: optical trap linewidths}. To leading order in the phonon degrees of freedom, these states couple to flexural phonons through the interaction Hamiltonian
\begin{align}
  \Hamilop_\atphonint &= \hbar \sum_{\phonindex \atnb \atn} \spare{ \atphong_{\phonindex  \atnb \atn} \phonaop_\phonindex \ket{\atnb}\bra{\atn} + \hc }, &
  \atphong_{ \phonindex \atnb \atn} &= -\frac{1}{2\sqrt{2}\pi} \frac{\atomoverlapFirstOrder_{\atnb\atn}}{\sqrt{\hbar\dens \phonfreq_\phonindex}\rad}.
\end{align}
The resulting depopulation rates can be calculated at first order in perturbation theory:
\begin{align}\label{eqn-S:linewidths optical trap}
  \depopulationRate_\atn &= \frac{1}{\sqrt{2}\pi} \frac{\boltzmann\temp}{\hbar^2\sqrt{\rad^5 \sqrt{\YoungE \dens^3}}} \sum_{\atnb \neq \atn} \frac{|\atomoverlapFirstOrder_{\atnb\atn}|^2}{\sqrt{|\transitionFreq|^5}}.
\end{align}
In deriving \cref{eqn-S:linewidths optical trap}, we assume a high thermal occupation $\nbar_\phonindex \gg 1$. We plot the depopulation rates for each state on the right-hand side of \cref{fig-S: optical trap linewidths}.

The potential is approximately harmonic for states close to the ground state of the optical trap at $\trappos=(\trapr,\trapphi,\trapz)$. The atom Hamiltonian can then be written as $\Hamilop_\motional = \sum_i \hbar \itrapfreq \hconj\ataop_i \ataop_i$ where we introduce bosonic creation and annihilation operators $\hconj\ataop_i$ and $\ataop_i$ for the harmonic motion of the atom in direction $i=\rpos,\phipos,\zpos$. The trap frequencies are $\itrapfreq = \sqrt{\partial_i^2 \pot_0(\trappos)/\atmass}$. The interaction between the phonons and the atomic motion is of the form
\begin{equation}\label{eqn-S:interaction Hamiltonian optical trap}
  \Hamilop_\atphonint \simeq \sum_{\phonindex i} \hbar (\ataop_i + \hconj\ataop_i)( \atphong_{\phonindex i} \phonaop_\phonindex + \cconj\atphong_{\phonindex i} \hconj\phonaop_\phonindex ).
\end{equation}
The coupling constants between the radial motion and flexural nanofiber phonons in particular is~\cite{S_hummer_heating_2019}
\begin{equation}
    \atphong_{\phonindex \rpos} = - \frac{1}{4\pi} \frac{1}{\rad} \sqrt{\frac{\atmass \rtrapfreq^3}{\dens \phonfreq_\phonindex}} e^{+\im (\phonl \phipos_0 + \phonk\zpos_0)}.
\end{equation}
We again denote the radial motional states by $\ket{\atn}$, where $\atn\in\N$ is the number of motional quanta. For each state $\ket{\atn}$, the spontaneous radiative decay rate is $\atLinewidth_{0} \equiv 2\pi \sum_{\phonindex} \DOS_{\phonindex} |\atphong_{\phonindex\rpos}|^2$. Here, the sum runs over the phonon modes $\phonindex$ resonant with the trap and $\DOS_{\phonindex} = \left|d\phonfreq_\phonindex/d\phonk\right|^{-1}$ is the phonon density of states. The depopulation rate is $\atLinewidth_{\atn} \simeq (2 \atn + 1) \nbar_\phonindex \atLinewidth_{0}$ if the thermal occupation $\nbar_\phonindex \gg 1$ of the resonant phonon modes is large. Hence, we obtain the following analytical expression for the phonon-induced depopulation rates of the radial motional states of an atom close to the ground state $\ket{0}$ of a nanofiber-based optical trap:
\begin{equation}\label{eqn-S:linewidths harmonic trap}
 \depopulationRate_\atn = \frac{(2 \atn + 1)}{2\sqrt{2}\pi} \frac{\boltzmann\temp \atmass}{\hbar}  \sqrt{\frac{\rtrapfreq}{\rad^5 \sqrt{\YoungE \dens^3}}}.
\end{equation}
We use this expression to verify our numerical methods: The numerical result $\depopulationRate_\atn = \SI{214}{\hertz}$ for the ground state $\ket{0}$ obtained using \cref{eqn-S:linewidths optical trap} and presented in \cref{fig-S: optical trap linewidths} agrees well with the rate $\depopulationRate_\atn = \SI{216}{\hertz}$ obtained analytically using \cref{eqn-S:linewidths harmonic trap}. These results are compatible with experimentally observed linewidths~\cite{S_reitz_coherence_2013,hummer_heating_2019,albrecht_fictitious_2016}.

\section{Heterodyne Fluorescence Spectroscopy}
\label{sec: spectroscopy}

In the paper, we propose heterodyne fluorescence spectroscopy to probe the quantized spectrum of surface-bound motional states. Under suitable conditions~\cite{S_cirac_spectrum_1993}, the resulting signal reveals Raman-type transitions between different states of the radial center-of-mass motion of atoms in their electronic ground state. This approach has advantages compared to the transmission~\cite{S_sague_cold-atom_2007,patterson_spectral_2018} or fluorescence excitation spectroscopy~\cite{S_nayak_optical_2007} used in previous experimental studies of surface-induced effects on atoms near optical nanofibers. These latter techniques probe surface-induced shifts between the ground state and a given excited electronic state of the atoms. In consequence, their resolution is limited by the natural linewidth of the excited electronic state. For the Raman spectroscopy technique proposed here, the surface-induced shifts only change the overall strength of the signal but not its shape. In consequence, the Raman spectroscopy is not limited by spectral width of the optically excited state and can provide access to the closely spaced energy levels shown in Fig.~2 of the paper.

To probe the radial motional states of atoms bound directly to the nanofiber surface, a circularly polarized probe laser with a frequency $\photfreqin$ detuned from resonance with the atom is coupled into the fiber as a traveling wave. The resulting polarization in the nanofiber region is quasi-circularly polarized, with azimuthal order $\photl=\pm1$; see \cref{sec: nanofiber photons}. The probe beam has a wavelength in the single-mode regime of the nanofiber, such that probe photons are guided on the $\HEmode_{11}$ band in the nanofiber region. We assume that the probe laser is sufficiently far detuned from resonance with transitions between the $6\text{S}$ and $6\text{P}$ manifolds of the cesium atom to treat the atom as an effective two-level system with ground state $\ket{g}$, excited state $\ket{e}$, and transition frequency $\atintfreq$. Those photons that are scattered by the atom back into the nanofiber in the forward direction are recombined with the local oscillator on a beam splitter. The frequency $\photfreqout$ of a scattered photon is changed to $\photfreqout$ when the atom simultaneously changes its motional state, leading to motional sidebands in the spectrum of the probe beam. The frequency difference between the probe beam and the local oscillator results in a beat that can be observed with a photodetector. The local oscillator is shifted by an offset $\LOshift$ such that the spectrum of the photocurrent contains sidebands at $\photfreqin+\LOshift - \photfreqout$. This shift separates the Stokes- and anti-Stokes sidebands in the final signal and to choose the optimal working point for the photodetector. Moreover, the polarization of the local oscillator is matched to the polarization of the probe beam. In consequence, the beat signal is predominantly due to photons that are scattered without changing their polarization. This specific choice of polarizations eliminates the contribution of changes of the atoms' azimuthal motional state to the spectroscopy signal, while the detection of light scattered in the forward direction minimizes the recoil in the axial motion of the atoms. As a result, the proposed spectroscopy configuration is only sensitive to the radial motion of the atoms, and the motional sidebands correspond to transitions $\atn \to \atnb$ between different radial motional states.

The atom-phonon-photon system can then be described by the Hamiltonian $\Hamilop' = \Hamilop_\motional + \Hamilop_\vibrational + \Hamilop_\atphonint + \Hamilop_\electronic + \Hamilop_\photonic + \Hamilop_\atphotint$ where the electronic structure of the atom is governed by $\Hamilop_\electronic = \hbar \atintfreq \ketbra{e}{e}$ and the atom interacts with the electric field through the dipole coupling $\Hamilop_\atphotint = - \dipolevecop \cdot \Efieldop(\atposop)$. Here, $\dipolevecop$ is the dipole moment of the atom. This model assumes that the probe laser is weak such that multiple scattering of a photon by several atoms can be neglected, and it is sufficient to treat every atom individually. To predict the spectral distribution of the power $\power(\photfreqoutin)$ of the scattered light as a function of the frequency difference $\photfreqoutin \equiv \photfreqout - \photfreqin$, one can calculate the steady-state of the system in the presence of a coherently driven laser mode and a thermal nanofiber phonon bath using a master equation approach~\cite{S_lindberg_resonance_1986,cirac_spectrum_1993}. There is, however, an alternative way to approximate the resulting spectrum that is sufficient for the purpose of this paper: The motional states we consider have lifetimes corresponding to $2\pi/\atLinewidth_\atn \sim \SI{1}{\milli\second}$ that are much longer than the time of $2\pi/\atintdecayrate \sim \SI{100}{\nano\second}$ it takes a probe photon to be absorbed and re-emitted by the atom. Here, $\atintdecayrate$ is the lifetime of states in the $6\text{P}$ manifold of cesium. We can, therefore, treat the motional states as eigenstates for the duration of the scattering process and neglect their coupling to the nanofiber phonons. This approximation allows us to employ scattering theory to obtain the position and relative weight of the motional sidebands in the spectrum $\power(\photfreqoutin)$. In a second step, we then account for the finite linewidth of transitions between the motional states.

We assume that the probe laser has a sufficiently low power such that the atom only interacts with one photon at a time. The relevant transitions are, therefore, between states where the atom starts in its internal ground state $\ket{g}$ and the motional state $\ket{\atindex}=\ket{\atn,\atl,\atk}$, and ends again in its ground state but with a different motional state $\ket{\atindexb}=\ket{\atnb,\atlb,\atkb}$. Simultaneously, a photon is scattered from the mode $\photindexin$ to the mode $\photindexout$. Since we detect only scattered photons that are still nanofiber-guided, propagate in the same direction, and have the same polarization, the modes $\photindexin$ and $\photindexout$ can only differ in their frequencies. Conservation of angular momentum then implies that $\photlb=\photl$. Moreover, we can neglect the change in kinetic energy of the atom due to recoil along the nanofiber axis, so $\atkb \simeq \atk$. Energy conservation hence requires the detected photon to have a frequency shifted by $\photfreqoutin = \atfreq_\atn - \atfreq_\atnb$. One can show using the resolvent ~\cite{S_cohen-tannoudji_atom-photon_1998} that the scattering matrix element for transitions $\atn \to \atnb$ while changing the frequency of the photon by $\photfreqoutin$ is
\begin{equation}\label{eqn-S:scattering matrix element}
    \Smatrixcomp_{\atnb\atn}(\photfreqoutin) \simeq \frac{2\pi \im}{\hbar^2} \dirac\pare{ \atfreq_{\atnb\atn} - \photfreqoutin } \frac{ (\dipolecomp/3)^2 \FCtensorcomp_{\atnb\atn} }{ \laserdetuning + \im \atintdecayrate/2}.
\end{equation}
Here, $\atfreq_{\atnb\atn} \equiv \atfreq_{\atnb} - \atfreq_\atn$ is the frequency difference between the initial and the final radial motional state of the atom and $\laserdetuning\equiv \photfreqin - \atintfreq$ is the detuning of the probe laser from resonance with the atom. Note that $\atintfreq$ and $\atintdecayrate$ are modified by the presence of the nanofiber compared to a cesium atom in free space. They depend on the distance between the atom and the nanofiber and hence on the radial motional state $\atn$. In the following, we assume that differences in the transition frequency and decay rate can be neglected over the limited range of motional states we consider. The relative weights of the sidebands in \cref{eqn-S:scattering matrix element} are determined by the Franck-Condon factors
\begin{equation}\label{eqn-S:Franck Condon factors}
  \FCtensorcomp_{\atnb\atn} \equiv \frac{\Efieldmodedens_\photindexout \Efieldmodedens_\photindexin}{(2\pi)^2}  \int_0^\infty \cconj\atradwf_{\atnb}(\rpos) \, \cconj{\emoder}_\photindexout(\rpos) \cdot \emoder_\photindexin(\rpos) \, \atradwf_{\atn}(\rpos) \dd \rpos.
\end{equation}
In deriving \cref{eqn-S:scattering matrix element}, we (i) exploit that the scattering of a probe photon by the atom is sufficiently fast such that the motional state of the atom does not decay in the meantime; (ii) assume that $|\laserdetuning| \gg |\atfreq_{\atnb\atn}|$, which is the case for the weakly bound states considered in the paper if the probe laser is detuned by a few $\si{\nano\meter}$; (iii) assume that the detuning is sufficiently large for the response of the atom to be isotropic, that is, $\braket{g|\dipolecompop^i \dipolecompop^j|g}= (\dipolecomp/3)^2 \kronecker^{ij}$ where $d\in\R$ and $\dipolecompop^i$ are components of the dipole moment $\dipolevecop$ of the atom.

The power of the scattered light is $\power(\photfreqoutin) \propto \sum_{\atn,\atnb\neq\atn}n(\atn) |\Smatrixcomp_{\atnb\atn}(\photfreqoutin)|$ where $n(\atn)$ is the number of atoms initially in the motional state $\atn$. In practice, the sharp sidebands in \cref{eqn-S:scattering matrix element} are broadened due to sources of noise and decoherence affecting either the laser or the motion of the atom. If the same laser source is used for both the probe beam and the reference beam, the frequency drift of the laser has no effect and the linewidths of the sidebands are determined by the decoherence of the motional atomic states. We can model the phonon-induced linewidths of the motional states by replacing the sharp sidebands in \cref{eqn-S:scattering matrix element} with Lorentzian resonances of the appropriate width $\atLinewidth_{\atnb\atn}$ and the same total power:
\begin{equation}
   \dirac\pare{ \atfreq_{\atnb\atn} - \photfreqoutin }  \to \frac{1}{\pi} \frac{ \atLinewidth_{\atnb\atn}/2 }{ \pare{ \atfreq_{\atnb\atn} - \photfreqoutin }^2 + \pare{\atLinewidth_{\atnb\atn}/2}^2  }.
\end{equation}
The motional states considered in the paper fall into a frequency interval that is small compared to the depth of the potential $\pot(\rpos)$. In consequence, we can approximate the occupation $n(\atn)$ of these states as constant. The power of the light scattered by the atom is therefore
\begin{equation}
  \power(\photfreqoutin) \propto \sum_{\atn,\atnb\neq\atn} \frac{ \atLinewidth_{\atnb\atn}/2 }{ \pare{ \atfreq_{\atnb\atn} - \photfreqoutin }^2 + \pare{\atLinewidth_{\atnb\atn}/2}^2 } \left| \FCtensorcomp_{\atnb\atn} \right|^2
\end{equation}
as a function of the frequency difference between probe photons and detected photons.

\begin{figure}
  \centering
  \includegraphics[height=158.26pt]{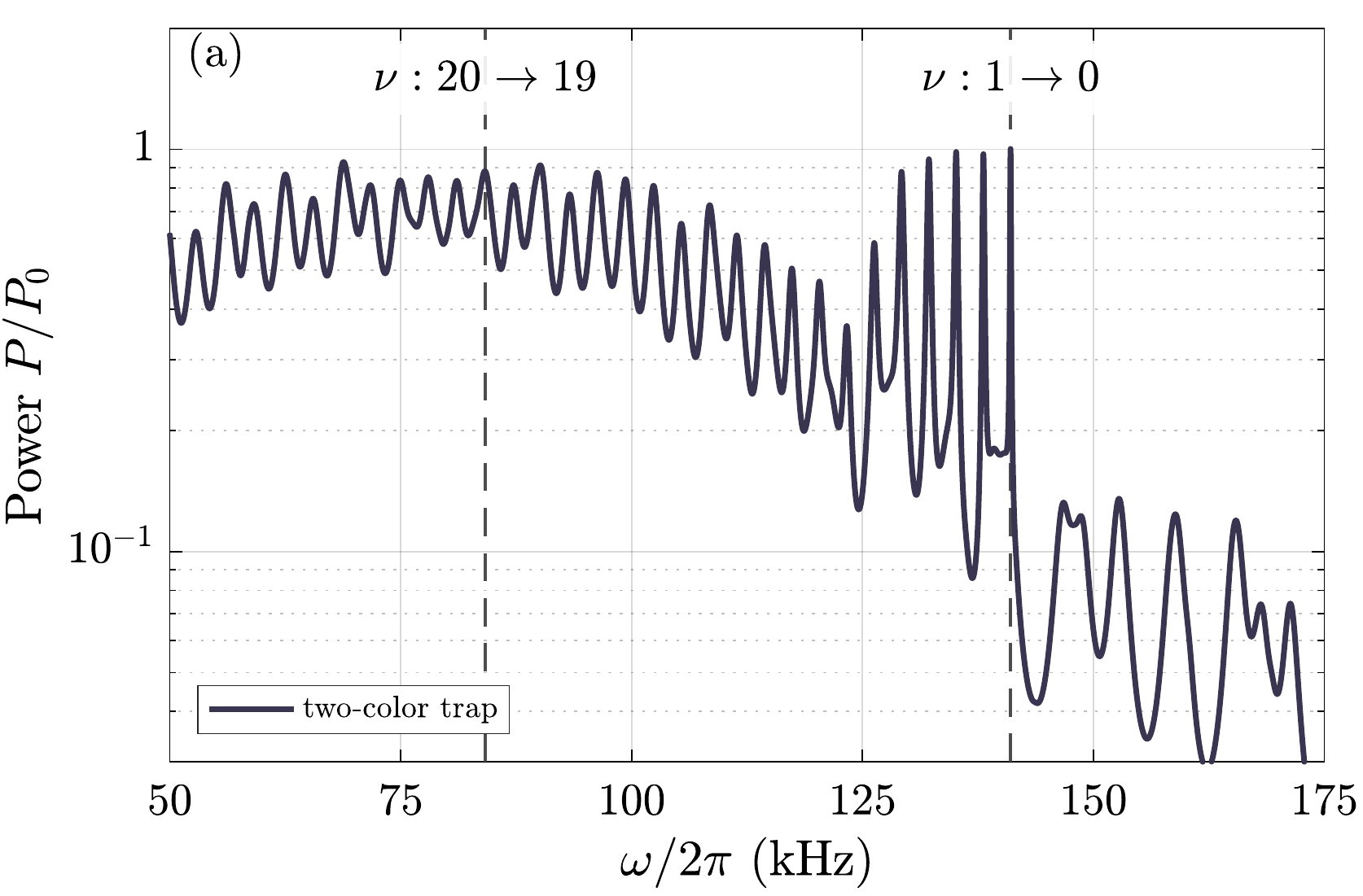}
  \hspace*{.6cm}
  \includegraphics[height=158.26pt]{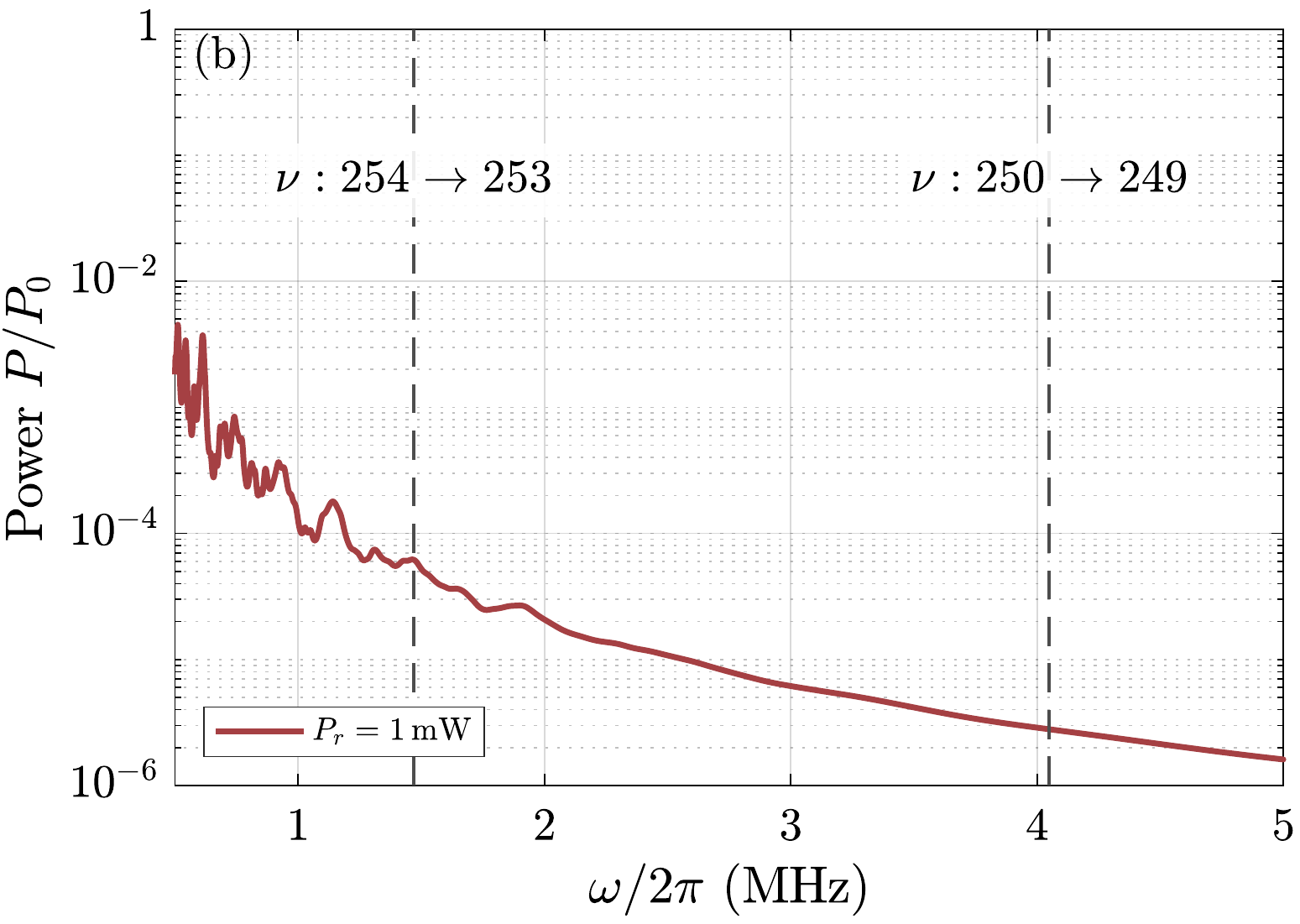}
  \caption{Sidebands in the fluorescence spectrum of an atom bound to an optical nanofiber. Panel (a) shows the spectrum for atoms in a two-color trap. The sidebands are due to transitions between the radial motional states shown in \cref{fig-S: optical trap linewidths}. The motion in azimuthal and axial direction leads to additional sidebands that are not represented here. We neglect the coupling between the motion in radial, azimuthal, and axial direction. Panel (b) corresponds Fig.~3b in the paper and shows a larger interval of the spectrum for adsorbed atoms. The indicated transitions involve the states $\atn=(253,249)$, which have frequencies $\atfreq_\atn = -2\pi \times (\num{8.9},\num{20})\,\si{\mega\hertz}$ and lie deeper than the states shown in Fig.~2a of the paper.}
  \label{fig-S: fluorescence spectra}
\end{figure}

In Fig.~3 of the paper, we show fluorescence spectra for adsorbed atoms and atoms in the hybrid light- and surface-induced potential. In \csubref{fig-S: fluorescence spectra}{a}, we plot the spectrum due to transitions between the optically trapped states shown in \cref{fig-S: optical trap linewidths}. We use \cref{eqn-S:linewidths optical trap} to calculate the linewidths, assuming that the linewidths of atoms trapped in two-color traps around a long nanofiber are limited by depopulation. We further approximate the population of the motional states as equal. In practice, the spectrum features additional sidebands from the motion in axial and azimuthal direction since the two-color trap confines the atom in all three spatial directions. These sidebands are omitted in \cref{fig-S: fluorescence spectra}. We use the power $\power_0$ of the sideband corresponding to transitions between the radial ground state $\atn=0$ and first excited state $\atn=1$ as a reference and plot all spectra in units of $P_0$.

\Csubref{fig-S: fluorescence spectra}{b} shows the fluorescence spectrum for adsorbed atoms in a larger frequency interval than in Fig.~3a in the paper, involving states with larger binding energies. The corresponding wave functions have a much smaller spatial extent, which results in smaller Franck-Condon factors. Atoms in these states are, therefore, much less likely to scatter a nanofiber-guided photon and are more difficult to probe. Moreover, transitions with larger frequencies can no longer be resolved due to their increasing linewidths. For these reasons, we focus on states with binding energies of a few $\si{\mega\hertz}$ and transition frequencies of a few hundred $\si{\kilo\hertz}$ in the paper.


\begin{thebibliography}{90}%
\makeatletter
\providecommand \@ifxundefined [1]{%
 \@ifx{#1\undefined}
}%
\providecommand \@ifnum [1]{%
 \ifnum #1\expandafter \@firstoftwo
 \else \expandafter \@secondoftwo
 \fi
}%
\providecommand \@ifx [1]{%
 \ifx #1\expandafter \@firstoftwo
 \else \expandafter \@secondoftwo
 \fi
}%
\providecommand \natexlab [1]{#1}%
\providecommand \enquote  [1]{``#1''}%
\providecommand \bibnamefont  [1]{#1}%
\providecommand \bibfnamefont [1]{#1}%
\providecommand \citenamefont [1]{#1}%
\providecommand \href@noop [0]{\@secondoftwo}%
\providecommand \href [0]{\begingroup \@sanitize@url \@href}%
\providecommand \@href[1]{\@@startlink{#1}\@@href}%
\providecommand \@@href[1]{\endgroup#1\@@endlink}%
\providecommand \@sanitize@url [0]{\catcode `\\12\catcode `\$12\catcode
  `\&12\catcode `\#12\catcode `\^12\catcode `\_12\catcode `\%12\relax}%
\providecommand \@@startlink[1]{}%
\providecommand \@@endlink[0]{}%
\providecommand \url  [0]{\begingroup\@sanitize@url \@url }%
\providecommand \@url [1]{\endgroup\@href {#1}{\urlprefix }}%
\providecommand \urlprefix  [0]{URL }%
\providecommand \Eprint [0]{\href }%
\providecommand \doibase [0]{http://dx.doi.org/}%
\providecommand \selectlanguage [0]{\@gobble}%
\providecommand \bibinfo  [0]{\@secondoftwo}%
\providecommand \bibfield  [0]{\@secondoftwo}%
\providecommand \translation [1]{[#1]}%
\providecommand \BibitemOpen [0]{}%
\providecommand \bibitemStop [0]{}%
\providecommand \bibitemNoStop [0]{.\EOS\space}%
\providecommand \EOS [0]{\spacefactor3000\relax}%
\providecommand \BibitemShut  [1]{\csname bibitem#1\endcsname}%
\let\auto@bib@innerbib\@empty
\bibitem [{\citenamefont {Chang}\ \emph {et~al.}(2018)\citenamefont {Chang},
  \citenamefont {Douglas}, \citenamefont {{Gonz{\'a}lez-Tudela}}, \citenamefont
  {Hung},\ and\ \citenamefont {Kimble}}]{chang_colloquium:_2018}%
  \BibitemOpen
  \bibfield  {author} {\bibinfo {author} {\bibfnamefont {D.~E.}\ \bibnamefont
  {Chang}}, \bibinfo {author} {\bibfnamefont {J.~S.}\ \bibnamefont {Douglas}},
  \bibinfo {author} {\bibfnamefont {A.}~\bibnamefont {{Gonz{\'a}lez-Tudela}}},
  \bibinfo {author} {\bibfnamefont {C.-L.}\ \bibnamefont {Hung}}, \ and\
  \bibinfo {author} {\bibfnamefont {H.~J.}\ \bibnamefont {Kimble}},\ }\href
  {\doibase 10.1103/RevModPhys.90.031002} {\bibfield  {journal} {\bibinfo
  {journal} {Rev. Mod. Phys.}\ }\textbf {\bibinfo {volume} {90}},\ \bibinfo
  {pages} {031002} (\bibinfo {year} {2018})}\BibitemShut {NoStop}%
\bibitem [{\citenamefont {Corzo}\ \emph {et~al.}(2019)\citenamefont {Corzo},
  \citenamefont {Raskop}, \citenamefont {Chandra}, \citenamefont {Sheremet},
  \citenamefont {Gouraud},\ and\ \citenamefont
  {Laurat}}]{corzo_waveguide-coupled_2019}%
  \BibitemOpen
  \bibfield  {author} {\bibinfo {author} {\bibfnamefont {N.~V.}\ \bibnamefont
  {Corzo}}, \bibinfo {author} {\bibfnamefont {J.}~\bibnamefont {Raskop}},
  \bibinfo {author} {\bibfnamefont {A.}~\bibnamefont {Chandra}}, \bibinfo
  {author} {\bibfnamefont {A.~S.}\ \bibnamefont {Sheremet}}, \bibinfo {author}
  {\bibfnamefont {B.}~\bibnamefont {Gouraud}}, \ and\ \bibinfo {author}
  {\bibfnamefont {J.}~\bibnamefont {Laurat}},\ }\href {\doibase
  10.1038/s41586-019-0902-3} {\bibfield  {journal} {\bibinfo  {journal}
  {Nature}\ }\textbf {\bibinfo {volume} {566}},\ \bibinfo {pages} {359}
  (\bibinfo {year} {2019})}\BibitemShut {NoStop}%
\bibitem [{\citenamefont {Dalvit}\ \emph {et~al.}(2011)\citenamefont {Dalvit},
  \citenamefont {Milonni}, \citenamefont {Roberts},\ and\ \citenamefont
  {da~Rosa}}]{dalvit_casimir_2011}%
  \BibitemOpen
  \bibinfo {editor} {\bibfnamefont {D.}~\bibnamefont {Dalvit}}, \bibinfo
  {editor} {\bibfnamefont {P.}~\bibnamefont {Milonni}}, \bibinfo {editor}
  {\bibfnamefont {D.}~\bibnamefont {Roberts}}, \ and\ \bibinfo {editor}
  {\bibfnamefont {F.}~\bibnamefont {da~Rosa}},\ eds.,\ \href@noop {} {\emph
  {\bibinfo {title} {Casimir {{Physics}}}}}\ (\bibinfo  {publisher}
  {{Springer}},\ \bibinfo {address} {{Berlin}},\ \bibinfo {year}
  {2011})\BibitemShut {NoStop}%
\bibitem [{\citenamefont {Gierling}\ \emph {et~al.}(2011)\citenamefont
  {Gierling}, \citenamefont {Schneeweiss}, \citenamefont {Visanescu},
  \citenamefont {Federsel}, \citenamefont {H{\"a}ffner}, \citenamefont {Kern},
  \citenamefont {Judd}, \citenamefont {G{\"u}nther},\ and\ \citenamefont
  {Fort{\'a}gh}}]{gierling_cold-atom_2011}%
  \BibitemOpen
  \bibfield  {author} {\bibinfo {author} {\bibfnamefont {M.}~\bibnamefont
  {Gierling}}, \bibinfo {author} {\bibfnamefont {P.}~\bibnamefont
  {Schneeweiss}}, \bibinfo {author} {\bibfnamefont {G.}~\bibnamefont
  {Visanescu}}, \bibinfo {author} {\bibfnamefont {P.}~\bibnamefont {Federsel}},
  \bibinfo {author} {\bibfnamefont {M.}~\bibnamefont {H{\"a}ffner}}, \bibinfo
  {author} {\bibfnamefont {D.~P.}\ \bibnamefont {Kern}}, \bibinfo {author}
  {\bibfnamefont {T.~E.}\ \bibnamefont {Judd}}, \bibinfo {author}
  {\bibfnamefont {A.}~\bibnamefont {G{\"u}nther}}, \ and\ \bibinfo {author}
  {\bibfnamefont {J.}~\bibnamefont {Fort{\'a}gh}},\ }\href {\doibase
  10.1038/nnano.2011.80} {\bibfield  {journal} {\bibinfo  {journal} {Nat.
  Nanotechnol.}\ }\textbf {\bibinfo {volume} {6}},\ \bibinfo {pages} {446}
  (\bibinfo {year} {2011})}\BibitemShut {NoStop}%
\bibitem [{\citenamefont {Schneeweiss}\ \emph {et~al.}(2012)\citenamefont
  {Schneeweiss}, \citenamefont {Gierling}, \citenamefont {Visanescu},
  \citenamefont {Kern}, \citenamefont {Judd}, \citenamefont {G{\"u}nther},\
  and\ \citenamefont {Fort{\'a}gh}}]{schneeweiss_dispersion_2012}%
  \BibitemOpen
  \bibfield  {author} {\bibinfo {author} {\bibfnamefont {P.}~\bibnamefont
  {Schneeweiss}}, \bibinfo {author} {\bibfnamefont {M.}~\bibnamefont
  {Gierling}}, \bibinfo {author} {\bibfnamefont {G.}~\bibnamefont {Visanescu}},
  \bibinfo {author} {\bibfnamefont {D.~P.}\ \bibnamefont {Kern}}, \bibinfo
  {author} {\bibfnamefont {T.~E.}\ \bibnamefont {Judd}}, \bibinfo {author}
  {\bibfnamefont {A.}~\bibnamefont {G{\"u}nther}}, \ and\ \bibinfo {author}
  {\bibfnamefont {J.}~\bibnamefont {Fort{\'a}gh}},\ }\href {\doibase
  10.1038/nnano.2012.93} {\bibfield  {journal} {\bibinfo  {journal} {Nat.
  Nanotechnol.}\ }\textbf {\bibinfo {volume} {7}},\ \bibinfo {pages} {515}
  (\bibinfo {year} {2012})}\BibitemShut {NoStop}%
\bibitem [{\citenamefont {Yang}\ \emph {et~al.}(2017)\citenamefont {Yang},
  \citenamefont {Koll{\'a}r}, \citenamefont {Taylor}, \citenamefont {Turner},\
  and\ \citenamefont {Lev}}]{yang_scanning_2017}%
  \BibitemOpen
  \bibfield  {author} {\bibinfo {author} {\bibfnamefont {F.}~\bibnamefont
  {Yang}}, \bibinfo {author} {\bibfnamefont {A.~J.}\ \bibnamefont
  {Koll{\'a}r}}, \bibinfo {author} {\bibfnamefont {S.~F.}\ \bibnamefont
  {Taylor}}, \bibinfo {author} {\bibfnamefont {R.~W.}\ \bibnamefont {Turner}},
  \ and\ \bibinfo {author} {\bibfnamefont {B.~L.}\ \bibnamefont {Lev}},\ }\href
  {\doibase 10.1103/PhysRevApplied.7.034026} {\bibfield  {journal} {\bibinfo
  {journal} {Phys. Rev. Appl.}\ }\textbf {\bibinfo {volume} {7}},\ \bibinfo
  {pages} {034026} (\bibinfo {year} {2017})}\BibitemShut {NoStop}%
\bibitem [{\citenamefont {Fichet}\ \emph {et~al.}(2007)\citenamefont {Fichet},
  \citenamefont {Dutier}, \citenamefont {Yarovitsky}, \citenamefont {Todorov},
  \citenamefont {Hamdi}, \citenamefont {Maurin}, \citenamefont {Saltiel},
  \citenamefont {Sarkisyan}, \citenamefont {{M.-P. Gorza}}, \citenamefont
  {Bloch},\ and\ \citenamefont {Ducloy}}]{fichet_exploring_2007}%
  \BibitemOpen
  \bibfield  {author} {\bibinfo {author} {\bibfnamefont {M.}~\bibnamefont
  {Fichet}}, \bibinfo {author} {\bibfnamefont {G.}~\bibnamefont {Dutier}},
  \bibinfo {author} {\bibfnamefont {A.}~\bibnamefont {Yarovitsky}}, \bibinfo
  {author} {\bibfnamefont {P.}~\bibnamefont {Todorov}}, \bibinfo {author}
  {\bibfnamefont {I.}~\bibnamefont {Hamdi}}, \bibinfo {author} {\bibfnamefont
  {I.}~\bibnamefont {Maurin}}, \bibinfo {author} {\bibfnamefont
  {S.}~\bibnamefont {Saltiel}}, \bibinfo {author} {\bibfnamefont
  {D.}~\bibnamefont {Sarkisyan}}, \bibinfo {author} {\bibnamefont {{M.-P.
  Gorza}}}, \bibinfo {author} {\bibfnamefont {D.}~\bibnamefont {Bloch}}, \ and\
  \bibinfo {author} {\bibfnamefont {M.}~\bibnamefont {Ducloy}},\ }\href
  {\doibase 10.1209/0295-5075/77/54001} {\bibfield  {journal} {\bibinfo
  {journal} {Europhys. Lett.}\ }\textbf {\bibinfo {volume} {77}},\ \bibinfo {pages} {54001}
  (\bibinfo {year} {2007})}\BibitemShut {NoStop}%
\bibitem [{\citenamefont {Peyrot}\ \emph {et~al.}(2019)\citenamefont {Peyrot},
  \citenamefont {{\v S}ibali{\'c}}, \citenamefont {Sortais}, \citenamefont
  {Browaeys}, \citenamefont {Sargsyan}, \citenamefont {Sarkisyan},
  \citenamefont {Hughes},\ and\ \citenamefont
  {Adams}}]{peyrot_measurement_2019}%
  \BibitemOpen
  \bibfield  {author} {\bibinfo {author} {\bibfnamefont {T.}~\bibnamefont
  {Peyrot}}, \bibinfo {author} {\bibfnamefont {N.}~\bibnamefont {{\v
  S}ibali{\'c}}}, \bibinfo {author} {\bibfnamefont {Y.~R.~P.}\ \bibnamefont
  {Sortais}}, \bibinfo {author} {\bibfnamefont {A.}~\bibnamefont {Browaeys}},
  \bibinfo {author} {\bibfnamefont {A.}~\bibnamefont {Sargsyan}}, \bibinfo
  {author} {\bibfnamefont {D.}~\bibnamefont {Sarkisyan}}, \bibinfo {author}
  {\bibfnamefont {I.~G.}\ \bibnamefont {Hughes}}, \ and\ \bibinfo {author}
  {\bibfnamefont {C.~S.}\ \bibnamefont {Adams}},\ }\href {\doibase
  10.1103/PhysRevA.100.022503} {\bibfield  {journal} {\bibinfo  {journal}
  {Phys. Rev. A}\ }\textbf {\bibinfo {volume} {100}},\ \bibinfo {pages}
  {022503} (\bibinfo {year} {2019})}\BibitemShut {NoStop}%
\bibitem [{\citenamefont {Onofrio}(2006)}]{onofrio_casimir_2006}%
  \BibitemOpen
  \bibfield  {author} {\bibinfo {author} {\bibfnamefont {R.}~\bibnamefont
  {Onofrio}},\ }\href {\doibase 10.1088/1367-2630/8/10/237} {\bibfield
  {journal} {\bibinfo  {journal} {New J. Phys.}\ }\textbf {\bibinfo {volume}
  {8}},\ \bibinfo {pages} {237} (\bibinfo {year} {2006})}\BibitemShut {NoStop}%
\bibitem [{\citenamefont {Intravaia}\ \emph {et~al.}(2015)\citenamefont
  {Intravaia}, \citenamefont {Mkrtchian}, \citenamefont {Buhmann},
  \citenamefont {Scheel}, \citenamefont {Dalvit},\ and\ \citenamefont
  {Henkel}}]{intravaia_friction_2015}%
  \BibitemOpen
  \bibfield  {author} {\bibinfo {author} {\bibfnamefont {F.}~\bibnamefont
  {Intravaia}}, \bibinfo {author} {\bibfnamefont {V.~E.}\ \bibnamefont
  {Mkrtchian}}, \bibinfo {author} {\bibfnamefont {S.~Y.}\ \bibnamefont
  {Buhmann}}, \bibinfo {author} {\bibfnamefont {S.}~\bibnamefont {Scheel}},
  \bibinfo {author} {\bibfnamefont {D.~A.~R.}\ \bibnamefont {Dalvit}}, \ and\
  \bibinfo {author} {\bibfnamefont {C.}~\bibnamefont {Henkel}},\ }\href
  {\doibase 10.1088/0953-8984/27/21/214020} {\bibfield  {journal} {\bibinfo
  {journal} {J. Phys.: Condens. Matter}\ }\textbf {\bibinfo {volume} {27}},\
  \bibinfo {pages} {214020} (\bibinfo {year} {2015})}\BibitemShut {NoStop}%
\bibitem [{\citenamefont {Hammes}\ \emph {et~al.}(2002)\citenamefont {Hammes},
  \citenamefont {Rychtarik}, \citenamefont {N{\"a}gerl},\ and\ \citenamefont
  {Grimm}}]{hammes_cold-atom_2002}%
  \BibitemOpen
  \bibfield  {author} {\bibinfo {author} {\bibfnamefont {M.}~\bibnamefont
  {Hammes}}, \bibinfo {author} {\bibfnamefont {D.}~\bibnamefont {Rychtarik}},
  \bibinfo {author} {\bibfnamefont {H.-C.}\ \bibnamefont {N{\"a}gerl}}, \ and\
  \bibinfo {author} {\bibfnamefont {R.}~\bibnamefont {Grimm}},\ }\href
  {\doibase 10.1103/PhysRevA.66.051401} {\bibfield  {journal} {\bibinfo
  {journal} {Phys. Rev. A}\ }\textbf {\bibinfo {volume} {66}},\ \bibinfo
  {pages} {051401(R)} (\bibinfo {year} {2002})}\BibitemShut {NoStop}%
\bibitem [{\citenamefont {Stehle}\ \emph {et~al.}(2011)\citenamefont {Stehle},
  \citenamefont {Bender}, \citenamefont {Zimmermann}, \citenamefont {Kern},
  \citenamefont {Fleischer},\ and\ \citenamefont
  {Slama}}]{stehle_plasmonically_2011}%
  \BibitemOpen
  \bibfield  {author} {\bibinfo {author} {\bibfnamefont {C.}~\bibnamefont
  {Stehle}}, \bibinfo {author} {\bibfnamefont {H.}~\bibnamefont {Bender}},
  \bibinfo {author} {\bibfnamefont {C.}~\bibnamefont {Zimmermann}}, \bibinfo
  {author} {\bibfnamefont {D.}~\bibnamefont {Kern}}, \bibinfo {author}
  {\bibfnamefont {M.}~\bibnamefont {Fleischer}}, \ and\ \bibinfo {author}
  {\bibfnamefont {S.}~\bibnamefont {Slama}},\ }\href {\doibase
  10.1038/nphoton.2011.159} {\bibfield  {journal} {\bibinfo  {journal} {Nat.
  Photonics}\ }\textbf {\bibinfo {volume} {5}},\ \bibinfo {pages} {494}
  (\bibinfo {year} {2011})}\BibitemShut {NoStop}%
\bibitem [{\citenamefont {Thompson}\ \emph {et~al.}(2013)\citenamefont
  {Thompson}, \citenamefont {Tiecke}, \citenamefont {de~Leon}, \citenamefont
  {Feist}, \citenamefont {Akimov}, \citenamefont {Gullans}, \citenamefont
  {Zibrov}, \citenamefont {Vuleti{\'c}},\ and\ \citenamefont
  {Lukin}}]{thompson_coupling_2013}%
  \BibitemOpen
  \bibfield  {author} {\bibinfo {author} {\bibfnamefont {J.~D.}\ \bibnamefont
  {Thompson}}, \bibinfo {author} {\bibfnamefont {T.~G.}\ \bibnamefont
  {Tiecke}}, \bibinfo {author} {\bibfnamefont {N.~P.}\ \bibnamefont {de~Leon}},
  \bibinfo {author} {\bibfnamefont {J.}~\bibnamefont {Feist}}, \bibinfo
  {author} {\bibfnamefont {A.~V.}\ \bibnamefont {Akimov}}, \bibinfo {author}
  {\bibfnamefont {M.}~\bibnamefont {Gullans}}, \bibinfo {author} {\bibfnamefont
  {A.~S.}\ \bibnamefont {Zibrov}}, \bibinfo {author} {\bibfnamefont
  {V.}~\bibnamefont {Vuleti{\'c}}}, \ and\ \bibinfo {author} {\bibfnamefont
  {M.~D.}\ \bibnamefont {Lukin}},\ }\href {\doibase 10.1126/science.1237125}
  {\bibfield  {journal} {\bibinfo  {journal} {Science}\ }\textbf {\bibinfo
  {volume} {340}},\ \bibinfo {pages} {1202} (\bibinfo {year}
  {2013})}\BibitemShut {NoStop}%
\bibitem [{\citenamefont {Goban}\ \emph {et~al.}(2015)\citenamefont {Goban},
  \citenamefont {Hung}, \citenamefont {Hood}, \citenamefont {Yu}, \citenamefont
  {Muniz}, \citenamefont {Painter},\ and\ \citenamefont
  {Kimble}}]{goban_superradiance_2015}%
  \BibitemOpen
  \bibfield  {author} {\bibinfo {author} {\bibfnamefont {A.}~\bibnamefont
  {Goban}}, \bibinfo {author} {\bibfnamefont {C.-L.}\ \bibnamefont {Hung}},
  \bibinfo {author} {\bibfnamefont {J.~D.}\ \bibnamefont {Hood}}, \bibinfo
  {author} {\bibfnamefont {S.-P.}\ \bibnamefont {Yu}}, \bibinfo {author}
  {\bibfnamefont {J.~A.}\ \bibnamefont {Muniz}}, \bibinfo {author}
  {\bibfnamefont {O.}~\bibnamefont {Painter}}, \ and\ \bibinfo {author}
  {\bibfnamefont {H.~J.}\ \bibnamefont {Kimble}},\ }\href {\doibase
  10.1103/PhysRevLett.115.063601} {\bibfield  {journal} {\bibinfo  {journal}
  {Phys. Rev. Lett.}\ }\textbf {\bibinfo {volume} {115}},\ \bibinfo {pages}
  {063601} (\bibinfo {year} {2015})}\BibitemShut {NoStop}%
\bibitem [{\citenamefont {Vetsch}\ \emph {et~al.}(2010)\citenamefont {Vetsch},
  \citenamefont {Reitz}, \citenamefont {Sagu{\'e}}, \citenamefont {Schmidt},
  \citenamefont {Dawkins},\ and\ \citenamefont
  {Rauschenbeutel}}]{vetsch_optical_2010}%
  \BibitemOpen
  \bibfield  {author} {\bibinfo {author} {\bibfnamefont {E.}~\bibnamefont
  {Vetsch}}, \bibinfo {author} {\bibfnamefont {D.}~\bibnamefont {Reitz}},
  \bibinfo {author} {\bibfnamefont {G.}~\bibnamefont {Sagu{\'e}}}, \bibinfo
  {author} {\bibfnamefont {R.}~\bibnamefont {Schmidt}}, \bibinfo {author}
  {\bibfnamefont {S.~T.}\ \bibnamefont {Dawkins}}, \ and\ \bibinfo {author}
  {\bibfnamefont {A.}~\bibnamefont {Rauschenbeutel}},\ }\href {\doibase
  10.1103/PhysRevLett.104.203603} {\bibfield  {journal} {\bibinfo  {journal}
  {Phys. Rev. Lett.}\ }\textbf {\bibinfo {volume} {104}},\ \bibinfo {pages}
  {203603} (\bibinfo {year} {2010})}\BibitemShut {NoStop}%
\bibitem [{\citenamefont {Goban}\ \emph {et~al.}(2012)\citenamefont {Goban},
  \citenamefont {Choi}, \citenamefont {Alton}, \citenamefont {Ding},
  \citenamefont {Lacro{\^u}te}, \citenamefont {Pototschnig}, \citenamefont
  {Thiele}, \citenamefont {Stern},\ and\ \citenamefont
  {Kimble}}]{goban_demonstration_2012}%
  \BibitemOpen
  \bibfield  {author} {\bibinfo {author} {\bibfnamefont {A.}~\bibnamefont
  {Goban}}, \bibinfo {author} {\bibfnamefont {K.~S.}\ \bibnamefont {Choi}},
  \bibinfo {author} {\bibfnamefont {D.~J.}\ \bibnamefont {Alton}}, \bibinfo
  {author} {\bibfnamefont {D.}~\bibnamefont {Ding}}, \bibinfo {author}
  {\bibfnamefont {C.}~\bibnamefont {Lacro{\^u}te}}, \bibinfo {author}
  {\bibfnamefont {M.}~\bibnamefont {Pototschnig}}, \bibinfo {author}
  {\bibfnamefont {T.}~\bibnamefont {Thiele}}, \bibinfo {author} {\bibfnamefont
  {N.~P.}\ \bibnamefont {Stern}}, \ and\ \bibinfo {author} {\bibfnamefont
  {H.~J.}\ \bibnamefont {Kimble}},\ }\href {\doibase
  10.1103/PhysRevLett.109.033603} {\bibfield  {journal} {\bibinfo  {journal}
  {Phys. Rev. Lett.}\ }\textbf {\bibinfo {volume} {109}},\ \bibinfo {pages}
  {033603} (\bibinfo {year} {2012})}\BibitemShut {NoStop}%
\bibitem [{\citenamefont {B{\'e}guin}\ \emph {et~al.}(2014)\citenamefont
  {B{\'e}guin}, \citenamefont {Bookjans}, \citenamefont {Christensen},
  \citenamefont {S{\o}rensen}, \citenamefont {M{\"u}ller}, \citenamefont
  {Polzik},\ and\ \citenamefont {Appel}}]{beguin_generation_2014}%
  \BibitemOpen
  \bibfield  {author} {\bibinfo {author} {\bibfnamefont {J.-B.}\ \bibnamefont
  {B{\'e}guin}}, \bibinfo {author} {\bibfnamefont {E.~M.}\ \bibnamefont
  {Bookjans}}, \bibinfo {author} {\bibfnamefont {S.~L.}\ \bibnamefont
  {Christensen}}, \bibinfo {author} {\bibfnamefont {H.~L.}\ \bibnamefont
  {S{\o}rensen}}, \bibinfo {author} {\bibfnamefont {J.~H.}\ \bibnamefont
  {M{\"u}ller}}, \bibinfo {author} {\bibfnamefont {E.~S.}\ \bibnamefont
  {Polzik}}, \ and\ \bibinfo {author} {\bibfnamefont {J.}~\bibnamefont
  {Appel}},\ }\href {\doibase 10.1103/PhysRevLett.113.263603} {\bibfield
  {journal} {\bibinfo  {journal} {Phys. Rev. Lett.}\ }\textbf {\bibinfo
  {volume} {113}},\ \bibinfo {pages} {263603} (\bibinfo {year}
  {2014})}\BibitemShut {NoStop}%
\bibitem [{\citenamefont {Kato}\ and\ \citenamefont
  {Aoki}(2015)}]{kato_strong_2015}%
  \BibitemOpen
  \bibfield  {author} {\bibinfo {author} {\bibfnamefont {S.}~\bibnamefont
  {Kato}}\ and\ \bibinfo {author} {\bibfnamefont {T.}~\bibnamefont {Aoki}},\
  }\href {\doibase 10.1103/PhysRevLett.115.093603} {\bibfield  {journal}
  {\bibinfo  {journal} {Phys. Rev. Lett.}\ }\textbf {\bibinfo {volume} {115}},\
  \bibinfo {pages} {093603} (\bibinfo {year} {2015})}\BibitemShut {NoStop}%
\bibitem [{\citenamefont {Lee}\ \emph {et~al.}(2015)\citenamefont {Lee},
  \citenamefont {Grover}, \citenamefont {Hoffman}, \citenamefont {Orozco},\
  and\ \citenamefont {Rolston}}]{lee_inhomogeneous_2015}%
  \BibitemOpen
  \bibfield  {author} {\bibinfo {author} {\bibfnamefont {J.}~\bibnamefont
  {Lee}}, \bibinfo {author} {\bibfnamefont {J.~A.}\ \bibnamefont {Grover}},
  \bibinfo {author} {\bibfnamefont {J.~E.}\ \bibnamefont {Hoffman}}, \bibinfo
  {author} {\bibfnamefont {L.~A.}\ \bibnamefont {Orozco}}, \ and\ \bibinfo
  {author} {\bibfnamefont {S.~L.}\ \bibnamefont {Rolston}},\ }\href {\doibase
  10.1088/0953-4075/48/16/165004} {\bibfield  {journal} {\bibinfo  {journal}
  {J. Phys. B}\ }\textbf {\bibinfo {volume} {48}},\ \bibinfo {pages} {165004}
  (\bibinfo {year} {2015})}\BibitemShut {NoStop}%
\bibitem [{\citenamefont {Corzo}\ \emph {et~al.}(2016)\citenamefont {Corzo},
  \citenamefont {Gouraud}, \citenamefont {Chandra}, \citenamefont {Goban},
  \citenamefont {Sheremet}, \citenamefont {Kupriyanov},\ and\ \citenamefont
  {Laurat}}]{corzo_large_2016}%
  \BibitemOpen
  \bibfield  {author} {\bibinfo {author} {\bibfnamefont {N.~V.}\ \bibnamefont
  {Corzo}}, \bibinfo {author} {\bibfnamefont {B.}~\bibnamefont {Gouraud}},
  \bibinfo {author} {\bibfnamefont {A.}~\bibnamefont {Chandra}}, \bibinfo
  {author} {\bibfnamefont {A.}~\bibnamefont {Goban}}, \bibinfo {author}
  {\bibfnamefont {A.~S.}\ \bibnamefont {Sheremet}}, \bibinfo {author}
  {\bibfnamefont {D.~V.}\ \bibnamefont {Kupriyanov}}, \ and\ \bibinfo {author}
  {\bibfnamefont {J.}~\bibnamefont {Laurat}},\ }\href {\doibase
  10.1103/PhysRevLett.117.133603} {\bibfield  {journal} {\bibinfo  {journal}
  {Phys. Rev. Lett.}\ }\textbf {\bibinfo {volume} {117}},\ \bibinfo {pages}
  {133603} (\bibinfo {year} {2016})}\BibitemShut {NoStop}%
\bibitem [{\citenamefont {Desjonqueres}\ and\ \citenamefont
  {Spanjaard}(2012)}]{desjonqueres_concepts_2012}%
  \BibitemOpen
  \bibfield  {author} {\bibinfo {author} {\bibfnamefont {M.-C.}\ \bibnamefont
  {Desjonqueres}}\ and\ \bibinfo {author} {\bibfnamefont {D.}~\bibnamefont
  {Spanjaard}},\ }\href@noop {} {\emph {\bibinfo {title} {Concepts in {{Surface
  Physics}}}}}\ (\bibinfo  {publisher} {{Springer}},\ \bibinfo {address}
  {{Berlin}},\ \bibinfo {year} {2012})\BibitemShut {NoStop}%
\bibitem [{\citenamefont {Hung}\ \emph {et~al.}(2013)\citenamefont {Hung},
  \citenamefont {Meenehan}, \citenamefont {Chang}, \citenamefont {Painter},\
  and\ \citenamefont {Kimble}}]{hung_trapped_2013}%
  \BibitemOpen
  \bibfield  {author} {\bibinfo {author} {\bibfnamefont {C.-L.}\ \bibnamefont
  {Hung}}, \bibinfo {author} {\bibfnamefont {S.~M.}\ \bibnamefont {Meenehan}},
  \bibinfo {author} {\bibfnamefont {D.~E.}\ \bibnamefont {Chang}}, \bibinfo
  {author} {\bibfnamefont {O.}~\bibnamefont {Painter}}, \ and\ \bibinfo
  {author} {\bibfnamefont {H.~J.}\ \bibnamefont {Kimble}},\ }\href {\doibase
  10.1088/1367-2630/15/8/083026} {\bibfield  {journal} {\bibinfo  {journal}
  {New J. Phys.}\ }\textbf {\bibinfo {volume} {15}},\ \bibinfo {pages} {083026}
  (\bibinfo {year} {2013})}\BibitemShut {NoStop}%
\bibitem [{\citenamefont {Chang}\ \emph {et~al.}(2014)\citenamefont {Chang},
  \citenamefont {Sinha}, \citenamefont {Taylor},\ and\ \citenamefont
  {Kimble}}]{chang_trapping_2014}%
  \BibitemOpen
  \bibfield  {author} {\bibinfo {author} {\bibfnamefont {D.~E.}\ \bibnamefont
  {Chang}}, \bibinfo {author} {\bibfnamefont {K.}~\bibnamefont {Sinha}},
  \bibinfo {author} {\bibfnamefont {J.~M.}\ \bibnamefont {Taylor}}, \ and\
  \bibinfo {author} {\bibfnamefont {H.~J.}\ \bibnamefont {Kimble}},\ }\href
  {\doibase 10.1038/ncomms5343} {\bibfield  {journal} {\bibinfo  {journal}
  {Nat. Commun.}\ }\textbf {\bibinfo {volume} {5}},\ \bibinfo {pages} {4343}
  (\bibinfo {year} {2014})}\BibitemShut {NoStop}%
\bibitem [{\citenamefont {{Gonz{\'a}lez-Tudela}}\ \emph
  {et~al.}(2015)\citenamefont {{Gonz{\'a}lez-Tudela}}, \citenamefont {Hung},
  \citenamefont {Chang}, \citenamefont {Cirac},\ and\ \citenamefont
  {Kimble}}]{gonzalez-tudela_subwavelength_2015}%
  \BibitemOpen
  \bibfield  {author} {\bibinfo {author} {\bibfnamefont {A.}~\bibnamefont
  {{Gonz{\'a}lez-Tudela}}}, \bibinfo {author} {\bibfnamefont {C.-L.}\
  \bibnamefont {Hung}}, \bibinfo {author} {\bibfnamefont {D.~E.}\ \bibnamefont
  {Chang}}, \bibinfo {author} {\bibfnamefont {J.~I.}\ \bibnamefont {Cirac}}, \
  and\ \bibinfo {author} {\bibfnamefont {H.~J.}\ \bibnamefont {Kimble}},\
  }\href {\doibase 10.1038/nphoton.2015.54} {\bibfield  {journal} {\bibinfo
  {journal} {Nat. Photonics}\ }\textbf {\bibinfo {volume} {9}},\ \bibinfo
  {pages} {320} (\bibinfo {year} {2015})}\BibitemShut {NoStop}%
\bibitem [{\citenamefont {Lima}\ \emph {et~al.}(2000)\citenamefont {Lima},
  \citenamefont {Chevrollier}, \citenamefont {Di~Lorenzo}, \citenamefont
  {Segundo},\ and\ \citenamefont {Ori{\'a}}}]{lima_long-range_2000}%
  \BibitemOpen
  \bibfield  {author} {\bibinfo {author} {\bibfnamefont {E.~G.}\ \bibnamefont
  {Lima}}, \bibinfo {author} {\bibfnamefont {M.}~\bibnamefont {Chevrollier}},
  \bibinfo {author} {\bibfnamefont {O.}~\bibnamefont {Di~Lorenzo}}, \bibinfo
  {author} {\bibfnamefont {P.~C.}\ \bibnamefont {Segundo}}, \ and\ \bibinfo
  {author} {\bibfnamefont {M.}~\bibnamefont {Ori{\'a}}},\ }\href {\doibase
  10.1103/PhysRevA.62.013410} {\bibfield  {journal} {\bibinfo  {journal} {Phys.
  Rev. A}\ }\textbf {\bibinfo {volume} {62}},\ \bibinfo {pages} {013410}
  (\bibinfo {year} {2000})}\BibitemShut {NoStop}%
\bibitem [{\citenamefont {de~Silans}\ \emph {et~al.}(2006)\citenamefont
  {de~Silans}, \citenamefont {Farias}, \citenamefont {Ori{\'a}},\ and\
  \citenamefont {Chevrollier}}]{silans_laser-induced_2006}%
  \BibitemOpen
  \bibfield  {author} {\bibinfo {author} {\bibfnamefont {T.~P.}\ \bibnamefont
  {de~Silans}}, \bibinfo {author} {\bibfnamefont {B.}~\bibnamefont {Farias}},
  \bibinfo {author} {\bibfnamefont {M.}~\bibnamefont {Ori{\'a}}}, \ and\
  \bibinfo {author} {\bibfnamefont {M.}~\bibnamefont {Chevrollier}},\ }\href
  {\doibase 10.1007/s00340-005-2007-y} {\bibfield  {journal} {\bibinfo
  {journal} {Appl. Phys. B}\ }\textbf {\bibinfo {volume} {82}},\ \bibinfo
  {pages} {367} (\bibinfo {year} {2006})}\BibitemShut {NoStop}%
\bibitem [{\citenamefont {Afanasiev}\ \emph {et~al.}(2007)\citenamefont
  {Afanasiev}, \citenamefont {Melentiev},\ and\ \citenamefont
  {Balykin}}]{afanasiev_laser-induced_2007}%
  \BibitemOpen
  \bibfield  {author} {\bibinfo {author} {\bibfnamefont {A.~E.}\ \bibnamefont
  {Afanasiev}}, \bibinfo {author} {\bibfnamefont {P.~N.}\ \bibnamefont
  {Melentiev}}, \ and\ \bibinfo {author} {\bibfnamefont {V.~I.}\ \bibnamefont
  {Balykin}},\ }\href {\doibase 10.1134/S0021364007150052} {\bibfield
  {journal} {\bibinfo  {journal} {J. Exp. Theor. Phys. Lett.}\ }\textbf
  {\bibinfo {volume} {86}},\ \bibinfo {pages} {172} (\bibinfo {year}
  {2007})}\BibitemShut {NoStop}%
\bibitem [{\citenamefont {Afanas'ev}\ \emph {et~al.}(2008)\citenamefont
  {Afanas'ev}, \citenamefont {Melent'ev},\ and\ \citenamefont
  {Balykin}}]{afanasev_atomic_2008}%
  \BibitemOpen
  \bibfield  {author} {\bibinfo {author} {\bibfnamefont {A.~E.}\ \bibnamefont
  {Afanas'ev}}, \bibinfo {author} {\bibfnamefont {P.~N.}\ \bibnamefont
  {Melent'ev}}, \ and\ \bibinfo {author} {\bibfnamefont {V.~I.}\ \bibnamefont
  {Balykin}},\ }\href {\doibase 10.3103/S1062873808050213} {\bibfield
  {journal} {\bibinfo  {journal} {Bull. Russ. Acad. Sci. Phys.}\ }\textbf
  {\bibinfo {volume} {72}},\ \bibinfo {pages} {664} (\bibinfo {year}
  {2008})}\BibitemShut {NoStop}%
\bibitem [{\citenamefont {Soares}\ \emph {et~al.}(2009)\citenamefont {Soares},
  \citenamefont {De~Silans}, \citenamefont {Ori{\'a}},\ and\ \citenamefont
  {Chevrollier}}]{soares_2d_2009}%
  \BibitemOpen
  \bibfield  {author} {\bibinfo {author} {\bibfnamefont {W.~M.}\ \bibnamefont
  {Soares}}, \bibinfo {author} {\bibfnamefont {T.~P.}\ \bibnamefont
  {De~Silans}}, \bibinfo {author} {\bibfnamefont {M.}~\bibnamefont {Ori{\'a}}},
  \ and\ \bibinfo {author} {\bibfnamefont {M.}~\bibnamefont {Chevrollier}},\
  }\href {\doibase 10.1142/S0217751X09045340} {\bibfield  {journal} {\bibinfo
  {journal} {Int. J. Mod. Phys. A}\ }\textbf {\bibinfo {volume} {24}},\
  \bibinfo {pages} {1764} (\bibinfo {year} {2009})}\BibitemShut {NoStop}%
\bibitem [{\citenamefont {Nayak}\ \emph {et~al.}(2007)\citenamefont {Nayak},
  \citenamefont {Melentiev}, \citenamefont {Morinaga}, \citenamefont {Kien},
  \citenamefont {Balykin},\ and\ \citenamefont {Hakuta}}]{nayak_optical_2007}%
  \BibitemOpen
  \bibfield  {author} {\bibinfo {author} {\bibfnamefont {K.~P.}\ \bibnamefont
  {Nayak}}, \bibinfo {author} {\bibfnamefont {P.~N.}\ \bibnamefont
  {Melentiev}}, \bibinfo {author} {\bibfnamefont {M.}~\bibnamefont {Morinaga}},
  \bibinfo {author} {\bibfnamefont {F.~L.}\ \bibnamefont {Kien}}, \bibinfo
  {author} {\bibfnamefont {V.~I.}\ \bibnamefont {Balykin}}, \ and\ \bibinfo
  {author} {\bibfnamefont {K.}~\bibnamefont {Hakuta}},\ }\href {\doibase
  10.1364/OE.15.005431} {\bibfield  {journal} {\bibinfo  {journal} {Opt.
  Express}\ }\textbf {\bibinfo {volume} {15}},\ \bibinfo {pages} {5431}
  (\bibinfo {year} {2007})}\BibitemShut {NoStop}%
\bibitem [{\citenamefont {Gortel}\ \emph {et~al.}(1980)\citenamefont {Gortel},
  \citenamefont {Kreuzer},\ and\ \citenamefont
  {Teshima}}]{gortel_desorption_1980}%
  \BibitemOpen
  \bibfield  {author} {\bibinfo {author} {\bibfnamefont {Z.~W.}\ \bibnamefont
  {Gortel}}, \bibinfo {author} {\bibfnamefont {H.~J.}\ \bibnamefont {Kreuzer}},
  \ and\ \bibinfo {author} {\bibfnamefont {R.}~\bibnamefont {Teshima}},\ }\href
  {\doibase 10.1103/PhysRevB.22.5655} {\bibfield  {journal} {\bibinfo
  {journal} {Phys. Rev. B}\ }\textbf {\bibinfo {volume} {22}},\ \bibinfo
  {pages} {5655} (\bibinfo {year} {1980})}\BibitemShut {NoStop}%
\bibitem [{\citenamefont {Kreuzer}\ and\ \citenamefont
  {Gortel}(1986)}]{kreuzer_physisorption_1986}%
  \BibitemOpen
  \bibfield  {author} {\bibinfo {author} {\bibfnamefont {H.~J.}\ \bibnamefont
  {Kreuzer}}\ and\ \bibinfo {author} {\bibfnamefont {Z.}~\bibnamefont
  {Gortel}},\ }\href@noop {} {\emph {\bibinfo {title} {Physisorption
  {{Kinetics}}}}}\ (\bibinfo  {publisher} {{Springer}},\ \bibinfo {address}
  {{Berlin}},\ \bibinfo {year} {1986})\BibitemShut {NoStop}%
\bibitem [{\citenamefont {Kien}\ \emph {et~al.}(2007)\citenamefont {Kien},
  \citenamefont {Dutta~Gupta},\ and\ \citenamefont
  {Hakuta}}]{kien_phonon-mediated_2007}%
  \BibitemOpen
  \bibfield  {author} {\bibinfo {author} {\bibfnamefont {F.~L.}\ \bibnamefont
  {Kien}}, \bibinfo {author} {\bibfnamefont {S.}~\bibnamefont {Dutta~Gupta}}, \
  and\ \bibinfo {author} {\bibfnamefont {K.}~\bibnamefont {Hakuta}},\ }\href
  {\doibase 10.1103/PhysRevA.75.062904} {\bibfield  {journal} {\bibinfo
  {journal} {Phys. Rev. A}\ }\textbf {\bibinfo {volume} {75}},\ \bibinfo
  {pages} {062904} (\bibinfo {year} {2007})}\BibitemShut {NoStop}%
\bibitem [{\citenamefont {Nieddu}\ \emph {et~al.}(2016)\citenamefont {Nieddu},
  \citenamefont {Gokhroo},\ and\ \citenamefont
  {Chormaic}}]{nieddu_optical_2016}%
  \BibitemOpen
  \bibfield  {author} {\bibinfo {author} {\bibfnamefont {T.}~\bibnamefont
  {Nieddu}}, \bibinfo {author} {\bibfnamefont {V.}~\bibnamefont {Gokhroo}}, \
  and\ \bibinfo {author} {\bibfnamefont {S.~N.}\ \bibnamefont {Chormaic}},\
  }\href {\doibase 10.1088/2040-8978/18/5/053001} {\bibfield  {journal}
  {\bibinfo  {journal} {J. Opt.}\ }\textbf {\bibinfo {volume} {18}},\ \bibinfo
  {pages} {053001} (\bibinfo {year} {2016})}\BibitemShut {NoStop}%
\bibitem [{\citenamefont {Solano}\ \emph
  {et~al.}(2017{\natexlab{a}})\citenamefont {Solano}, \citenamefont {Grover},
  \citenamefont {Hoffman}, \citenamefont {Ravets}, \citenamefont {Fatemi},
  \citenamefont {Orozco},\ and\ \citenamefont {Rolston}}]{solano_chapter_2017}%
  \BibitemOpen
  \bibfield  {author} {\bibinfo {author} {\bibfnamefont {P.}~\bibnamefont
  {Solano}}, \bibinfo {author} {\bibfnamefont {J.~A.}\ \bibnamefont {Grover}},
  \bibinfo {author} {\bibfnamefont {J.~E.}\ \bibnamefont {Hoffman}}, \bibinfo
  {author} {\bibfnamefont {S.}~\bibnamefont {Ravets}}, \bibinfo {author}
  {\bibfnamefont {F.~K.}\ \bibnamefont {Fatemi}}, \bibinfo {author}
  {\bibfnamefont {L.~A.}\ \bibnamefont {Orozco}}, \ and\ \bibinfo {author}
  {\bibfnamefont {S.~L.}\ \bibnamefont {Rolston}},\ }in\ \href {\doibase
  10.1016/bs.aamop.2017.02.003} {\emph {\bibinfo {booktitle} {Advances {{In
  Atomic}}, {{Molecular}}, and {{Optical Physics}}}}},\ Vol.~\bibinfo {volume}
  {66},\ \bibinfo {editor} {edited by\ \bibinfo {editor} {\bibfnamefont
  {E.}~\bibnamefont {Arimondo}}, \bibinfo {editor} {\bibfnamefont {C.~C.}\
  \bibnamefont {Lin}}, \ and\ \bibinfo {editor} {\bibfnamefont {S.~F.}\
  \bibnamefont {Yelin}}}\ (\bibinfo  {publisher} {{Academic Press}},\ \bibinfo
  {year} {2017})\ pp.\ \bibinfo {pages} {439--505}\BibitemShut {NoStop}%
\bibitem [{\citenamefont {Nayak}\ \emph {et~al.}(2018)\citenamefont {Nayak},
  \citenamefont {Sadgrove}, \citenamefont {Yalla}, \citenamefont {Kien},\ and\
  \citenamefont {Hakuta}}]{nayak_nanofiber_2018}%
  \BibitemOpen
  \bibfield  {author} {\bibinfo {author} {\bibfnamefont {K.~P.}\ \bibnamefont
  {Nayak}}, \bibinfo {author} {\bibfnamefont {M.}~\bibnamefont {Sadgrove}},
  \bibinfo {author} {\bibfnamefont {R.}~\bibnamefont {Yalla}}, \bibinfo
  {author} {\bibfnamefont {F.~L.}\ \bibnamefont {Kien}}, \ and\ \bibinfo
  {author} {\bibfnamefont {K.}~\bibnamefont {Hakuta}},\ }\href {\doibase
  10.1088/2040-8986/aac35e} {\bibfield  {journal} {\bibinfo  {journal} {J.
  Opt.}\ }\textbf {\bibinfo {volume} {20}},\ \bibinfo {pages} {073001}
  (\bibinfo {year} {2018})}\BibitemShut {NoStop}%
\bibitem [{\citenamefont {Stephens}\ \emph {et~al.}(1994)\citenamefont
  {Stephens}, \citenamefont {Rhodes},\ and\ \citenamefont
  {Wieman}}]{stephens_study_1994}%
  \BibitemOpen
  \bibfield  {author} {\bibinfo {author} {\bibfnamefont {M.}~\bibnamefont
  {Stephens}}, \bibinfo {author} {\bibfnamefont {R.}~\bibnamefont {Rhodes}}, \
  and\ \bibinfo {author} {\bibfnamefont {C.}~\bibnamefont {Wieman}},\ }\href
  {\doibase 10.1063/1.358502} {\bibfield  {journal} {\bibinfo  {journal} {J.
  Appl. Phys.}\ }\textbf {\bibinfo {volume} {76}},\ \bibinfo {pages} {3479}
  (\bibinfo {year} {1994})}\BibitemShut {NoStop}%
\bibitem [{\citenamefont {Bouchiat}\ \emph {et~al.}(1999)\citenamefont
  {Bouchiat}, \citenamefont {Gu{\'e}na}, \citenamefont {Jacquier},
  \citenamefont {Lintz},\ and\ \citenamefont
  {Papoyan}}]{bouchiat_electrical_1999}%
  \BibitemOpen
  \bibfield  {author} {\bibinfo {author} {\bibfnamefont {M.~A.}\ \bibnamefont
  {Bouchiat}}, \bibinfo {author} {\bibfnamefont {J.}~\bibnamefont {Gu{\'e}na}},
  \bibinfo {author} {\bibfnamefont {P.}~\bibnamefont {Jacquier}}, \bibinfo
  {author} {\bibfnamefont {M.}~\bibnamefont {Lintz}}, \ and\ \bibinfo {author}
  {\bibfnamefont {A.~V.}\ \bibnamefont {Papoyan}},\ }\href {\doibase
  10.1007/s003400050752} {\bibfield  {journal} {\bibinfo  {journal} {Appl.
  Phys. B}\ }\textbf {\bibinfo {volume} {68}},\ \bibinfo {pages} {1109}
  (\bibinfo {year} {1999})}\BibitemShut {NoStop}%
\bibitem [{\citenamefont {de~Freitas}\ \emph {et~al.}(2002)\citenamefont
  {de~Freitas}, \citenamefont {Oria},\ and\ \citenamefont
  {Chevrollier}}]{freitas_spectroscopy_2002}%
  \BibitemOpen
  \bibfield  {author} {\bibinfo {author} {\bibfnamefont {H.~N.}\ \bibnamefont
  {de~Freitas}}, \bibinfo {author} {\bibfnamefont {M.}~\bibnamefont {Oria}}, \
  and\ \bibinfo {author} {\bibfnamefont {M.}~\bibnamefont {Chevrollier}},\
  }\href {\doibase 10.1007/s00340-002-1029-y} {\bibfield  {journal} {\bibinfo
  {journal} {Appl. Phys. B}\ }\textbf {\bibinfo {volume} {75}},\ \bibinfo
  {pages} {703} (\bibinfo {year} {2002})}\BibitemShut {NoStop}%
\bibitem [{\citenamefont {H{\"u}mmer}\ \emph {et~al.}(2019)\citenamefont
  {H{\"u}mmer}, \citenamefont {Schneeweiss}, \citenamefont {Rauschenbeutel},\
  and\ \citenamefont {{Romero-Isart}}}]{hummer_heating_2019}%
  \BibitemOpen
  \bibfield  {author} {\bibinfo {author} {\bibfnamefont {D.}~\bibnamefont
  {H{\"u}mmer}}, \bibinfo {author} {\bibfnamefont {P.}~\bibnamefont
  {Schneeweiss}}, \bibinfo {author} {\bibfnamefont {A.}~\bibnamefont
  {Rauschenbeutel}}, \ and\ \bibinfo {author} {\bibfnamefont {O.}~\bibnamefont
  {{Romero-Isart}}},\ }\href {\doibase 10.1103/PhysRevX.9.041034} {\bibfield
  {journal} {\bibinfo  {journal} {Phys. Rev. X}\ }\textbf {\bibinfo {volume}
  {9}},\ \bibinfo {pages} {041034} (\bibinfo {year} {2019})}\BibitemShut
  {NoStop}%
\bibitem [{\citenamefont {Pennetta}\ \emph {et~al.}(2016)\citenamefont
  {Pennetta}, \citenamefont {Xie},\ and\ \citenamefont
  {Russell}}]{pennetta_tapered_2016}%
  \BibitemOpen
  \bibfield  {author} {\bibinfo {author} {\bibfnamefont {R.}~\bibnamefont
  {Pennetta}}, \bibinfo {author} {\bibfnamefont {S.}~\bibnamefont {Xie}}, \
  and\ \bibinfo {author} {\bibfnamefont {P.~S.~J.}\ \bibnamefont {Russell}},\
  }\href {\doibase 10.1103/PhysRevLett.117.273901} {\bibfield  {journal}
  {\bibinfo  {journal} {Phys. Rev. Lett.}\ }\textbf {\bibinfo {volume} {117}},\
  \bibinfo {pages} {273901} (\bibinfo {year} {2016})}\BibitemShut {NoStop}%
\bibitem [{\citenamefont {Solano}\ \emph
  {et~al.}(2017{\natexlab{b}})\citenamefont {Solano}, \citenamefont
  {{Barberis-Blostein}}, \citenamefont {Fatemi}, \citenamefont {Orozco},\ and\
  \citenamefont {Rolston}}]{solano_super-radiance_2017}%
  \BibitemOpen
  \bibfield  {author} {\bibinfo {author} {\bibfnamefont {P.}~\bibnamefont
  {Solano}}, \bibinfo {author} {\bibfnamefont {P.}~\bibnamefont
  {{Barberis-Blostein}}}, \bibinfo {author} {\bibfnamefont {F.~K.}\
  \bibnamefont {Fatemi}}, \bibinfo {author} {\bibfnamefont {L.~A.}\
  \bibnamefont {Orozco}}, \ and\ \bibinfo {author} {\bibfnamefont {S.~L.}\
  \bibnamefont {Rolston}},\ }\href {\doibase 10.1038/s41467-017-01994-3}
  {\bibfield  {journal} {\bibinfo  {journal} {Nat. Commun.}\ }\textbf {\bibinfo
  {volume} {8}},\ \bibinfo {pages} {1857} (\bibinfo {year}
  {2017}{\natexlab{b}})}\BibitemShut {NoStop}%
\bibitem [{\citenamefont {Le~Kien}\ and\ \citenamefont
  {Rauschenbeutel}(2017)}]{le_kien_nanofiber-mediated_2017}%
  \BibitemOpen
  \bibfield  {author} {\bibinfo {author} {\bibfnamefont {F.}~\bibnamefont
  {Le~Kien}}\ and\ \bibinfo {author} {\bibfnamefont {A.}~\bibnamefont
  {Rauschenbeutel}},\ }\href {\doibase 10.1103/PhysRevA.95.023838} {\bibfield
  {journal} {\bibinfo  {journal} {Phys. Rev. A}\ }\textbf {\bibinfo {volume}
  {95}},\ \bibinfo {pages} {023838} (\bibinfo {year} {2017})}\BibitemShut
  {NoStop}%
\bibitem [{\citenamefont {Olmos}\ \emph {et~al.}(2020)\citenamefont {Olmos},
  \citenamefont {Buonaiuto}, \citenamefont {Schneeweiss},\ and\ \citenamefont
  {Lesanovsky}}]{olmos_interaction_2020}%
  \BibitemOpen
  \bibfield  {author} {\bibinfo {author} {\bibfnamefont {B.}~\bibnamefont
  {Olmos}}, \bibinfo {author} {\bibfnamefont {G.}~\bibnamefont {Buonaiuto}},
  \bibinfo {author} {\bibfnamefont {P.}~\bibnamefont {Schneeweiss}}, \ and\
  \bibinfo {author} {\bibfnamefont {I.}~\bibnamefont {Lesanovsky}},\ }\href
  {\doibase 10.1103/PhysRevA.102.043711} {\bibfield  {journal} {\bibinfo
  {journal} {Phys. Rev. A}\ }\textbf {\bibinfo {volume} {102}},\ \bibinfo
  {pages} {043711} (\bibinfo {year} {2020})}\BibitemShut {NoStop}%
\bibitem [{\citenamefont {Dowling}\ and\ \citenamefont
  {{Gea-Banacloche}}(1996)}]{dowling_evanescent_1996}%
  \BibitemOpen
  \bibfield  {author} {\bibinfo {author} {\bibfnamefont {J.~P.}\ \bibnamefont
  {Dowling}}\ and\ \bibinfo {author} {\bibfnamefont {J.}~\bibnamefont
  {{Gea-Banacloche}}},\ }\href {\doibase 10.1016/S1049-250X(08)60098-1}
  {\bibfield  {journal} {\bibinfo  {journal} {Adv. At. Mol. Opt. Phys.}\
  }\textbf {\bibinfo {volume} {37}},\ \bibinfo {pages} {1} (\bibinfo {year}
  {1996})}\BibitemShut {NoStop}%
\bibitem [{\citenamefont {Le~Kien}\ \emph
  {et~al.}(2004{\natexlab{a}})\citenamefont {Le~Kien}, \citenamefont
  {Balykin},\ and\ \citenamefont {Hakuta}}]{le_kien_atom_2004}%
  \BibitemOpen
  \bibfield  {author} {\bibinfo {author} {\bibfnamefont {F.}~\bibnamefont
  {Le~Kien}}, \bibinfo {author} {\bibfnamefont {V.~I.}\ \bibnamefont
  {Balykin}}, \ and\ \bibinfo {author} {\bibfnamefont {K.}~\bibnamefont
  {Hakuta}},\ }\href {\doibase 10.1103/PhysRevA.70.063403} {\bibfield
  {journal} {\bibinfo  {journal} {Phys. Rev. A}\ }\textbf {\bibinfo {volume}
  {70}},\ \bibinfo {pages} {063403} (\bibinfo {year}
  {2004}{\natexlab{a}})}\BibitemShut {NoStop}%
\bibitem [{\citenamefont {Buhmann}(2012)}]{buhmann_dispersion_2012}%
  \BibitemOpen
  \bibfield  {author} {\bibinfo {author} {\bibfnamefont {S.~Y.}\ \bibnamefont
  {Buhmann}},\ }\href@noop {} {\emph {\bibinfo {title} {Dispersion {{Forces
  I}}: {{Macroscopic Quantum Electrodynamics}} and {{Ground}}-{{State
  Casimir}}, {{Casimir}}\textendash{{Polder}} and van der {{Waals Forces}}}}}\
  (\bibinfo  {publisher} {{Springer}},\ \bibinfo {address} {{Berlin}},\
  \bibinfo {year} {2012})\BibitemShut {NoStop}%
\bibitem [{\citenamefont {Fuchs}\ \emph {et~al.}(2018)\citenamefont {Fuchs},
  \citenamefont {Bennett}, \citenamefont {Krems},\ and\ \citenamefont
  {Buhmann}}]{fuchs_nonadditivity_2018}%
  \BibitemOpen
  \bibfield  {author} {\bibinfo {author} {\bibfnamefont {S.}~\bibnamefont
  {Fuchs}}, \bibinfo {author} {\bibfnamefont {R.}~\bibnamefont {Bennett}},
  \bibinfo {author} {\bibfnamefont {R.~V.}\ \bibnamefont {Krems}}, \ and\
  \bibinfo {author} {\bibfnamefont {S.~Y.}\ \bibnamefont {Buhmann}},\ }\href
  {\doibase 10.1103/PhysRevLett.121.083603} {\bibfield  {journal} {\bibinfo
  {journal} {Phys. Rev. Lett.}\ }\textbf {\bibinfo {volume} {121}},\ \bibinfo
  {pages} {083603} (\bibinfo {year} {2018})}\BibitemShut {NoStop}%
\bibitem [{\citenamefont {Le~Kien}\ \emph
  {et~al.}(2013{\natexlab{a}})\citenamefont {Le~Kien}, \citenamefont
  {Schneeweiss},\ and\ \citenamefont
  {Rauschenbeutel}}]{le_kien_dynamical_2013}%
  \BibitemOpen
  \bibfield  {author} {\bibinfo {author} {\bibfnamefont {F.}~\bibnamefont
  {Le~Kien}}, \bibinfo {author} {\bibfnamefont {P.}~\bibnamefont
  {Schneeweiss}}, \ and\ \bibinfo {author} {\bibfnamefont {A.}~\bibnamefont
  {Rauschenbeutel}},\ }\href {\doibase 10.1140/epjd/e2013-30729-x} {\bibfield
  {journal} {\bibinfo  {journal} {Eur. Phys. J. D}\ }\textbf {\bibinfo {volume}
  {67}},\ \bibinfo {pages} {92} (\bibinfo {year}
  {2013}{\natexlab{a}})}\BibitemShut {NoStop}%
\bibitem [{\citenamefont {Le~Kien}\ \emph
  {et~al.}(2013{\natexlab{b}})\citenamefont {Le~Kien}, \citenamefont
  {Schneeweiss},\ and\ \citenamefont
  {Rauschenbeutel}}]{le_kien_state-dependent_2013}%
  \BibitemOpen
  \bibfield  {author} {\bibinfo {author} {\bibfnamefont {F.}~\bibnamefont
  {Le~Kien}}, \bibinfo {author} {\bibfnamefont {P.}~\bibnamefont
  {Schneeweiss}}, \ and\ \bibinfo {author} {\bibfnamefont {A.}~\bibnamefont
  {Rauschenbeutel}},\ }\href {\doibase 10.1103/PhysRevA.88.033840} {\bibfield
  {journal} {\bibinfo  {journal} {Phys. Rev. A}\ }\textbf {\bibinfo {volume}
  {88}},\ \bibinfo {pages} {033840} (\bibinfo {year}
  {2013}{\natexlab{b}})}\BibitemShut {NoStop}%
\bibitem [{\citenamefont {Zaremba}\ and\ \citenamefont
  {Kohn}(1977)}]{zaremba_theory_1977}%
  \BibitemOpen
  \bibfield  {author} {\bibinfo {author} {\bibfnamefont {E.}~\bibnamefont
  {Zaremba}}\ and\ \bibinfo {author} {\bibfnamefont {W.}~\bibnamefont {Kohn}},\
  }\href {\doibase 10.1103/PhysRevB.15.1769} {\bibfield  {journal} {\bibinfo
  {journal} {Phys. Rev. B}\ }\textbf {\bibinfo {volume} {15}},\ \bibinfo
  {pages} {1769} (\bibinfo {year} {1977})}\BibitemShut {NoStop}%
\bibitem [{\citenamefont {Zangwill}(1988)}]{zangwill_physics_1988}%
  \BibitemOpen
  \bibfield  {author} {\bibinfo {author} {\bibfnamefont {A.}~\bibnamefont
  {Zangwill}},\ }\href@noop {} {\emph {\bibinfo {title} {Physics at
  {{Surfaces}}}}}\ (\bibinfo  {publisher} {{Cambridge University Press}},\
  \bibinfo {address} {{Cambridge}},\ \bibinfo {year} {1988})\BibitemShut
  {NoStop}%
\bibitem [{\citenamefont {Hoinkes}(1980)}]{hoinkes_physical_1980}%
  \BibitemOpen
  \bibfield  {author} {\bibinfo {author} {\bibfnamefont {H.}~\bibnamefont
  {Hoinkes}},\ }\href {\doibase 10.1103/RevModPhys.52.933} {\bibfield
  {journal} {\bibinfo  {journal} {Rev. Mod. Phys.}\ }\textbf {\bibinfo {volume}
  {52}},\ \bibinfo {pages} {933} (\bibinfo {year} {1980})}\BibitemShut
  {NoStop}%
\bibitem [{Note1()}]{Note1}%
  \BibitemOpen
  \bibinfo {note} {Precise calculations of the dispersion force need to account
  for the full complexity and imperfections of the atom-surface system~\cite
  {klimchitskaya_casimir_2009}. While it is possible to calculate the exact
  form of the dispersion force between an atom and a dielectric cylinder from
  first principles~\cite
  {schmeits_physical_1977,boustimi_van_2002,nabutovskii_interaction_1979}, we
  are here mainly interested in scenarios where the dispersion force is only
  dominant at atom-surface separations smaller than the radius of the
  nanofiber. In this limit, the exact solution can be approximated by the
  nonretarded dispersion force near a half-space~\cite
  {boustimi_van_2002,le_kien_atom_2004}.}\BibitemShut {Stop}%
\bibitem [{\citenamefont {Klimchitskaya}\ \emph {et~al.}(2009)\citenamefont
  {Klimchitskaya}, \citenamefont {Mohideen},\ and\ \citenamefont
  {Mostepanenko}}]{klimchitskaya_casimir_2009}%
  \BibitemOpen
  \bibfield  {author} {\bibinfo {author} {\bibfnamefont {G.~L.}\ \bibnamefont
  {Klimchitskaya}}, \bibinfo {author} {\bibfnamefont {U.}~\bibnamefont
  {Mohideen}}, \ and\ \bibinfo {author} {\bibfnamefont {V.~M.}\ \bibnamefont
  {Mostepanenko}},\ }\href {\doibase 10.1103/RevModPhys.81.1827} {\bibfield
  {journal} {\bibinfo  {journal} {Rev. Mod. Phys.}\ }\textbf {\bibinfo {volume}
  {81}},\ \bibinfo {pages} {1827} (\bibinfo {year} {2009})}\BibitemShut
  {NoStop}%
\bibitem [{\citenamefont {Schmeits}\ and\ \citenamefont
  {Lucas}(1977)}]{schmeits_physical_1977}%
  \BibitemOpen
  \bibfield  {author} {\bibinfo {author} {\bibfnamefont {M.}~\bibnamefont
  {Schmeits}}\ and\ \bibinfo {author} {\bibfnamefont {A.~A.}\ \bibnamefont
  {Lucas}},\ }\href {\doibase 10.1016/0039-6028(77)90265-5} {\bibfield
  {journal} {\bibinfo  {journal} {Surf. Sci.}\ }\textbf {\bibinfo {volume}
  {64}},\ \bibinfo {pages} {176} (\bibinfo {year} {1977})}\BibitemShut
  {NoStop}%
\bibitem [{\citenamefont {Boustimi}\ \emph {et~al.}(2002)\citenamefont
  {Boustimi}, \citenamefont {Baudon}, \citenamefont {Candori},\ and\
  \citenamefont {Robert}}]{boustimi_van_2002}%
  \BibitemOpen
  \bibfield  {author} {\bibinfo {author} {\bibfnamefont {M.}~\bibnamefont
  {Boustimi}}, \bibinfo {author} {\bibfnamefont {J.}~\bibnamefont {Baudon}},
  \bibinfo {author} {\bibfnamefont {P.}~\bibnamefont {Candori}}, \ and\
  \bibinfo {author} {\bibfnamefont {J.}~\bibnamefont {Robert}},\ }\href
  {\doibase 10.1103/PhysRevB.65.155402} {\bibfield  {journal} {\bibinfo
  {journal} {Phys. Rev. B}\ }\textbf {\bibinfo {volume} {65}},\ \bibinfo
  {pages} {155402} (\bibinfo {year} {2002})}\BibitemShut {NoStop}%
\bibitem [{\citenamefont {Nabutovskii}\ \emph {et~al.}(1979)\citenamefont
  {Nabutovskii}, \citenamefont {Belosludov},\ and\ \citenamefont
  {Korotkikh}}]{nabutovskii_interaction_1979}%
  \BibitemOpen
  \bibfield  {author} {\bibinfo {author} {\bibfnamefont {V.~M.}\ \bibnamefont
  {Nabutovskii}}, \bibinfo {author} {\bibfnamefont {V.~R.}\ \bibnamefont
  {Belosludov}}, \ and\ \bibinfo {author} {\bibfnamefont {A.~M.}\ \bibnamefont
  {Korotkikh}},\ }\href@noop {} {\bibfield  {journal} {\bibinfo  {journal} {J.
  Exp. Theor. Phys.}\ }\textbf {\bibinfo {volume} {50}},\ \bibinfo {pages}
  {352} (\bibinfo {year} {1979})}\BibitemShut {NoStop}%
\bibitem [{\citenamefont {McLachlan}(1964)}]{mclachlan_van_1964}%
  \BibitemOpen
  \bibfield  {author} {\bibinfo {author} {\bibfnamefont {A.~D.}\ \bibnamefont
  {McLachlan}},\ }\href {\doibase 10.1080/00268976300101141} {\bibfield
  {journal} {\bibinfo  {journal} {Mol. Phys.}\ }\textbf {\bibinfo {volume}
  {7}},\ \bibinfo {pages} {381} (\bibinfo {year} {1964})}\BibitemShut {NoStop}%
\bibitem [{\citenamefont {Wylie}\ and\ \citenamefont
  {Sipe}(1984)}]{wylie_quantum_1984}%
  \BibitemOpen
  \bibfield  {author} {\bibinfo {author} {\bibfnamefont {J.~M.}\ \bibnamefont
  {Wylie}}\ and\ \bibinfo {author} {\bibfnamefont {J.~E.}\ \bibnamefont
  {Sipe}},\ }\href {\doibase 10.1103/PhysRevA.30.1185} {\bibfield  {journal}
  {\bibinfo  {journal} {Phys. Rev. A}\ }\textbf {\bibinfo {volume} {30}},\
  \bibinfo {pages} {1185} (\bibinfo {year} {1984})}\BibitemShut {NoStop}%
\bibitem [{\citenamefont {Stern}\ \emph {et~al.}(2011)\citenamefont {Stern},
  \citenamefont {Alton},\ and\ \citenamefont
  {Kimble}}]{stern_simulations_2011}%
  \BibitemOpen
  \bibfield  {author} {\bibinfo {author} {\bibfnamefont {N.~P.}\ \bibnamefont
  {Stern}}, \bibinfo {author} {\bibfnamefont {D.~J.}\ \bibnamefont {Alton}}, \
  and\ \bibinfo {author} {\bibfnamefont {H.~J.}\ \bibnamefont {Kimble}},\
  }\href {\doibase 10.1088/1367-2630/13/8/085004} {\bibfield  {journal}
  {\bibinfo  {journal} {New J. Phys.}\ }\textbf {\bibinfo {volume} {13}},\
  \bibinfo {pages} {085004} (\bibinfo {year} {2011})}\BibitemShut {NoStop}%
\bibitem [{Note2()}]{Note2}%
  \BibitemOpen
  \bibinfo {note} {Our findings do not change appreciably when using an
  exponential barrier $D\protect \qopname \relax o{exp}\left [ { - (r-R) p }
  \right ]$ instead of the polynomial in \protect \cref {eqn: adsorption
  potential}, for instance with $C/h = \SI {1.56}{\tera \hertz \protect \,\nano
  \meter ^3}$, $V_\protect \text {min}/h = \SI {-128}{\tera \hertz }$,
  repulsive amplitude $D/h = \SI {1.6e6}{\tera \hertz }$, and decay length $p =
  \SI {53}{\nano \meter ^{-1}}$ as suggested in Ref.~\cite
  {kien_phonon-mediated_2007}.}\BibitemShut {Stop}%
\bibitem [{Note3()}]{Note3}%
  \BibitemOpen
  \bibinfo {note} {See Supplemental Material appended at the end of this article for an extended discussion of the photonic and phononic nanofiber modes, the calculation of the motional linewidths, and the heterodyne fluorescence spectroscopy scheme, which includes Refs.~\cite{achenbach_wave_1973,gurtin_linear_1984,cohen-tannoudji_photons_2004,armenakas_free_1969,glauber_quantum_1991,messiah_quantum_2014,cirac_laser_1992,breuer_theory_2002,reitz_coherence_2013,albrecht_fictitious_2016,cirac_spectrum_1993,sague_cold-atom_2007,patterson_spectral_2018,lindberg_resonance_1986,cohen-tannoudji_atom-photon_1998}.}\BibitemShut
  {Stop}%
\bibitem [{\citenamefont {Achenbach}(1973)}]{achenbach_wave_1973}%
  \BibitemOpen
  \bibfield  {author} {\bibinfo {author} {\bibfnamefont {J.~D.}\ \bibnamefont
  {Achenbach}},\ }\href@noop {} {\emph {\bibinfo {title} {Wave {{Propagation}}
  in {{Elastic Solids}}}}}\ (\bibinfo  {publisher} {{North-Holland
  Publishing}},\ \bibinfo {address} {{Amsterdam}},\ \bibinfo {year}
  {1973})\BibitemShut {NoStop}%
\bibitem [{\citenamefont {Gurtin}(1984)}]{gurtin_linear_1984}%
  \BibitemOpen
  \bibfield  {author} {\bibinfo {author} {\bibfnamefont {M.~E.}\ \bibnamefont
  {Gurtin}},\ }in\ \href@noop {} {\emph {\bibinfo {booktitle} {Linear
  {{Theories}} of {{Elasticity}} and {{Thermoelasticity}}, {{Linear}} and
  {{Nonlinear Theories}} of {{Rods}}, {{Plates}}, and {{Shells}}}}},\ \bibinfo
  {series} {Mechanics of {{Solids}}}, Vol.~\bibinfo {volume} {2},\ \bibinfo
  {editor} {edited by\ \bibinfo {editor} {\bibfnamefont {C.}~\bibnamefont
  {Truesdell}}}\ (\bibinfo  {publisher} {{Springer}},\ \bibinfo {address}
  {{Berlin}},\ \bibinfo {year} {1984})\BibitemShut {NoStop}%
\bibitem [{\citenamefont {{Cohen-Tannoudji}}\ \emph {et~al.}(2004)\citenamefont
  {{Cohen-Tannoudji}}, \citenamefont {{Dupont-Roc}},\ and\ \citenamefont
  {Grynberg}}]{cohen-tannoudji_photons_2004}%
  \BibitemOpen
  \bibfield  {author} {\bibinfo {author} {\bibfnamefont {C.}~\bibnamefont
  {{Cohen-Tannoudji}}}, \bibinfo {author} {\bibfnamefont {J.}~\bibnamefont
  {{Dupont-Roc}}}, \ and\ \bibinfo {author} {\bibfnamefont {G.}~\bibnamefont
  {Grynberg}},\ }\href@noop {} {\emph {\bibinfo {title} {Photons and {{Atoms}}:
  {{Introduction}} to {{Quantum Electrodynamics}}}}}\ (\bibinfo  {publisher}
  {{Wiley-VCH}},\ \bibinfo {address} {{Weinheim}},\ \bibinfo {year}
  {2004})\BibitemShut {NoStop}%
\bibitem [{\citenamefont {Armen{\`a}kas}\ \emph {et~al.}(1969)\citenamefont
  {Armen{\`a}kas}, \citenamefont {Gazis},\ and\ \citenamefont
  {Herrmann}}]{armenakas_free_1969}%
  \BibitemOpen
  \bibfield  {author} {\bibinfo {author} {\bibfnamefont {A.~E.}\ \bibnamefont
  {Armen{\`a}kas}}, \bibinfo {author} {\bibfnamefont {D.~C.}\ \bibnamefont
  {Gazis}}, \ and\ \bibinfo {author} {\bibfnamefont {G.}~\bibnamefont
  {Herrmann}},\ }\href@noop {} {\emph {\bibinfo {title} {Free {{Vibrations}} of
  {{Circular Cylindrical Shells}}}}}\ (\bibinfo  {publisher} {{Pergamon
  Press}},\ \bibinfo {address} {{Oxford}},\ \bibinfo {year} {1969})\BibitemShut
  {NoStop}%
\bibitem [{\citenamefont {Glauber}\ and\ \citenamefont
  {Lewenstein}(1991)}]{glauber_quantum_1991}%
  \BibitemOpen
  \bibfield  {author} {\bibinfo {author} {\bibfnamefont {R.~J.}\ \bibnamefont
  {Glauber}}\ and\ \bibinfo {author} {\bibfnamefont {M.}~\bibnamefont
  {Lewenstein}},\ }\href {\doibase 10.1103/PhysRevA.43.467} {\bibfield
  {journal} {\bibinfo  {journal} {Phys. Rev. A}\ }\textbf {\bibinfo {volume}
  {43}},\ \bibinfo {pages} {467} (\bibinfo {year} {1991})}\BibitemShut
  {NoStop}%
\bibitem [{\citenamefont {Messiah}(2014)}]{messiah_quantum_2014}%
  \BibitemOpen
  \bibfield  {author} {\bibinfo {author} {\bibfnamefont {A.}~\bibnamefont
  {Messiah}},\ }\href@noop {} {\emph {\bibinfo {title} {Quantum
  {{Mechanics}}}}}\ (\bibinfo  {publisher} {{Dover Publications}},\ \bibinfo
  {address} {{New York}},\ \bibinfo {year} {2014})\BibitemShut {NoStop}%
\bibitem [{\citenamefont {Cirac}\ \emph {et~al.}(1992)\citenamefont {Cirac},
  \citenamefont {Blatt}, \citenamefont {Zoller},\ and\ \citenamefont
  {Phillips}}]{cirac_laser_1992}%
  \BibitemOpen
  \bibfield  {author} {\bibinfo {author} {\bibfnamefont {J.~I.}\ \bibnamefont
  {Cirac}}, \bibinfo {author} {\bibfnamefont {R.}~\bibnamefont {Blatt}},
  \bibinfo {author} {\bibfnamefont {P.}~\bibnamefont {Zoller}}, \ and\ \bibinfo
  {author} {\bibfnamefont {W.~D.}\ \bibnamefont {Phillips}},\ }\href {\doibase
  10.1103/PhysRevA.46.2668} {\bibfield  {journal} {\bibinfo  {journal} {Phys.
  Rev. A}\ }\textbf {\bibinfo {volume} {46}},\ \bibinfo {pages} {2668}
  (\bibinfo {year} {1992})}\BibitemShut {NoStop}%
\bibitem [{\citenamefont {Breuer}\ and\ \citenamefont
  {Petruccione}(2002)}]{breuer_theory_2002}%
  \BibitemOpen
  \bibfield  {author} {\bibinfo {author} {\bibfnamefont {H.-P.}\ \bibnamefont
  {Breuer}}\ and\ \bibinfo {author} {\bibfnamefont {F.}~\bibnamefont
  {Petruccione}},\ }\href@noop {} {\emph {\bibinfo {title} {The {{Theory}} of
  {{Open Quantum Systems}}}}}\ (\bibinfo  {publisher} {{Oxford University
  Press}},\ \bibinfo {address} {{Oxford}},\ \bibinfo {year} {2002})\BibitemShut
  {NoStop}%
\bibitem [{\citenamefont {Reitz}\ \emph {et~al.}(2013)\citenamefont {Reitz},
  \citenamefont {Sayrin}, \citenamefont {Mitsch}, \citenamefont {Schneeweiss},\
  and\ \citenamefont {Rauschenbeutel}}]{reitz_coherence_2013}%
  \BibitemOpen
  \bibfield  {author} {\bibinfo {author} {\bibfnamefont {D.}~\bibnamefont
  {Reitz}}, \bibinfo {author} {\bibfnamefont {C.}~\bibnamefont {Sayrin}},
  \bibinfo {author} {\bibfnamefont {R.}~\bibnamefont {Mitsch}}, \bibinfo
  {author} {\bibfnamefont {P.}~\bibnamefont {Schneeweiss}}, \ and\ \bibinfo
  {author} {\bibfnamefont {A.}~\bibnamefont {Rauschenbeutel}},\ }\href
  {\doibase 10.1103/PhysRevLett.110.243603} {\bibfield  {journal} {\bibinfo
  {journal} {Phys. Rev. Lett.}\ }\textbf {\bibinfo {volume} {110}},\ \bibinfo
  {pages} {243603} (\bibinfo {year} {2013})}\BibitemShut {NoStop}%
\bibitem [{\citenamefont {Albrecht}\ \emph {et~al.}(2016)\citenamefont
  {Albrecht}, \citenamefont {Meng}, \citenamefont {Clausen}, \citenamefont
  {Dareau}, \citenamefont {Schneeweiss},\ and\ \citenamefont
  {Rauschenbeutel}}]{albrecht_fictitious_2016}%
  \BibitemOpen
  \bibfield  {author} {\bibinfo {author} {\bibfnamefont {B.}~\bibnamefont
  {Albrecht}}, \bibinfo {author} {\bibfnamefont {Y.}~\bibnamefont {Meng}},
  \bibinfo {author} {\bibfnamefont {C.}~\bibnamefont {Clausen}}, \bibinfo
  {author} {\bibfnamefont {A.}~\bibnamefont {Dareau}}, \bibinfo {author}
  {\bibfnamefont {P.}~\bibnamefont {Schneeweiss}}, \ and\ \bibinfo {author}
  {\bibfnamefont {A.}~\bibnamefont {Rauschenbeutel}},\ }\href {\doibase
  10.1103/PhysRevA.94.061401} {\bibfield  {journal} {\bibinfo  {journal} {Phys.
  Rev. A}\ }\textbf {\bibinfo {volume} {94}},\ \bibinfo {pages} {061401(R)}
  (\bibinfo {year} {2016})}\BibitemShut {NoStop}%
\bibitem [{\citenamefont {Cirac}\ \emph {et~al.}(1993)\citenamefont {Cirac},
  \citenamefont {Blatt}, \citenamefont {Parkins},\ and\ \citenamefont
  {Zoller}}]{cirac_spectrum_1993}%
  \BibitemOpen
  \bibfield  {author} {\bibinfo {author} {\bibfnamefont {J.~I.}\ \bibnamefont
  {Cirac}}, \bibinfo {author} {\bibfnamefont {R.}~\bibnamefont {Blatt}},
  \bibinfo {author} {\bibfnamefont {A.~S.}\ \bibnamefont {Parkins}}, \ and\
  \bibinfo {author} {\bibfnamefont {P.}~\bibnamefont {Zoller}},\ }\href
  {\doibase 10.1103/PhysRevA.48.2169} {\bibfield  {journal} {\bibinfo
  {journal} {Phys. Rev. A}\ }\textbf {\bibinfo {volume} {48}},\ \bibinfo
  {pages} {2169} (\bibinfo {year} {1993})}\BibitemShut {NoStop}%
\bibitem [{\citenamefont {Sagu{\'e}}\ \emph {et~al.}(2007)\citenamefont
  {Sagu{\'e}}, \citenamefont {Vetsch}, \citenamefont {Alt}, \citenamefont
  {Meschede},\ and\ \citenamefont {Rauschenbeutel}}]{sague_cold-atom_2007}%
  \BibitemOpen
  \bibfield  {author} {\bibinfo {author} {\bibfnamefont {G.}~\bibnamefont
  {Sagu{\'e}}}, \bibinfo {author} {\bibfnamefont {E.}~\bibnamefont {Vetsch}},
  \bibinfo {author} {\bibfnamefont {W.}~\bibnamefont {Alt}}, \bibinfo {author}
  {\bibfnamefont {D.}~\bibnamefont {Meschede}}, \ and\ \bibinfo {author}
  {\bibfnamefont {A.}~\bibnamefont {Rauschenbeutel}},\ }\href {\doibase
  10.1103/PhysRevLett.99.163602} {\bibfield  {journal} {\bibinfo  {journal}
  {Phys. Rev. Lett.}\ }\textbf {\bibinfo {volume} {99}},\ \bibinfo {pages}
  {163602} (\bibinfo {year} {2007})}\BibitemShut {NoStop}%
\bibitem [{\citenamefont {Patterson}\ \emph {et~al.}(2018)\citenamefont
  {Patterson}, \citenamefont {Solano}, \citenamefont {Julienne}, \citenamefont
  {Orozco},\ and\ \citenamefont {Rolston}}]{patterson_spectral_2018}%
  \BibitemOpen
  \bibfield  {author} {\bibinfo {author} {\bibfnamefont {B.~D.}\ \bibnamefont
  {Patterson}}, \bibinfo {author} {\bibfnamefont {P.}~\bibnamefont {Solano}},
  \bibinfo {author} {\bibfnamefont {P.~S.}\ \bibnamefont {Julienne}}, \bibinfo
  {author} {\bibfnamefont {L.~A.}\ \bibnamefont {Orozco}}, \ and\ \bibinfo
  {author} {\bibfnamefont {S.~L.}\ \bibnamefont {Rolston}},\ }\href {\doibase
  10.1103/PhysRevA.97.032509} {\bibfield  {journal} {\bibinfo  {journal} {Phys.
  Rev. A}\ }\textbf {\bibinfo {volume} {97}},\ \bibinfo {pages} {032509}
  (\bibinfo {year} {2018})}\BibitemShut {NoStop}%
\bibitem [{\citenamefont {Lindberg}(1986)}]{lindberg_resonance_1986}%
  \BibitemOpen
  \bibfield  {author} {\bibinfo {author} {\bibfnamefont {M.}~\bibnamefont
  {Lindberg}},\ }\href {\doibase 10.1103/PhysRevA.34.3178} {\bibfield
  {journal} {\bibinfo  {journal} {Phys. Rev. A}\ }\textbf {\bibinfo {volume}
  {34}},\ \bibinfo {pages} {3178} (\bibinfo {year} {1986})}\BibitemShut
  {NoStop}%
\bibitem [{\citenamefont {{Cohen-Tannoudji}}\ \emph {et~al.}(1998)\citenamefont
  {{Cohen-Tannoudji}}, \citenamefont {{Dupont-Roc}},\ and\ \citenamefont
  {Grynberg}}]{cohen-tannoudji_atom-photon_1998}%
  \BibitemOpen
  \bibfield  {author} {\bibinfo {author} {\bibfnamefont {C.}~\bibnamefont
  {{Cohen-Tannoudji}}}, \bibinfo {author} {\bibfnamefont {J.}~\bibnamefont
  {{Dupont-Roc}}}, \ and\ \bibinfo {author} {\bibfnamefont {G.}~\bibnamefont
  {Grynberg}},\ }\href@noop {} {\emph {\bibinfo {title} {Atom-{{Photon
  Interactions}}: {{Basic Processes}} and {{Applications}}}}}\ (\bibinfo
  {publisher} {{Wiley-VCH}},\ \bibinfo {address} {{Weinheim}},\ \bibinfo {year}
  {1998})\BibitemShut {NoStop}%
\bibitem [{\citenamefont {Bass}\ \emph {et~al.}(2001)\citenamefont {Bass},
  \citenamefont {Van~Stryland}, \citenamefont {Williams},\ and\ \citenamefont
  {Wolfe}}]{bass_handbook_2001}%
  \BibitemOpen
  \bibinfo {editor} {\bibfnamefont {M.}~\bibnamefont {Bass}}, \bibinfo {editor}
  {\bibfnamefont {E.~W.}\ \bibnamefont {Van~Stryland}}, \bibinfo {editor}
  {\bibfnamefont {D.~R.}\ \bibnamefont {Williams}}, \ and\ \bibinfo {editor}
  {\bibfnamefont {W.~L.}\ \bibnamefont {Wolfe}},\ eds.,\ \href@noop {} {\emph
  {\bibinfo {title} {Handbook of {{Optics}}: {{Devices}}, Measurements, and
  Properties}}},\ \bibinfo {edition} {2nd}\ ed.,\ Vol.~\bibinfo {volume} {2}\
  (\bibinfo  {publisher} {{McGraw-Hill}},\ \bibinfo {address} {{New York}},\
  \bibinfo {year} {2001})\BibitemShut {NoStop}%
\bibitem [{Note4()}]{Note4}%
  \BibitemOpen
  \bibinfo {note} {This is the largest radius compatible with the single-mode
  regime for the light fields of the two-color trap}\BibitemShut {NoStop}%
\bibitem [{Note5()}]{Note5}%
  \BibitemOpen
  \bibinfo {note} {We use the commercial COMSOL Multiphysics\protect
  \textsuperscript {\textregistered } software package~\cite
  {comsol_inc_comsol_2016}. Unbound states are obtained by approximating free
  space with an interval sufficiently large so as not to influence any of the
  results presented in this Letter.}\BibitemShut {Stop}%
\bibitem [{\citenamefont {{COMSOL Inc}}(2016)}]{comsol_inc_comsol_2016}%
  \BibitemOpen
  \bibfield  {author} {\bibinfo {author} {\bibnamefont {{COMSOL Inc}}},\
  }\href@noop {} {\emph {\bibinfo {title} {{{COMSOL Multiphysics Reference
  Manual}}, Version 5.2a}}}\ (\bibinfo {address} {{Stockholm}},\ \bibinfo
  {year} {2016})\BibitemShut {NoStop}%
\bibitem [{\citenamefont {Meija}\ \emph {et~al.}(2016)\citenamefont {Meija},
  \citenamefont {Coplen}, \citenamefont {Berglund}, \citenamefont {Brand},
  \citenamefont {Bi{\`e}vre}, \citenamefont {Gr{\"o}ning}, \citenamefont
  {Holden}, \citenamefont {Irrgeher}, \citenamefont {Loss}, \citenamefont
  {Walczyk},\ and\ \citenamefont {Prohaska}}]{meija_atomic_2016}%
  \BibitemOpen
  \bibfield  {author} {\bibinfo {author} {\bibfnamefont {J.}~\bibnamefont
  {Meija}}, \bibinfo {author} {\bibfnamefont {T.~B.}\ \bibnamefont {Coplen}},
  \bibinfo {author} {\bibfnamefont {M.}~\bibnamefont {Berglund}}, \bibinfo
  {author} {\bibfnamefont {W.~A.}\ \bibnamefont {Brand}}, \bibinfo {author}
  {\bibfnamefont {P.~D.}\ \bibnamefont {Bi{\`e}vre}}, \bibinfo {author}
  {\bibfnamefont {M.}~\bibnamefont {Gr{\"o}ning}}, \bibinfo {author}
  {\bibfnamefont {N.~E.}\ \bibnamefont {Holden}}, \bibinfo {author}
  {\bibfnamefont {J.}~\bibnamefont {Irrgeher}}, \bibinfo {author}
  {\bibfnamefont {R.~D.}\ \bibnamefont {Loss}}, \bibinfo {author}
  {\bibfnamefont {T.}~\bibnamefont {Walczyk}}, \ and\ \bibinfo {author}
  {\bibfnamefont {T.}~\bibnamefont {Prohaska}},\ }\href {\doibase
  10.1515/pac-2015-0305} {\bibfield  {journal} {\bibinfo  {journal} {Pure Appl.
  Chem.}\ }\textbf {\bibinfo {volume} {88}},\ \bibinfo {pages} {265} (\bibinfo
  {year} {2016})}\BibitemShut {NoStop}%
\bibitem [{\citenamefont {Meeker}\ and\ \citenamefont
  {Meitzler}(1964)}]{meeker_guided_1964}%
  \BibitemOpen
  \bibfield  {author} {\bibinfo {author} {\bibfnamefont {T.~R.}\ \bibnamefont
  {Meeker}}\ and\ \bibinfo {author} {\bibfnamefont {A.~H.}\ \bibnamefont
  {Meitzler}},\ }in\ \href@noop {} {\emph {\bibinfo {booktitle} {Methods and
  {{Devices}}, {{Part A}}}}},\ \bibinfo {series} {Physical {{Acoustics}}:
  {{Principles}} and {{Methods}}}, Vol.\ \bibinfo {volume} {I A},\ \bibinfo
  {editor} {edited by\ \bibinfo {editor} {\bibfnamefont {W.~P.}\ \bibnamefont
  {Mason}}}\ (\bibinfo  {publisher} {{Academic Press}},\ \bibinfo {address}
  {{New York}},\ \bibinfo {year} {1964})\ pp.\ \bibinfo {pages}
  {111--167}\BibitemShut {NoStop}%
\bibitem [{\citenamefont {Jessen}\ \emph {et~al.}(1992)\citenamefont {Jessen},
  \citenamefont {Gerz}, \citenamefont {Lett}, \citenamefont {Phillips},
  \citenamefont {Rolston}, \citenamefont {Spreeuw},\ and\ \citenamefont
  {Westbrook}}]{jessen_observation_1992}%
  \BibitemOpen
  \bibfield  {author} {\bibinfo {author} {\bibfnamefont {P.~S.}\ \bibnamefont
  {Jessen}}, \bibinfo {author} {\bibfnamefont {C.}~\bibnamefont {Gerz}},
  \bibinfo {author} {\bibfnamefont {P.~D.}\ \bibnamefont {Lett}}, \bibinfo
  {author} {\bibfnamefont {W.~D.}\ \bibnamefont {Phillips}}, \bibinfo {author}
  {\bibfnamefont {S.~L.}\ \bibnamefont {Rolston}}, \bibinfo {author}
  {\bibfnamefont {R.~J.~C.}\ \bibnamefont {Spreeuw}}, \ and\ \bibinfo {author}
  {\bibfnamefont {C.~I.}\ \bibnamefont {Westbrook}},\ }\href {\doibase
  10.1103/PhysRevLett.69.49} {\bibfield  {journal} {\bibinfo  {journal} {Phys.
  Rev. Lett.}\ }\textbf {\bibinfo {volume} {69}},\ \bibinfo {pages} {49}
  (\bibinfo {year} {1992})}\BibitemShut {NoStop}%
\bibitem [{\citenamefont {Marcuse}(1982)}]{marcuse_light_1982}%
  \BibitemOpen
  \bibfield  {author} {\bibinfo {author} {\bibfnamefont {D.}~\bibnamefont
  {Marcuse}},\ }\href@noop {} {\emph {\bibinfo {title} {Light {{Transmission
  Optics}}}}}\ (\bibinfo  {publisher} {{Van Nostrand Reinhold}},\ \bibinfo
  {address} {{New York}},\ \bibinfo {year} {1982})\BibitemShut {NoStop}%
\bibitem [{\citenamefont {Snyder}\ and\ \citenamefont
  {Love}(2012)}]{snyder_optical_2012}%
  \BibitemOpen
  \bibfield  {author} {\bibinfo {author} {\bibfnamefont {A.~W.}\ \bibnamefont
  {Snyder}}\ and\ \bibinfo {author} {\bibfnamefont {J.}~\bibnamefont {Love}},\
  }\href@noop {} {\emph {\bibinfo {title} {Optical {{Waveguide Theory}}}}}\
  (\bibinfo  {publisher} {{Springer}},\ \bibinfo {address} {{New York}},\
  \bibinfo {year} {2012})\BibitemShut {NoStop}%
\bibitem [{\citenamefont {Le~Kien}\ \emph
  {et~al.}(2004{\natexlab{b}})\citenamefont {Le~Kien}, \citenamefont {Liang},
  \citenamefont {Hakuta},\ and\ \citenamefont {Balykin}}]{le_kien_field_2004}%
  \BibitemOpen
  \bibfield  {author} {\bibinfo {author} {\bibfnamefont {F.}~\bibnamefont
  {Le~Kien}}, \bibinfo {author} {\bibfnamefont {J.~Q.}\ \bibnamefont {Liang}},
  \bibinfo {author} {\bibfnamefont {K.}~\bibnamefont {Hakuta}}, \ and\ \bibinfo
  {author} {\bibfnamefont {V.~I.}\ \bibnamefont {Balykin}},\ }\href {\doibase
  10.1016/j.optcom.2004.08.044} {\bibfield  {journal} {\bibinfo  {journal}
  {Opt. Commun.}\ }\textbf {\bibinfo {volume} {242}},\ \bibinfo {pages} {445}
  (\bibinfo {year} {2004}{\natexlab{b}})}\BibitemShut {NoStop}%
\bibitem [{\citenamefont {Wuttke}\ and\ \citenamefont
  {Rauschenbeutel}(2013)}]{wuttke_thermalization_2013}%
  \BibitemOpen
  \bibfield  {author} {\bibinfo {author} {\bibfnamefont {C.}~\bibnamefont
  {Wuttke}}\ and\ \bibinfo {author} {\bibfnamefont {A.}~\bibnamefont
  {Rauschenbeutel}},\ }\href {\doibase 10.1103/PhysRevLett.111.024301}
  {\bibfield  {journal} {\bibinfo  {journal} {Phys. Rev. Lett.}\ }\textbf
  {\bibinfo {volume} {111}},\ \bibinfo {pages} {024301} (\bibinfo {year}
  {2013})}\BibitemShut {NoStop}%
\bibitem [{\citenamefont {Meng}\ \emph {et~al.}(2018)\citenamefont {Meng},
  \citenamefont {Dareau}, \citenamefont {Schneeweiss},\ and\ \citenamefont
  {Rauschenbeutel}}]{meng_near-ground-state_2018}%
  \BibitemOpen
  \bibfield  {author} {\bibinfo {author} {\bibfnamefont {Y.}~\bibnamefont
  {Meng}}, \bibinfo {author} {\bibfnamefont {A.}~\bibnamefont {Dareau}},
  \bibinfo {author} {\bibfnamefont {P.}~\bibnamefont {Schneeweiss}}, \ and\
  \bibinfo {author} {\bibfnamefont {A.}~\bibnamefont {Rauschenbeutel}},\ }\href
  {\doibase 10.1103/PhysRevX.8.031054} {\bibfield  {journal} {\bibinfo
  {journal} {Phys. Rev. X}\ }\textbf {\bibinfo {volume} {8}},\ \bibinfo {pages}
  {031054} (\bibinfo {year} {2018})}\BibitemShut {NoStop}%
\end{thebibliography}

\begin{thebibliography}{23}%
\makeatletter
\providecommand \@ifxundefined [1]{%
 \@ifx{#1\undefined}
}%
\providecommand \@ifnum [1]{%
 \ifnum #1\expandafter \@firstoftwo
 \else \expandafter \@secondoftwo
 \fi
}%
\providecommand \@ifx [1]{%
 \ifx #1\expandafter \@firstoftwo
 \else \expandafter \@secondoftwo
 \fi
}%
\providecommand \natexlab [1]{#1}%
\providecommand \enquote  [1]{``#1''}%
\providecommand \bibnamefont  [1]{#1}%
\providecommand \bibfnamefont [1]{#1}%
\providecommand \citenamefont [1]{#1}%
\providecommand \href@noop [0]{\@secondoftwo}%
\providecommand \href [0]{\begingroup \@sanitize@url \@href}%
\providecommand \@href[1]{\@@startlink{#1}\@@href}%
\providecommand \@@href[1]{\endgroup#1\@@endlink}%
\providecommand \@sanitize@url [0]{\catcode `\\12\catcode `\$12\catcode
  `\&12\catcode `\#12\catcode `\^12\catcode `\_12\catcode `\%12\relax}%
\providecommand \@@startlink[1]{}%
\providecommand \@@endlink[0]{}%
\providecommand \url  [0]{\begingroup\@sanitize@url \@url }%
\providecommand \@url [1]{\endgroup\@href {#1}{\urlprefix }}%
\providecommand \urlprefix  [0]{URL }%
\providecommand \Eprint [0]{\href }%
\providecommand \doibase [0]{https://doi.org/}%
\providecommand \selectlanguage [0]{\@gobble}%
\providecommand \bibinfo  [0]{\@secondoftwo}%
\providecommand \bibfield  [0]{\@secondoftwo}%
\providecommand \translation [1]{[#1]}%
\providecommand \BibitemOpen [0]{}%
\providecommand \bibitemStop [0]{}%
\providecommand \bibitemNoStop [0]{.\EOS\space}%
\providecommand \EOS [0]{\spacefactor3000\relax}%
\providecommand \BibitemShut  [1]{\csname bibitem#1\endcsname}%
\let\auto@bib@innerbib\@empty
\bibitem [{\citenamefont {Achenbach}(1973)}]{S_achenbach_wave_1973}%
  \BibitemOpen
  \bibfield  {author} {\bibinfo {author} {\bibfnamefont {J.~D.}\ \bibnamefont
  {Achenbach}},\ }\href@noop {} {\emph {\bibinfo {title} {Wave {{Propagation}}
  in {{Elastic Solids}}}}}\ (\bibinfo  {publisher} {{North-Holland
  Publishing}},\ \bibinfo {address} {{Amsterdam}},\ \bibinfo {year}
  {1973})\BibitemShut {NoStop}%
\bibitem [{\citenamefont {Gurtin}(1984)}]{S_gurtin_linear_1984}%
  \BibitemOpen
  \bibfield  {author} {\bibinfo {author} {\bibfnamefont {M.~E.}\ \bibnamefont
  {Gurtin}},\ }\bibfield  {title} {\bibinfo {title} {The {{Linear Theory}} of
  {{Elasticity}}},\ }in\ \href@noop {} {\emph {\bibinfo {booktitle} {Linear
  {{Theories}} of {{Elasticity}} and {{Thermoelasticity}}, {{Linear}} and
  {{Nonlinear Theories}} of {{Rods}}, {{Plates}}, and {{Shells}}}}},\ \bibinfo
  {series} {Mechanics of {{Solids}}}, Vol.~\bibinfo {volume} {2},\ \bibinfo
  {editor} {edited by\ \bibinfo {editor} {\bibfnamefont {C.}~\bibnamefont
  {Truesdell}}}\ (\bibinfo  {publisher} {{Springer}},\ \bibinfo {address}
  {{Berlin}},\ \bibinfo {year} {1984})\BibitemShut {NoStop}%
\bibitem [{\citenamefont {{Cohen-Tannoudji}}\ \emph {et~al.}(2004)\citenamefont
  {{Cohen-Tannoudji}}, \citenamefont {{Dupont-Roc}},\ and\ \citenamefont
  {Grynberg}}]{S_cohen-tannoudji_photons_2004}%
  \BibitemOpen
  \bibfield  {author} {\bibinfo {author} {\bibfnamefont {C.}~\bibnamefont
  {{Cohen-Tannoudji}}}, \bibinfo {author} {\bibfnamefont {J.}~\bibnamefont
  {{Dupont-Roc}}},\ and\ \bibinfo {author} {\bibfnamefont {G.}~\bibnamefont
  {Grynberg}},\ }\href@noop {} {\emph {\bibinfo {title} {Photons and {{Atoms}}:
  {{Introduction}} to {{Quantum Electrodynamics}}}}}\ (\bibinfo  {publisher}
  {{Wiley-VCH}},\ \bibinfo {address} {{Weinheim}},\ \bibinfo {year}
  {2004})\BibitemShut {NoStop}%
\bibitem [{\citenamefont {Meeker}\ and\ \citenamefont
  {Meitzler}(1964)}]{S_meeker_guided_1964}%
  \BibitemOpen
  \bibfield  {author} {\bibinfo {author} {\bibfnamefont {T.~R.}\ \bibnamefont
  {Meeker}}\ and\ \bibinfo {author} {\bibfnamefont {A.~H.}\ \bibnamefont
  {Meitzler}},\ }\bibfield  {title} {\bibinfo {title} {Guided {{Wave
  Propagation}} in {{Elongated Cylinders}} and {{Plates}}},\ }in\ \href@noop {}
  {\emph {\bibinfo {booktitle} {Methods and {{Devices}}, {{Part A}}}}},\
  \bibinfo {series} {Physical {{Acoustics}}: {{Principles}} and {{Methods}}},
  Vol.\ \bibinfo {volume} {I A},\ \bibinfo {editor} {edited by\ \bibinfo
  {editor} {\bibfnamefont {W.~P.}\ \bibnamefont {Mason}}}\ (\bibinfo
  {publisher} {{Academic Press}},\ \bibinfo {address} {{New York}},\ \bibinfo
  {year} {1964})\ pp.\ \bibinfo {pages} {111--167}\BibitemShut {NoStop}%
\bibitem [{\citenamefont {Armen{\`a}kas}\ \emph {et~al.}(1969)\citenamefont
  {Armen{\`a}kas}, \citenamefont {Gazis},\ and\ \citenamefont
  {Herrmann}}]{S_armenakas_free_1969}%
  \BibitemOpen
  \bibfield  {author} {\bibinfo {author} {\bibfnamefont {A.~E.}\ \bibnamefont
  {Armen{\`a}kas}}, \bibinfo {author} {\bibfnamefont {D.~C.}\ \bibnamefont
  {Gazis}},\ and\ \bibinfo {author} {\bibfnamefont {G.}~\bibnamefont
  {Herrmann}},\ }\href@noop {} {\emph {\bibinfo {title} {Free {{Vibrations}} of
  {{Circular Cylindrical Shells}}}}}\ (\bibinfo  {publisher} {{Pergamon
  Press}},\ \bibinfo {address} {{Oxford}},\ \bibinfo {year} {1969})\BibitemShut
  {NoStop}%
\bibitem [{\citenamefont {H{\"u}mmer}\ \emph {et~al.}(2019)\citenamefont
  {H{\"u}mmer}, \citenamefont {Schneeweiss}, \citenamefont {Rauschenbeutel},\
  and\ \citenamefont {{Romero-Isart}}}]{S_hummer_heating_2019}%
  \BibitemOpen
  \bibfield  {author} {\bibinfo {author} {\bibfnamefont {D.}~\bibnamefont
  {H{\"u}mmer}}, \bibinfo {author} {\bibfnamefont {P.}~\bibnamefont
  {Schneeweiss}}, \bibinfo {author} {\bibfnamefont {A.}~\bibnamefont
  {Rauschenbeutel}},\ and\ \bibinfo {author} {\bibfnamefont {O.}~\bibnamefont
  {{Romero-Isart}}},\ }\bibfield  {title} {\bibinfo {title} {Heating in
  {{Nanophotonic Traps}} for {{Cold Atoms}}},\ }\href
  {https://doi.org/10.1103/PhysRevX.9.041034} {\bibfield  {journal} {\bibinfo
  {journal} {Phys. Rev. X}\ }\textbf {\bibinfo {volume} {9}},\ \bibinfo {pages}
  {041034} (\bibinfo {year} {2019})}\BibitemShut {NoStop}%
\bibitem [{\citenamefont {Bass}\ \emph {et~al.}(2001)\citenamefont {Bass},
  \citenamefont {Van~Stryland}, \citenamefont {Williams},\ and\ \citenamefont
  {Wolfe}}]{S_bass_handbook_2001}%
  \BibitemOpen
  \bibinfo {editor} {\bibfnamefont {M.}~\bibnamefont {Bass}}, \bibinfo {editor}
  {\bibfnamefont {E.~W.}\ \bibnamefont {Van~Stryland}}, \bibinfo {editor}
  {\bibfnamefont {D.~R.}\ \bibnamefont {Williams}},\ and\ \bibinfo {editor}
  {\bibfnamefont {W.~L.}\ \bibnamefont {Wolfe}},\ eds.,\ \href@noop {} {\emph
  {\bibinfo {title} {Handbook of {{Optics}}: {{Devices}}, Measurements, and
  Properties}}},\ \bibinfo {edition} {2nd}\ ed.,\ Vol.~\bibinfo {volume} {2}\
  (\bibinfo  {publisher} {{McGraw-Hill}},\ \bibinfo {address} {{New York}},\
  \bibinfo {year} {2001})\BibitemShut {NoStop}%
\bibitem [{\citenamefont {Glauber}\ and\ \citenamefont
  {Lewenstein}(1991)}]{S_glauber_quantum_1991}%
  \BibitemOpen
  \bibfield  {author} {\bibinfo {author} {\bibfnamefont {R.~J.}\ \bibnamefont
  {Glauber}}\ and\ \bibinfo {author} {\bibfnamefont {M.}~\bibnamefont
  {Lewenstein}},\ }\bibfield  {title} {\bibinfo {title} {Quantum optics of
  dielectric media},\ }\href {https://doi.org/10.1103/PhysRevA.43.467}
  {\bibfield  {journal} {\bibinfo  {journal} {Phys. Rev. A}\ }\textbf {\bibinfo
  {volume} {43}},\ \bibinfo {pages} {467} (\bibinfo {year} {1991})}\BibitemShut
  {NoStop}%
\bibitem [{\citenamefont {Marcuse}(1982)}]{S_marcuse_light_1982}%
  \BibitemOpen
  \bibfield  {author} {\bibinfo {author} {\bibfnamefont {D.}~\bibnamefont
  {Marcuse}},\ }\href@noop {} {\emph {\bibinfo {title} {Light {{Transmission
  Optics}}}}}\ (\bibinfo  {publisher} {{Van Nostrand Reinhold}},\ \bibinfo
  {address} {{New York}},\ \bibinfo {year} {1982})\BibitemShut {NoStop}%
\bibitem [{\citenamefont {Snyder}\ and\ \citenamefont
  {Love}(2012)}]{S_snyder_optical_2012}%
  \BibitemOpen
  \bibfield  {author} {\bibinfo {author} {\bibfnamefont {A.~W.}\ \bibnamefont
  {Snyder}}\ and\ \bibinfo {author} {\bibfnamefont {J.}~\bibnamefont {Love}},\
  }\href@noop {} {\emph {\bibinfo {title} {Optical {{Waveguide Theory}}}}}\
  (\bibinfo  {publisher} {{Springer}},\ \bibinfo {address} {{New York}},\
  \bibinfo {year} {2012})\BibitemShut {NoStop}%
\bibitem [{\citenamefont {Le~Kien}\ \emph {et~al.}(2004)\citenamefont
  {Le~Kien}, \citenamefont {Liang}, \citenamefont {Hakuta},\ and\ \citenamefont
  {Balykin}}]{S_le_kien_field_2004}%
  \BibitemOpen
  \bibfield  {author} {\bibinfo {author} {\bibfnamefont {F.}~\bibnamefont
  {Le~Kien}}, \bibinfo {author} {\bibfnamefont {J.~Q.}\ \bibnamefont {Liang}},
  \bibinfo {author} {\bibfnamefont {K.}~\bibnamefont {Hakuta}},\ and\ \bibinfo
  {author} {\bibfnamefont {V.~I.}\ \bibnamefont {Balykin}},\ }\bibfield
  {title} {\bibinfo {title} {Field intensity distributions and polarization
  orientations in a vacuum-clad subwavelength-diameter optical fiber},\ }\href
  {https://doi.org/10.1016/j.optcom.2004.08.044} {\bibfield  {journal}
  {\bibinfo  {journal} {Opt. Commun.}\ }\textbf {\bibinfo {volume} {242}},\
  \bibinfo {pages} {445} (\bibinfo {year} {2004})}\BibitemShut {NoStop}%
\bibitem [{\citenamefont {Messiah}(2014)}]{S_messiah_quantum_2014}%
  \BibitemOpen
  \bibfield  {author} {\bibinfo {author} {\bibfnamefont {A.}~\bibnamefont
  {Messiah}},\ }\href@noop {} {\emph {\bibinfo {title} {Quantum
  {{Mechanics}}}}}\ (\bibinfo  {publisher} {{Dover Publications}},\ \bibinfo
  {address} {{New York}},\ \bibinfo {year} {2014})\BibitemShut {NoStop}%
\bibitem [{\citenamefont {Cirac}\ \emph {et~al.}(1992)\citenamefont {Cirac},
  \citenamefont {Blatt}, \citenamefont {Zoller},\ and\ \citenamefont
  {Phillips}}]{S_cirac_laser_1992}%
  \BibitemOpen
  \bibfield  {author} {\bibinfo {author} {\bibfnamefont {J.~I.}\ \bibnamefont
  {Cirac}}, \bibinfo {author} {\bibfnamefont {R.}~\bibnamefont {Blatt}},
  \bibinfo {author} {\bibfnamefont {P.}~\bibnamefont {Zoller}},\ and\ \bibinfo
  {author} {\bibfnamefont {W.~D.}\ \bibnamefont {Phillips}},\ }\bibfield
  {title} {\bibinfo {title} {Laser cooling of trapped ions in a standing
  wave},\ }\href {https://doi.org/10.1103/PhysRevA.46.2668} {\bibfield
  {journal} {\bibinfo  {journal} {Phys. Rev. A}\ }\textbf {\bibinfo {volume}
  {46}},\ \bibinfo {pages} {2668} (\bibinfo {year} {1992})}\BibitemShut
  {NoStop}%
\bibitem [{\citenamefont {Breuer}\ and\ \citenamefont
  {Petruccione}(2002)}]{S_breuer_theory_2002}%
  \BibitemOpen
  \bibfield  {author} {\bibinfo {author} {\bibfnamefont {H.-P.}\ \bibnamefont
  {Breuer}}\ and\ \bibinfo {author} {\bibfnamefont {F.}~\bibnamefont
  {Petruccione}},\ }\href@noop {} {\emph {\bibinfo {title} {The {{Theory}} of
  {{Open Quantum Systems}}}}}\ (\bibinfo  {publisher} {{Oxford University
  Press}},\ \bibinfo {address} {{Oxford}},\ \bibinfo {year} {2002})\BibitemShut
  {NoStop}%
\bibitem [{\citenamefont {Wuttke}\ and\ \citenamefont
  {Rauschenbeutel}(2013)}]{S_wuttke_thermalization_2013}%
  \BibitemOpen
  \bibfield  {author} {\bibinfo {author} {\bibfnamefont {C.}~\bibnamefont
  {Wuttke}}\ and\ \bibinfo {author} {\bibfnamefont {A.}~\bibnamefont
  {Rauschenbeutel}},\ }\bibfield  {title} {\bibinfo {title} {Thermalization via
  {{Heat Radiation}} of an {{Individual Object Thinner}} than the {{Thermal
  Wavelength}}},\ }\href {https://doi.org/10.1103/PhysRevLett.111.024301}
  {\bibfield  {journal} {\bibinfo  {journal} {Phys. Rev. Lett.}\ }\textbf
  {\bibinfo {volume} {111}},\ \bibinfo {pages} {024301} (\bibinfo {year}
  {2013})}\BibitemShut {NoStop}%
\bibitem [{\citenamefont {Reitz}\ \emph {et~al.}(2013)\citenamefont {Reitz},
  \citenamefont {Sayrin}, \citenamefont {Mitsch}, \citenamefont {Schneeweiss},\
  and\ \citenamefont {Rauschenbeutel}}]{S_reitz_coherence_2013}%
  \BibitemOpen
  \bibfield  {author} {\bibinfo {author} {\bibfnamefont {D.}~\bibnamefont
  {Reitz}}, \bibinfo {author} {\bibfnamefont {C.}~\bibnamefont {Sayrin}},
  \bibinfo {author} {\bibfnamefont {R.}~\bibnamefont {Mitsch}}, \bibinfo
  {author} {\bibfnamefont {P.}~\bibnamefont {Schneeweiss}},\ and\ \bibinfo
  {author} {\bibfnamefont {A.}~\bibnamefont {Rauschenbeutel}},\ }\bibfield
  {title} {\bibinfo {title} {Coherence {{Properties}} of
  {{Nanofiber}}-{{Trapped Cesium Atoms}}},\ }\href
  {https://doi.org/10.1103/PhysRevLett.110.243603} {\bibfield  {journal}
  {\bibinfo  {journal} {Phys. Rev. Lett.}\ }\textbf {\bibinfo {volume} {110}},\
  \bibinfo {pages} {243603} (\bibinfo {year} {2013})}\BibitemShut {NoStop}%
\bibitem [{\citenamefont {Albrecht}\ \emph {et~al.}(2016)\citenamefont
  {Albrecht}, \citenamefont {Meng}, \citenamefont {Clausen}, \citenamefont
  {Dareau}, \citenamefont {Schneeweiss},\ and\ \citenamefont
  {Rauschenbeutel}}]{S_albrecht_fictitious_2016}%
  \BibitemOpen
  \bibfield  {author} {\bibinfo {author} {\bibfnamefont {B.}~\bibnamefont
  {Albrecht}}, \bibinfo {author} {\bibfnamefont {Y.}~\bibnamefont {Meng}},
  \bibinfo {author} {\bibfnamefont {C.}~\bibnamefont {Clausen}}, \bibinfo
  {author} {\bibfnamefont {A.}~\bibnamefont {Dareau}}, \bibinfo {author}
  {\bibfnamefont {P.}~\bibnamefont {Schneeweiss}},\ and\ \bibinfo {author}
  {\bibfnamefont {A.}~\bibnamefont {Rauschenbeutel}},\ }\bibfield  {title}
  {\bibinfo {title} {Fictitious magnetic-field gradients in optical microtraps
  as an experimental tool for interrogating and manipulating cold atoms},\
  }\href {https://doi.org/10.1103/PhysRevA.94.061401} {\bibfield  {journal}
  {\bibinfo  {journal} {Phys. Rev. A}\ }\textbf {\bibinfo {volume} {94}},\
  \bibinfo {pages} {061401} (\bibinfo {year} {2016})}\BibitemShut {NoStop}%
\bibitem [{\citenamefont {Cirac}\ \emph {et~al.}(1993)\citenamefont {Cirac},
  \citenamefont {Blatt}, \citenamefont {Parkins},\ and\ \citenamefont
  {Zoller}}]{S_cirac_spectrum_1993}%
  \BibitemOpen
  \bibfield  {author} {\bibinfo {author} {\bibfnamefont {J.~I.}\ \bibnamefont
  {Cirac}}, \bibinfo {author} {\bibfnamefont {R.}~\bibnamefont {Blatt}},
  \bibinfo {author} {\bibfnamefont {A.~S.}\ \bibnamefont {Parkins}},\ and\
  \bibinfo {author} {\bibfnamefont {P.}~\bibnamefont {Zoller}},\ }\bibfield
  {title} {\bibinfo {title} {Spectrum of resonance fluorescence from a single
  trapped ion},\ }\href {https://doi.org/10.1103/PhysRevA.48.2169} {\bibfield
  {journal} {\bibinfo  {journal} {Phys. Rev. A}\ }\textbf {\bibinfo {volume}
  {48}},\ \bibinfo {pages} {2169} (\bibinfo {year} {1993})}\BibitemShut
  {NoStop}%
\bibitem [{\citenamefont {Sagu{\'e}}\ \emph {et~al.}(2007)\citenamefont
  {Sagu{\'e}}, \citenamefont {Vetsch}, \citenamefont {Alt}, \citenamefont
  {Meschede},\ and\ \citenamefont {Rauschenbeutel}}]{S_sague_cold-atom_2007}%
  \BibitemOpen
  \bibfield  {author} {\bibinfo {author} {\bibfnamefont {G.}~\bibnamefont
  {Sagu{\'e}}}, \bibinfo {author} {\bibfnamefont {E.}~\bibnamefont {Vetsch}},
  \bibinfo {author} {\bibfnamefont {W.}~\bibnamefont {Alt}}, \bibinfo {author}
  {\bibfnamefont {D.}~\bibnamefont {Meschede}},\ and\ \bibinfo {author}
  {\bibfnamefont {A.}~\bibnamefont {Rauschenbeutel}},\ }\bibfield  {title}
  {\bibinfo {title} {Cold-{{Atom Physics Using Ultrathin Optical Fibers}}:
  {{Light}}-{{Induced Dipole Forces}} and {{Surface Interactions}}},\ }\href
  {https://doi.org/10.1103/PhysRevLett.99.163602} {\bibfield  {journal}
  {\bibinfo  {journal} {Phys. Rev. Lett.}\ }\textbf {\bibinfo {volume} {99}},\
  \bibinfo {pages} {163602} (\bibinfo {year} {2007})}\BibitemShut {NoStop}%
\bibitem [{\citenamefont {Patterson}\ \emph {et~al.}(2018)\citenamefont
  {Patterson}, \citenamefont {Solano}, \citenamefont {Julienne}, \citenamefont
  {Orozco},\ and\ \citenamefont {Rolston}}]{S_patterson_spectral_2018}%
  \BibitemOpen
  \bibfield  {author} {\bibinfo {author} {\bibfnamefont {B.~D.}\ \bibnamefont
  {Patterson}}, \bibinfo {author} {\bibfnamefont {P.}~\bibnamefont {Solano}},
  \bibinfo {author} {\bibfnamefont {P.~S.}\ \bibnamefont {Julienne}}, \bibinfo
  {author} {\bibfnamefont {L.~A.}\ \bibnamefont {Orozco}},\ and\ \bibinfo
  {author} {\bibfnamefont {S.~L.}\ \bibnamefont {Rolston}},\ }\bibfield
  {title} {\bibinfo {title} {Spectral asymmetry of atoms in the van der
  {{Waals}} potential of an optical nanofiber},\ }\href
  {https://doi.org/10.1103/PhysRevA.97.032509} {\bibfield  {journal} {\bibinfo
  {journal} {Phys. Rev. A}\ }\textbf {\bibinfo {volume} {97}},\ \bibinfo
  {pages} {032509} (\bibinfo {year} {2018})}\BibitemShut {NoStop}%
\bibitem [{\citenamefont {Nayak}\ \emph {et~al.}(2007)\citenamefont {Nayak},
  \citenamefont {Melentiev}, \citenamefont {Morinaga}, \citenamefont {Kien},
  \citenamefont {Balykin},\ and\ \citenamefont {Hakuta}}]{S_nayak_optical_2007}%
  \BibitemOpen
  \bibfield  {author} {\bibinfo {author} {\bibfnamefont {K.~P.}\ \bibnamefont
  {Nayak}}, \bibinfo {author} {\bibfnamefont {P.~N.}\ \bibnamefont
  {Melentiev}}, \bibinfo {author} {\bibfnamefont {M.}~\bibnamefont {Morinaga}},
  \bibinfo {author} {\bibfnamefont {F.~L.}\ \bibnamefont {Kien}}, \bibinfo
  {author} {\bibfnamefont {V.~I.}\ \bibnamefont {Balykin}},\ and\ \bibinfo
  {author} {\bibfnamefont {K.}~\bibnamefont {Hakuta}},\ }\bibfield  {title}
  {\bibinfo {title} {Optical nanofiber as an efficient tool for manipulating
  and probing atomic {{Fluorescence}}},\ }\href
  {https://doi.org/10.1364/OE.15.005431} {\bibfield  {journal} {\bibinfo
  {journal} {Opt. Express}\ }\textbf {\bibinfo {volume} {15}},\ \bibinfo
  {pages} {5431} (\bibinfo {year} {2007})}\BibitemShut {NoStop}%
\bibitem [{\citenamefont {Lindberg}(1986)}]{S_lindberg_resonance_1986}%
  \BibitemOpen
  \bibfield  {author} {\bibinfo {author} {\bibfnamefont {M.}~\bibnamefont
  {Lindberg}},\ }\bibfield  {title} {\bibinfo {title} {Resonance fluorescence
  of a laser-cooled trapped ion in the {{Lamb}}-{{Dicke}} limit},\ }\href
  {https://doi.org/10.1103/PhysRevA.34.3178} {\bibfield  {journal} {\bibinfo
  {journal} {Phys. Rev. A}\ }\textbf {\bibinfo {volume} {34}},\ \bibinfo
  {pages} {3178} (\bibinfo {year} {1986})}\BibitemShut {NoStop}%
\bibitem [{\citenamefont {{Cohen-Tannoudji}}\ \emph {et~al.}(1998)\citenamefont
  {{Cohen-Tannoudji}}, \citenamefont {{Dupont-Roc}},\ and\ \citenamefont
  {Grynberg}}]{S_cohen-tannoudji_atom-photon_1998}%
  \BibitemOpen
  \bibfield  {author} {\bibinfo {author} {\bibfnamefont {C.}~\bibnamefont
  {{Cohen-Tannoudji}}}, \bibinfo {author} {\bibfnamefont {J.}~\bibnamefont
  {{Dupont-Roc}}},\ and\ \bibinfo {author} {\bibfnamefont {G.}~\bibnamefont
  {Grynberg}},\ }\href@noop {} {\emph {\bibinfo {title} {Atom-{{Photon
  Interactions}}: {{Basic Processes}} and {{Applications}}}}}\ (\bibinfo
  {publisher} {{Wiley-VCH}},\ \bibinfo {address} {{Weinheim}},\ \bibinfo {year}
  {1998})\BibitemShut {NoStop}%
\end{thebibliography}
\end{document}